\pdfoutput=1
\documentclass[fleqn,usenatbib,useAMS]{mnras}

\usepackage{newtxtext,newtxmath}
\usepackage{amsmath,bm}
\usepackage{subcaption}
\usepackage{graphicx}
\usepackage{rotating}

\usepackage[T1]{fontenc}

\DeclareRobustCommand{\VAN}[3]{#2}
\let\VANthebibliography\thebibliography
\def\thebibliography{\DeclareRobustCommand{\VAN}[3]{##3}\VANthebibliography}

\usepackage{graphicx}	
\usepackage{amsmath}	

\title[Hydrodynamic shielding in galactic winds]{Hydrodynamic shielding in radiative multicloud outflows within multiphase galactic winds}

\author[A.~S.~Villares, W.~E.~Banda-Barrag\'{a}n, and C.~Rojas]{
A. S. Villares$^{1}$\thanks{E-mail: andres.villares@yachaytech.edu.ec (ASV)}, W.~E.~Banda-Barrag\'{a}n$^{1,2}$\thanks{E-mail: wbandabarragan@gmail.com (WEBB)}, and C.~Rojas$^{1}$
\\
$^{1}$Escuela de Ciencias F\'isicas y Nanotecnolog\'ia, Universidad Yachay Tech, Hacienda San Jos\'e S/N, 100119 Urcuqu\'i, Ecuador\\
$^{2}$Hamburger Sternwarte, University of Hamburg, Gojenbergsweg 112, 21029 Hamburg, Germany\\
}

\date{Accepted XXX. Received YYY; in original form ZZZ}

\pubyear{2024}

\begin{document}
\label{firstpage}
\pagerange{\pageref{firstpage}--\pageref{lastpage}}
\maketitle

\begin{abstract}
Stellar-driven galactic winds regulate the mass and energy content of star-forming galaxies. Emission- and absorption-line spectroscopy shows that these outflows are multiphase and comprised of dense gas clouds embedded in much hotter winds. Explaining the presence of cold gas in such environments is a challenging endeavour that requires numerical modelling. In this paper we report a set of 3D hydrodynamical simulations of supersonic winds interacting with radiative and adiabatic multicloud systems, in which clouds are placed along a stream and separated by different distances. As a complement to previous adiabatic, subsonic studies, we demonstrate that hydrodynamic shielding is also triggered in supersonic winds and operates differently in adiabatic and radiative regimes. We find that the condensation of warm, mixed gas in between clouds facilitates hydrodynamic shielding by replenishing dense gas along the stream, provided that its cooling length is shorter than the cloud radius. Small separation distances between clouds also favour hydrodynamic shielding by reducing drag forces and the extent of the mixing region around the clouds. In contrast, large separation distances promote mixing and dense gas destruction via dynamical instabilities. The transition between shielding and no-shielding scenarios across different cloud separation distances is smooth in radiative supersonic models, as opposed to their adiabatic counterparts for which clouds need to be in close proximity. Overall, hydrodynamic shielding and re-condensation are effective mechanisms for preserving cold gas in multiphase flows for several cloud-crushing times, and thus can help understand cold gas survival in galactic winds.
\end{abstract}


























\section{Introduction}

Galactic winds are complex, multifaceted outflows of matter and energy that arise in star-forming galaxies (see a recent review by \citealt{2023ARA&A..61..131F}). These winds are made of molecular, atomic, and ionised components, which despite having specific densities and temperatures, coexist along the outflow (e.g., see \citealt{2020Natur.584..364D,2021A&A...646A..66P,2023A&A...674L..15V} for the nuclear wind in our Galaxy, and \citealt{2008MNRAS.383..864W,2021MNRAS.508.4902V,2022ApJ...933..139L,2022ApJ...941..163R,2023MNRAS.525.6170M,2023ApJ...958..109L} for winds in starburst galaxies). Galactic winds are of great interest because they are one of the main mechanisms by which galaxies lose gas and metals, which in turn has a significant impact on how galaxies evolve (\citealt{2020MNRAS.493.3081R,2020A&ARv..28....2V}). Emission- and (quasar) absorption-line observations of star-forming galaxies, particularly starburst galaxies, reveal the presence of multiphase outflows within galactic winds (see a review by \citealt{2018Galax...6..138R}), in which a cold, dense gas component is embedded within much hotter and more diffuse gas phases \citep[e.g.][]{Shopbell_98, Tripp_11, Pfuhl_2015, Di_18, Salak_18,2022ApJ...936..171R}.\par


As they propagate outwards, galactic winds, launched by stellar feedback processes, encounter a diverse range of clumps with varying sizes in the interstellar and the circumgalactic media (see \citealt{2000MNRAS.314..511S,2008ApJ...674..157C,Schneider_20} for examples of simulations). Interstellar clumps are typically constituted by dense gas and dust clouds. The interaction between diffuse winds and these clouds results in significant alterations in the physical and chemical properties of both the hot winds and the cold clouds (see \citealt{2022MNRAS.511..859G,2022arXiv221015679A,2022MNRAS.515.1815P,2023MNRAS.522.4161J} for recent simulation work). For example, when the outflowing material interacts with dense clouds, the ram pressure exerted by the wind material compresses the clouds, leading to their disruption \citep[see e.g.][]{Klein_94,Bruggen_16,2019MNRAS.486.4526B}. Moreover, the net force that results from the momentum transfer of the wind material to clouds can cause acceleration of the clouds \citep[see][]{2004ApJ...604...74F,Cottle_20}. This process acts upon the upstream side of the cloud, pushing dense material downstream and leading to its fragmentation \citep{Gregori_2000,Schneider_17,2018MNRAS.473.5407M}.\par

Considering the significance of galactic winds for the ecology of the interstellar medium (ISM) and the circumgalactic medium (CGM), and for the global evolution of galaxies, it is crucial to gain a comprehensive understanding of the underlying physical processes that shape them (\citealt{Tumlinson_11}). Numerical simulations are essential tools to better understand the observations and the physical characteristics of the gas in such outflows. Indeed, a long history of numerical simulation studies of galactic winds exists. These studies encompass a wide range of hydrodynamic and magnetohydrodynamic models with different resolutions and domain sizes. Depending on the length scales that are targeted by the models, simulations can be classified in wind-cloud (e.g., \citealt{Schneider_17,2020MNRAS.499.4261S}), wind-multicloud (e.g., \citealt{Aluzas_12,Banda_20}), wind-launching (e.g,. \citealt{2022ApJ...936..133C,2023MNRAS.522.1843R}), and disc-wind (e.g., \citealt{Schneider_20,2021ApJ...913...68Z}) models.\par 

Wind-cloud simulations have demonstrated that the hot wind has the ability to erode and remove material from the clouds, eventually leading to their destruction via dynamical instabilities \citep[see][]{Nakamura_2006, Banda_16}. These models show that clouds can be easily destroyed in galactic wind environments. This occurs not only in adiabatic or inefficient-cooling scenarios (\citealt{2016MNRAS.457.4470P,2017MNRAS.470.2427G}), but also in models for which cooling is important (i.e. models for which the cooling time is smaller than the cloud-crushing time, e.g. \citealt{Cooper_2009,Schneider_17}). The fact that clouds are destroyed (before they can reach high speeds) over a wide range of initial conditions has posed some tension between simulations and astronomical observations (\citealt{2017MNRAS.468.4801Z}), which show the co-existence of cold clouds at very large distances from the galactic planes (e.g., see \citealt{2015ApJ...813...46B,2022ApJ...936..171R,2022ApJ...927..147T} and a review by \citealt{2018Galax...6..138R}). Recently, however, \cite{2018MNRAS.480L.111G,2020MNRAS.492.1970G} identified a region of the parameter space of wind-cloud interactions that facilitate cloud survival and even promote dense-gas mass growth. When the cooling time of the mixed warm gas (which forms via cloud erosion) is smaller than the cloud-crushing time, fast-moving cold gas reforms from the mixed phase (see also \citealt{2010MNRAS.404.1464M,2020MNRAS.492.1841L,2021MNRAS.501.1143K}).\par

While small-scale wind-cloud simulations have played an important role in understanding the entrainment of cold gas in hot ambient winds, they are still idealised models as they consider clouds as isolated entities. In reality, ISM and CGM clouds reside in multi-phase outflows in which inter-cloud interactions are also expected (e.g. see \citealt{2013ApJ...770L...4M,2018NatAs...2..901M,2023A&A...674L..15V} for our Galaxy's galactic wind, where individual outflowing clouds can be somewhat resolved). Can such interactions help explain the presence of dense cold gas in galactic winds? As we show in this paper, the answer is yes and hydrodynamic shielding is responsible for aiding dense gas survival. Hydrodynamic shielding is the process by which multiple clouds along a gas stream mutually protect each other from disruption via drag forces. Previous wind-multicloud simulations of groups of adiabatic clouds placed along a stream or part of turbulent density fields have demonstrated that hydrodynamical shielding can be an effective mechanism for extending their lifespan \citep[see][]{Forbes_2019, Banda_20, Aluzas_12}. In particular, \citep{Forbes_2019} studied adiabatic wind-multicloud systems with subsonic and transonic winds, showing that clouds can shield themselves provided that the separation distance between them is initially small.\par

The effects of hydrodynamic shielding has also been spotted in adiabatic simulations by \citep{Banda_20} in the form of warm shells of gas that result from inter-cloud interactions. Radiative wind-multicloud simulations (e.g. \citealt{Banda_21}), large-scale wind-launching models (e.g. \citealt{2020ApJ...903L..34K}), and disc-wind simulations (e.g. \citealt{Schneider_18}) also capture the overall effects of inter-cloud interactions, but the numerical resolution and the complexity of the resulting outflows makes it hard to isolate the effects of hydrodynamic shielding in all these models. In addition, radiative wind-multicloud scenarios with supersonic winds have not been characterised before. Thus, in this paper we assess the ability of systems of identical clouds, travelling along a straight trajectory with an initial separation distance ($\delta$, in cloud radius units) between them, to protect themselves (i.e. show signs of hydrodynamic shielding) against hydrodynamic drag and dynamical instabilities arising from their interactions with a hot supersonic wind gas. We explore the differences that arise when radiative cooling and supersonic winds are included in the simulations, compared to their subsonic and adiabatic counterparts.\par

This paper is structured as follows: Section \ref{Sec2} outlines the methodology and initial conditions employed in our simulations, including the description of the software and computational tools used for simulations and data analysis. Section \ref{Sec3} presents the results obtained and provides a comprehensive discussion on hydrodynamic shielding. Section \ref{sec:discussion} discusses the implications of our work, the effects of numerical resolution, and the limitations of our work. Finally, Section \ref{Sec5} summarises the key findings of this paper and provides final remarks.





\section{Methods and Simulations}\label{Sec2}

\subsection{Simulation code}
We investigate the interaction between streams of dense gas clouds embedded in a diffuse galactic wind. Our simulations were conducted using the PLUTO v4.3 code \citep[][]{Mignone_2007}. The simulations were carried out by simultaneously solving the ideal hydrodynamic (HD) equations in a three-dimensional Cartesian coordinate system $(X, Y, Z)$. The mass, momentum, and energy conservation laws are:
\begin{equation}
    \frac{\partial \rho}{\partial t} + \nabla \cdot \left[ \rho \bm{v} \right] = 0,
\end{equation}

\begin{equation}
    \frac{\partial  \left[ \rho \bm{v} \right]}{\partial t} + \nabla \cdot\left[ \rho \bm{v}\bm{v}+ \bm{I}P \right]=0,
\end{equation}

\begin{equation}\label{eq:energy_cons}
    \frac{\partial E}{\partial t} + \nabla \cdot\left[ (E+P) \bm{v} ) \right]= \Lambda,
\end{equation}

where $\rho=\mu m_{u} n$ is the mass density, $\mu$ is the mean particle mass, $m_{u}$ is the atomic mass unit, $n$ is the gas number density, $\bm{v}$ is the velocity, $P=(\gamma-1)\rho \epsilon$ is the ideal gas thermal pressure, $\gamma = \frac{5}{3}$ is the polytropic index, $E=\rho\epsilon+\frac{1}{2}\rho\bm{v^2}$ is the total energy density, $\epsilon$ represents the specific internal energy of the gas, and $\Lambda$ is the volumetric cooling rate. In addition, we include the additional advection equation of the form:
\begin{equation}
    \frac{\partial \rho C}{\partial t} + \nabla  \cdot \left[ \rho C \bm{v} \right] = 0,
\end{equation}

where $C$ is a Lagrangian scalar, which we used to track the evolution of gas initially contained in cloud material, i.e., $C=1$ for the gas inside the multi-cloud system and $C=0$ everywhere else.\par

The simulations were performed using the HLLC approximate Riemann solver \citep[][]{toro1994}, jointly with a third-order Runge-Kutta method (RK3) as the time-marching algorithm and a second-order parabolic method for spatial reconstruction. Together these algorithms constitute a robust numerical scheme that adequately captures shocks and discontinuities in compressible flows \citep[see][]{Mignone_14} and provides a good balance between accuracy and computational cost. Also, to ensure the stability of the numerical simulations, we set the Courant-Friedrichs-Lewy (CFL) number to 0.33.


\subsection{Computational set-up}
\begin{figure}
\begin{center}
  \begin{tabular}{c}
       \hspace{-0.7cm}\resizebox{!}{67mm}{\includegraphics{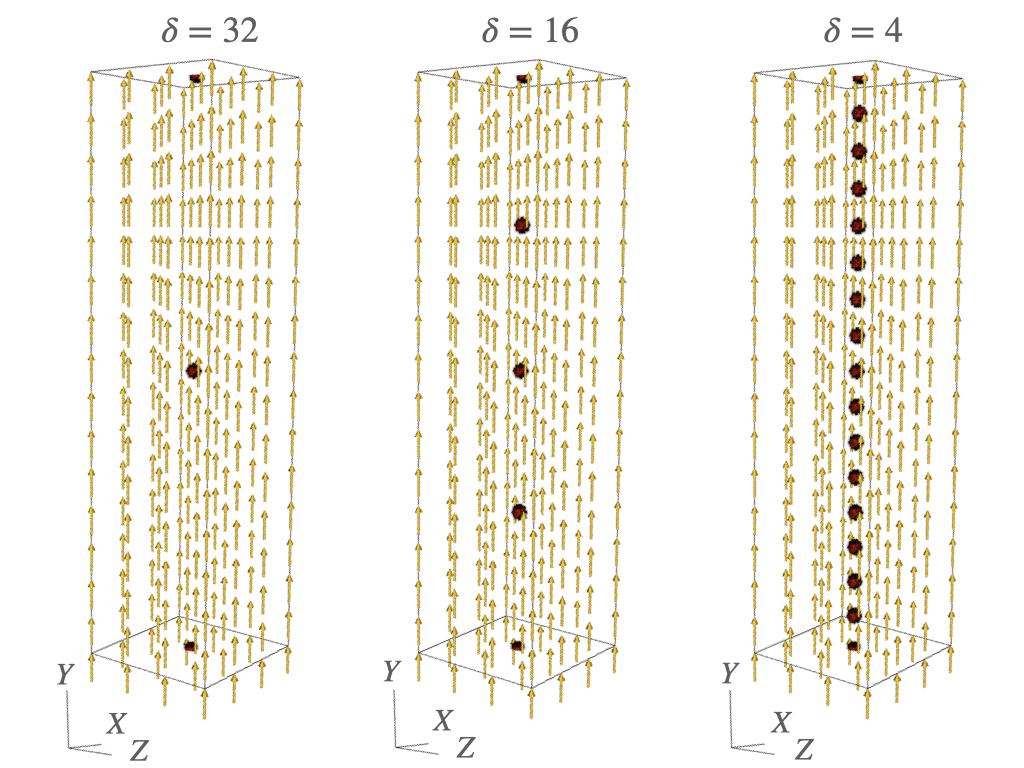}} 
  \end{tabular}
  \caption{Three-dimensional setups of three of our wind-multicloud models featuring the initial positions of a collection of spherical clouds, each separated from one another by a uniform distance of $\delta$ = 32, 16, and 4, respectively. The arrows represent the uniform velocity field of the galactic wind.} 
  \label{figics}
\end{center}
\end{figure}

Our study involves the simulation of an idealised two-phase medium, composed of spherical cold clouds and a hot ambient wind (see Figure \ref{figics}). Following \citep{Forbes_2019}, the clouds are arranged along a stream, surrounded by a hot supersonic wind with a uniform velocity field pointing in the $Y$ direction. The simulation domain is a rectangular prism with a $1:4:1$ aspect ratio for the $X:Y:Z$ axes. In physical units, the domain has a volume of $(100\times 400\times 100)\,\rm pc^{3}$. To keep all the cloud gas inside the computational domain during the entire duration of the simulations and prevent biases due to material leaving it, we set up periodic boundary conditions on all sides of the computational domain.\par

\subsection{Models}\label{models}

This study comprises a total of $20$ models aimed at investigating the impact of hydrodynamic shielding in a multicloud gas stream (see Table~\ref{table_simul}). All simulations cover a time-scale of $t_{\rm sim} \sim 5\,\rm Myr$, and include spherical clouds with identical radii $r_{\rm cl}=6.25\,\rm pc$, density $\rho_{\rm cl}=1.11 \times 10^{-24}\,\rm g\,cm^{-3}$, number density $n_{\rm cl}=1\rm cm^{-3}$, and temperature $T_{\rm cl}= 10^{4}\, \rm K$, placed in a row along the $Y$ direction. Depending on the model, the centres of the clouds are separated by different distances, $d_{\rm sep}$, which we report using a dimensionless parameter $\delta=d_{\rm sep}/r_{\rm cl}$ (see column 8 of Table \ref{table_simul} and \citealt{Forbes_2019}). In all cases the wind properties are fixed according to the CC85 model (\citealt{Chevalier}) and assuming the multicloud systems are at distances of $ \rm r ^{*}\sim 2\, \rm kpc$ from the central starburst (see Appendix \ref{a1}). Thus, the wind has a density of $\rho_{\rm wind}=1.11 \times 10^{-26}\,\rm g\,cm^{-3}$, number density $n_{\rm wind}=0.01 \rm cm^{-3}$, temperature $T_{\rm wind}=10^{6}\, \rm K$ and Mach number ${\cal M}_{\rm wind}=v_{\rm wind}/c_{\rm wind}=3.5$, where the wind speed is $v_{\rm wind}=500\,\rm km\,s^{-1}$ and its sound speed is $c_{\rm wind}=143.8\,\rm km\,s^{-1}$. Initially, the cloud-to-wind density contrast is $\chi=10^2$, and both the wind and the multicloud system are set in pressure equilibrium at $P/k_{\rm b}= 10^{4}\,\rm K\,cm^{-3}$.\par

In terms of numerical resolution, we have standard-resolution models with $(256\times 1024 \times 256)$ cells along $(X\times Y \times Z)$, i.e., with $16$ cells covering a cloud radius ($R_{16}$), high-resolution models with $32$ cells covering a cloud radius ($R_{32}$), and low-resolution models with $8$ cells covering a cloud radius ($R_{8}$). Our control adiabatic and radiative runs have $\delta=16$. In the first part of the study (presented in Section \ref{Sec4.1}), we compare our control runs to study the overall evolution of wind-multicloud systems and the effects of radiative cooling. In the second part of our study (presented in Section \ref{Sec4.2}), we compare 5 adiabatic and 5 radiative numerical simulations at $R_{16}$ with different cloud separation distances ($\delta$) to study the effects of hydrodynamic shielding and how $\delta$ influences the evolution of cold clouds. In the last part of the study (see Section \ref{convergence1}), we contrast the evolution of 4 adiabatic and 4 radiative simulations with higher and lower resolutions to study numerical convergence in both scenarios.\par

\begin{table*}
\caption{Overview of the simulation parameters chosen for different models. Columns 1 and 2 respectively indicate the type of thermodynamic model and the numerical resolution in units of $X$ number of cells per cloud radius with the standard notation, ($R_{x}$). Columns 3, 4, 5, and 6 indicate the number of cells in the 3D domain, the size of the domain in physical units, the density contrast between the clouds and the wind, and the adiabatic index, respectively. Column 7 indicates the wind Mach number, ${\cal M}_{\rm wind}=v_{\rm wind}/c_{\rm wind}$. Columns 8 and 9 indicate the cooling floor and the initial separation distance between clouds given by the dimensionless parameter $\delta=d_{\rm sep}/r_{\rm cloud}$. The fixed parameters for all models include the cloud number density $n_{\rm cl}=1\rm cm^{-3}$, cloud temperature $T_{\rm cl}= 10^{4}\, \rm K$, wind number density $n_{\rm wind}=0.01 \rm cm^{-3}$, wind temperature $T_{\rm wind}=10^{6}\, \rm K$, wind speed $v_{\rm wind}=500\,\rm km\,s^{-1}$, and its sound speed $c_{\rm wind}=143.8\,\rm km\,s^{-1}$.} 
\centering 
\begin{tabular}{c c c c c c c c c} 
\hline 
(1) & (2) & (3) & (4) & (5) & (6)  & (7) & (8) & (9)\\ [0.5ex] 
Model & Resolution & Number of cells &  Computational domain $[\text{pc}^{3}]$ & $\chi$ & $\gamma$ & $\mathcal{M}_{\rm wind}$ & Cooling floor & $\delta$\\ [0.5ex] 
\hline 
Adiabatic & $R_{16}$ & $(256\times1024\times 256)$ & $(100 \times 400 \times 100)$ & $10^{2}$ & 5/3 & 3.5  & - & 2, 4, 8, 16, 32, 64\\
Radiative & $R_{16}$ & $(256\times1024\times 256)$ & $(100 \times 400 \times 100)$ & $10^{2}$ & 5/3 & 3.5  & $10^4$ & 2, 4, 8, 16, 32, 64\\ [0.5ex]
Adiabatic &  $R_{32}$ &$(512\times2048\times 512)$ & $(100 \times 400 \times 100)$ & $10^{2}$ & 5/3 & 3.5 & - & 8, 16\\
Adiabatic & $R_{8}$ &  $(128\times512\times 128)$ & $(100 \times 400 \times 100)$ & $10^{2}$ & 5/3 & 3.5 & - & 8, 16 \\ [0.5ex]
Radiative &  $R_{32}$ &$(512\times2048\times 512)$ & $(100 \times 400 \times 100)$ & $10^{2}$ & 5/3 & 3.5 & $10^4$ & 8, 16\\
Radiative & $R_{8}$ &  $(128\times512\times 128)$ & $(100 \times 400 \times 100)$ & $10^{2}$ & 5/3 & 3.5 & $10^4$ & 8, 16 \\
\hline 
\end{tabular} \par
\bigskip
\label{table_simul} 
\end{table*}

\subsection{Radiative cooling}

We study adiabatic models ($\Lambda = 0$ in equation \ref{eq:energy_cons}) and radiative models ($\Lambda = n^2\,\Tilde{\Lambda}$ in equation \ref{eq:energy_cons}). In the latter, the cooling rates, $\Tilde{\Lambda}$, are read and interpolated from a table (\citealt{2008A&A...488..429T}), pre-compiled using the \textsc{CLOUDY} code (\citealt{1998PASP..110..761F}) for a solar mix of atomic gas at redshift zero. To prevent gas from cooling below $10^4\,\rm K$, we impose a cooling floor at that temperature. This approach mimics the effects of heating processes in the CGM below $10^4\,\rm K$, and also allows us to compare our results with previous studies for which similar (e.g., \citealt{Bruggen_16,2018MNRAS.480L.111G,Casavecchia_23}) or slightly lower (e.g., \citealt{Schneider_17, Banda_21,Casavecchia_23}) cooling floors are imposed.

\subsection{Diagnostics}

To study the evolution of a set of clouds in a wind-multicloud model, several diagnostics are calculated from the simulated data. Following \citep{Banda_16}, the mass-weighted average of any variable $\mathcal{G}$ is computed by:
\begin{equation}\label{masswe}
    \langle \mathcal{G} \rangle =  \frac{\int \mathcal{G} \rho C dV}{M_{\rm cl}} = \frac{\int \mathcal{G} \rho C dV}{\int \rho C dV},
\end{equation}

\noindent where $V$ is the volume, $C$ is the tracer of cloud material, and  $\text{M}_{\rm cl}$ is the time-dependent cloud mass. In a similar way, the volume-weighted average of any variable $\mathcal{F}$ is given by:
\begin{equation}\label{vowe}
    \left[\mathcal{F}\right]  =  \frac{\int \mathcal{F} C dV}{V_{\rm cl}} = \frac{\int \mathcal{F} C dV}{\int C dV},
\end{equation}

\noindent where $V_{\rm cl}$ is the time-dependent cloud volume.\par

The degree of mixing between the cloud and the surrounding wind is defined as:
\begin{equation}\label{mixfrac}
    f_{\rm mix} = \frac{\int \rho\,C^{*}\,dV}{M_{\rm cl,0}},
\end{equation}

\noindent where the numerator represents the mass of mixed gas with $C^{*}=C$ when $0.01<C<0.99$, and $M_{\rm cl,0}$ represents the mass of the cloud material at time $t=0$ (\citealt{1995ApJ...454..172X}). Additionally, we define the dense-gas mass fraction \citep[][]{Heyer_2022} as the ratio of cloud mass with densities greater than a threshold density, $\rho_{\rm th}$, to the total cloud mass:
\begin{equation}\label{densefrac}
    f_{\rm dense} = \frac{M_{\rm cl}(\rho_{\rm cl}>\rho_{\rm th})}{M_{\rm cl, 0}},
\end{equation}


\noindent where $\rho_{\rm th}=\rho_{\rm cl,0}/2$ to focus on the cloud material with densities higher than half the initial cloud density. For comparison, we also compute the mass of cold gas, which is defined as the mass of material whose temperature is below $10\,T_{\rm cl, 0}$, where $T_{\rm cl,0}$ represents the initial temperature of the cloud:
\begin{equation}\label{coldgassfrac}
    f_{\rm cold} =  \frac{M_{\rm cl}(T_{\rm cl}<10\,T_{\rm cl,0})}{M_{\rm cl, 0}},
\end{equation}

This fraction corresponds to a temperature in between the initial temperatures of the cloud and the wind \citep[][]{Forbes_2019}, since the simulation is initialised in thermal pressure equilibrium with a density contrast of $\chi = 10^{2}$.\par

We introduce the concept of cloud-crossing time \citep[][]{Forbes_2019}, which represents the duration for the flow around the cloud to cross one cloud radius. This is defined as follows:
\begin{equation}
t_{\rm cross} = \frac{r_{\rm cl}}{v_{\rm wind}}=12.2\,\rm kyr,
\end{equation}

\noindent where $r_{\rm cl}$ represents the radius of the cloud, while $v_{\rm wind}$ represents the initial relative velocity of the wind with respect to the cloud. The simulation time is then $t_{\rm sim}=400 \,t_{\rm cross}$. Another relevant time scale is the cloud-crushing time, defined in \cite*{Klein_94} as:
\begin{equation}
t_{\rm cc}=\frac{\chi^{\frac{1}{2}}\,r_{\rm cl}}{{\cal M}_{\rm wind} c_{\rm wind}}=122.2\,\rm kyr,
\label{eq:CloudCrushing}
\end{equation}

\noindent which indicates the time it takes for an internal transmitted shock to move across the cloud at speed $v_{\rm ts}={\cal M_{\rm wind}} c_{\rm wind}/\chi^{0.5}$. It is the relevant dynamical time-scale to study cloud disruption.


In addition, we define the cloud cooling time, which characterises the time required for a system to dissipate its thermal energy due to radiative cooling. This time-scale is calculated as follows:
\begin{equation}
    t_{\rm cool,cl}\,=\,\frac{3\,k_{\rm B}\rm T_{\rm cl}}{2\,n_{\rm cl}\,\Tilde{\Lambda}}=8.3\,\rm kyr.
\label{eq:CoolTime}
\end{equation}

Similarly, we estimate the mixed-gas cooling time as defined in \citep{2018MNRAS.480L.111G}:
\begin{equation}
    t_{\rm cool,mix}=\chi\frac{\Tilde{\Lambda}(T_{\rm cl})}{\Tilde{\Lambda}(T_{\rm mix})}\,t_{\rm cool,cl}=9.4\,\rm kyr,
\label{eq:MixedCoolTime}
\end{equation}

\noindent where $T_{\rm mix}=(T_{\rm wind}\,T_{\rm cl})^{0.5}=10^5\,\rm K$ in the temperature of the mixed warm gas. From $t_{\rm cool,mix}$, we can define the cloud cooling length following \citep{2010ApJ...722..412Y,2013ApJ...766...45J}:
\begin{equation}
\ell_{\rm cool, cl}=v_{\rm ts}\,t_{\rm cool, cl}=0.43\,\rm pc,
\label{eq:CloudCoolingLength}
\end{equation}

\noindent and in a similar fashion the mixed-gas cooling length, $\ell_{\rm cool, mix}=v_{\rm ts}\,t_{\rm cool, mix}=0.48\,\rm pc$, which together result in a global cooling length of cold and warm gas of $\ell_{\rm cool}=\ell_{\rm cool, cl}+\ell_{\rm cool, mix}=0.91\,\rm pc$.

\section{Results}\label{Sec3}

\subsection{Evolution of wind-multicloud systems}\label{Sec4.1}

\begin{figure*}
\begin{center}
  \begin{tabular}{c c c c c c}
       \multicolumn{6}{l}{\hspace{-2mm}(a) Adiabatic ($\delta=16$)}\\
       \multicolumn{1}{c}{$t=0$} & \multicolumn{1}{c}{$32.7\,t_{\rm cross}=0.4\,\rm Myr$} & \multicolumn{1}{c}{$89.9\,t_{\rm cross}=1.1\,\rm Myr$} & \multicolumn{1}{c}{$155.3\,t_{\rm cross}=1.9\,\rm Myr$} & \multicolumn{1}{c}{$212.5\,t_{\rm cross}=2.6\,\rm Myr$} & \\   
       \hspace{-0.3cm}\resizebox{!}{87mm}{\includegraphics{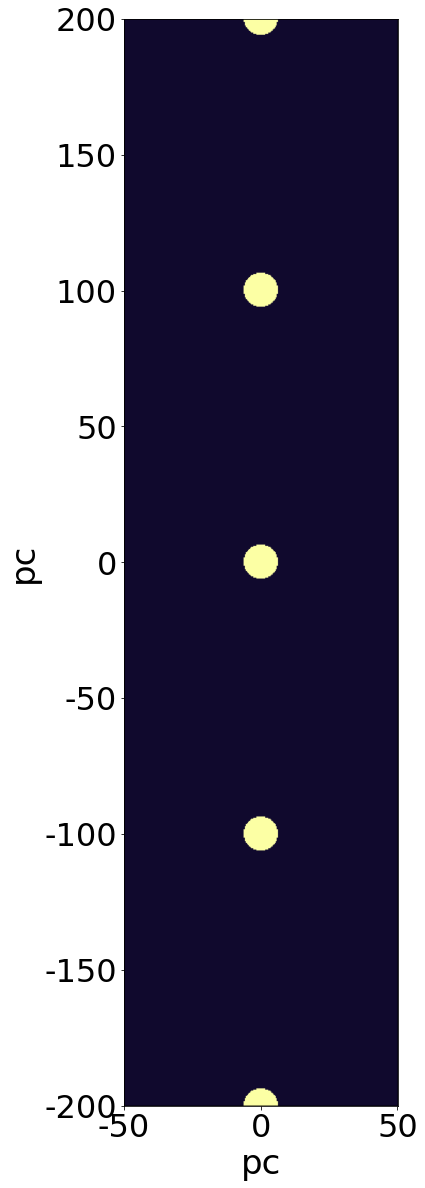}} & \hspace{-0.3cm}\resizebox{!}{87mm}{\includegraphics{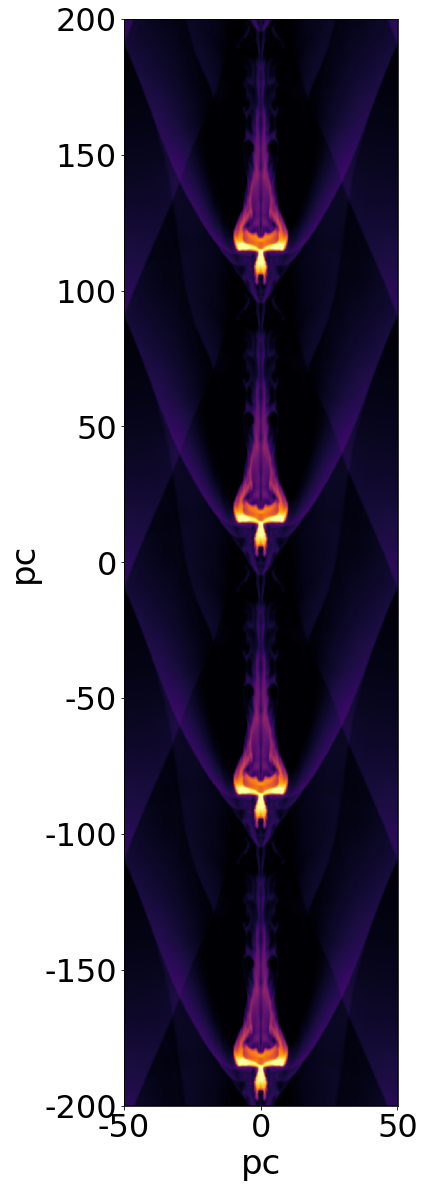}} & \hspace{-0.3cm}\resizebox{!}{87mm}{\includegraphics{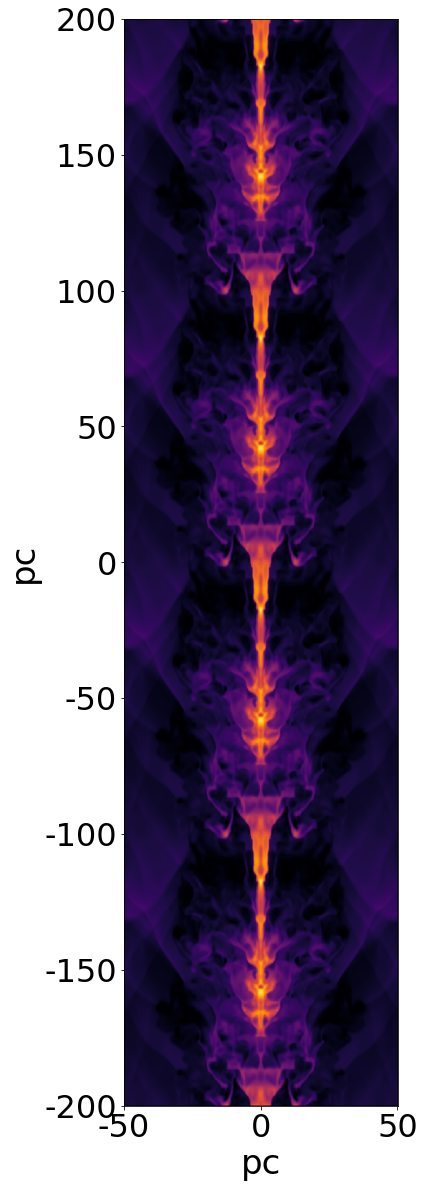}} & \hspace{-0.3cm}\resizebox{!}{87mm}{\includegraphics{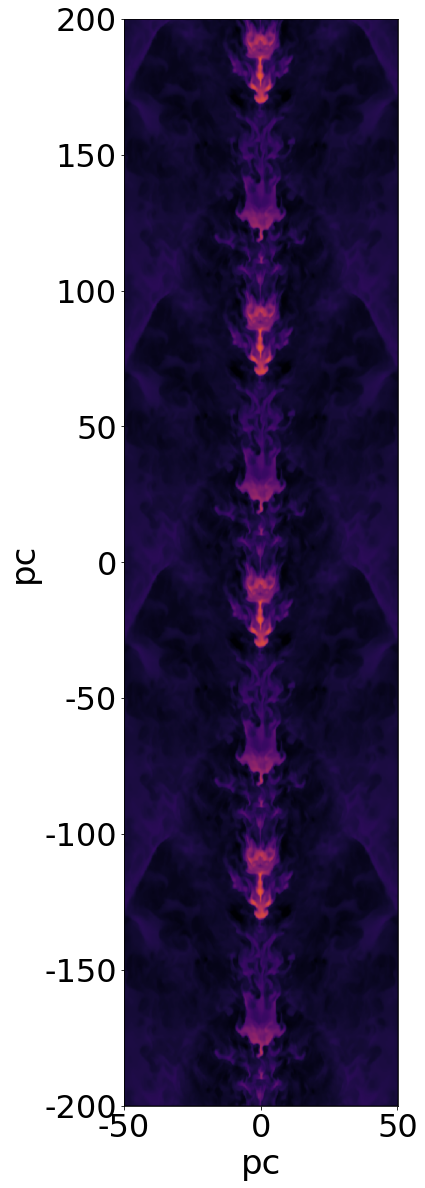}} & \hspace{-0.3cm}\resizebox{!}{87mm}{\includegraphics{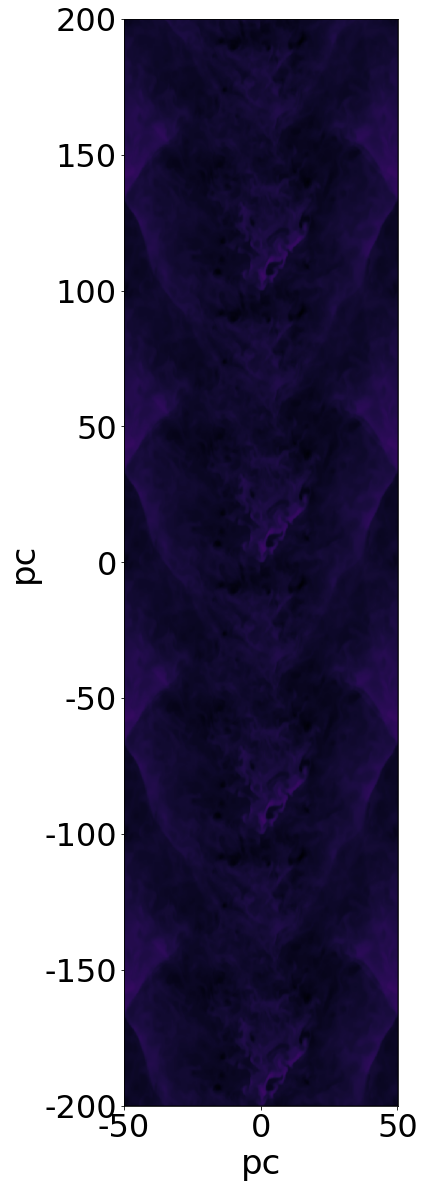}} \\
       \multicolumn{6}{l}{\hspace{-2mm}(b) Radiative ($\delta=16$)}\\
       \multicolumn{1}{c}{$t=0$} & \multicolumn{1}{c}{$32.7\,t_{\rm cross}=0.4\,\rm Myr$} & \multicolumn{1}{c}{$89.9\,t_{\rm cross}=1.1\,\rm Myr$} & \multicolumn{1}{c}{$155.3\,t_{\rm cross}=1.9\,\rm Myr$} & \multicolumn{1}{c}{$212.5\,t_{\rm cross}=2.6\,\rm Myr$} & \\     
       \hspace{-0.3cm}\resizebox{!}{87mm}{\includegraphics{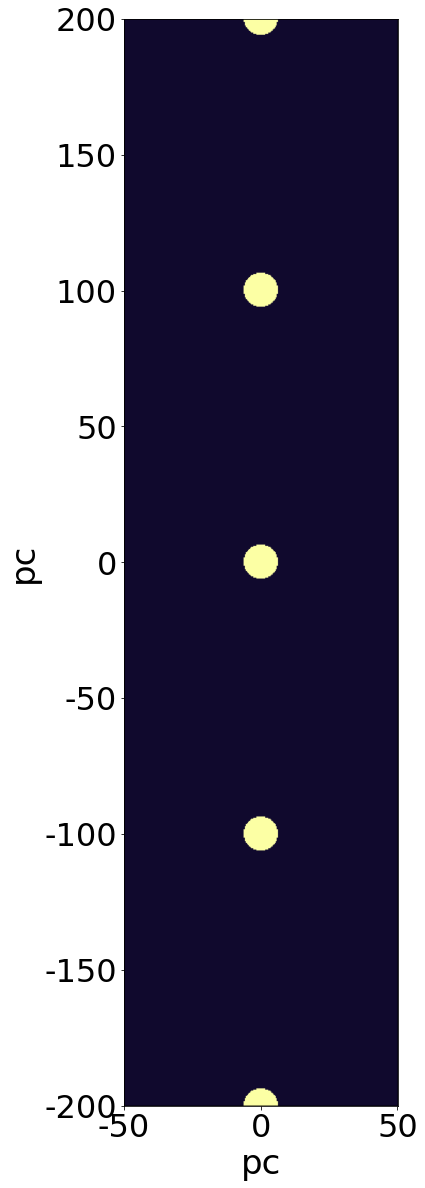}} & \hspace{-0.3cm}\resizebox{!}{87mm}{\includegraphics{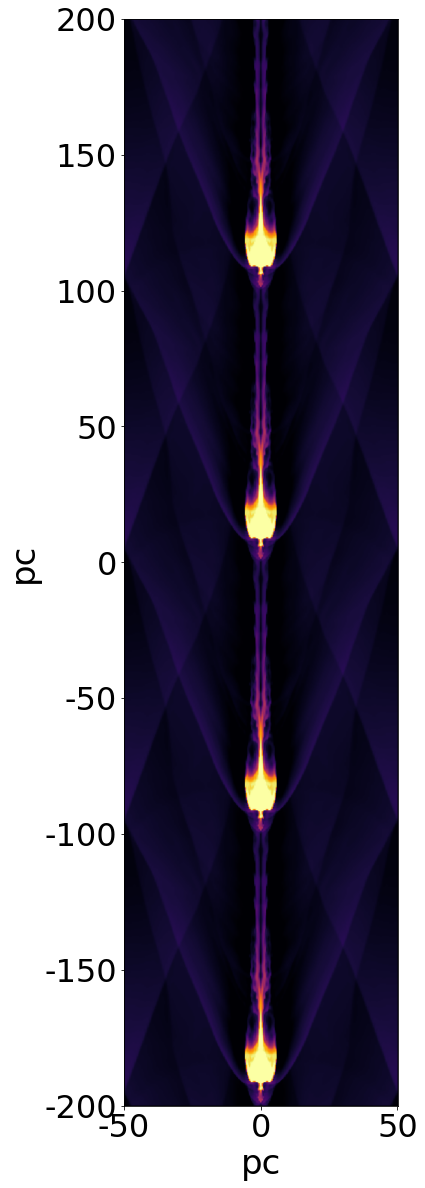}} & \hspace{-0.3cm}\resizebox{!}{87mm}{\includegraphics{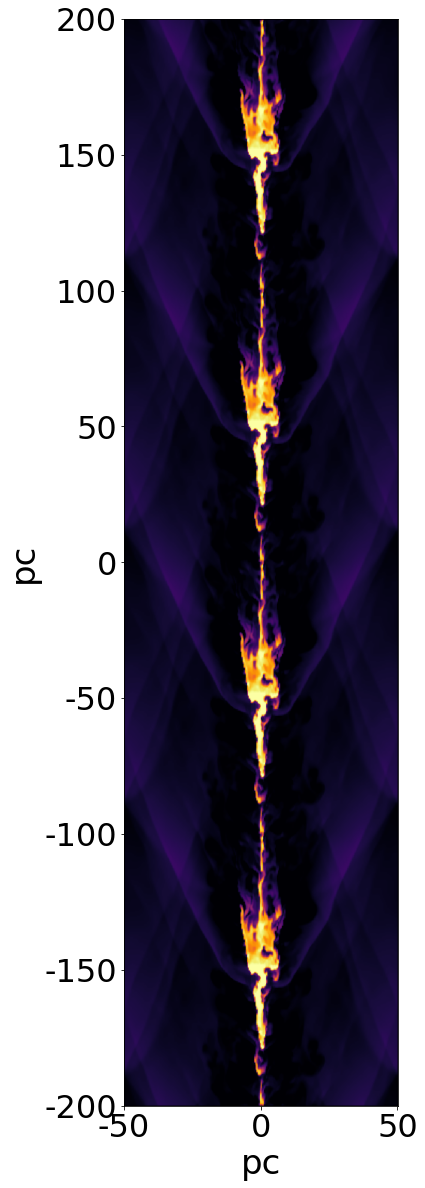}} & \hspace{-0.3cm}\resizebox{!}{87mm}{\includegraphics{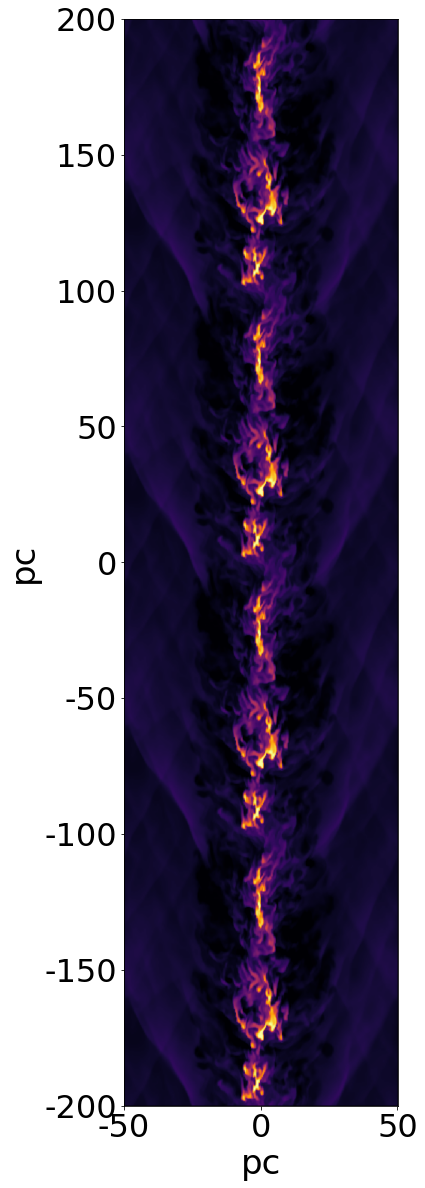}} & \hspace{-0.3cm}\resizebox{!}{87mm}{\includegraphics{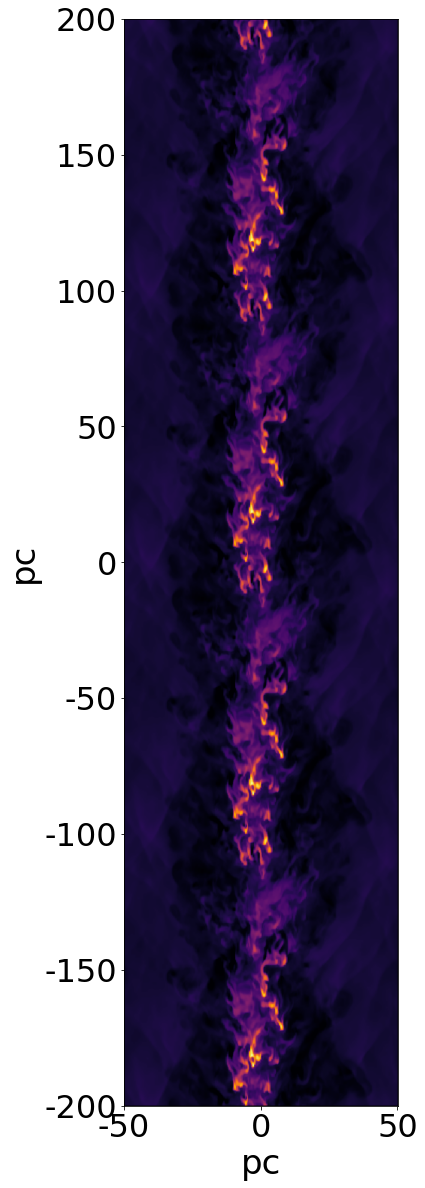}}\\
  \end{tabular}
  \begin{tabular}{c@{\hspace{-.035cm}}c}
  \resizebox{95mm}{!}{\includegraphics{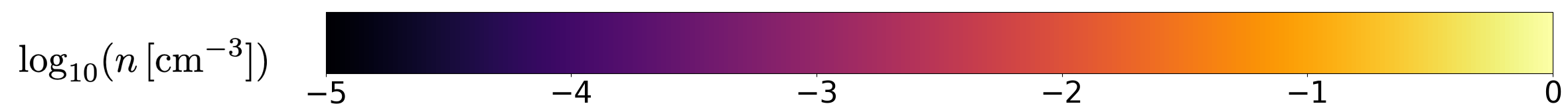}}
  \end{tabular}
  \caption{2D slices at $Z=0$ of the gas number density in the adiabatic (top panel a) and radiative (bottom panel b) control models at five different times (shown in columns) throughout the simulation. The evolution of the models begins almost identically with the formation of a bow shock as the wind interacts with the front surface of each cloud. The subsequent injection of thermal energy in the adiabatic model increases the temperature of the cloud gas and contributes to the expansion and mixing of cloud and wind gas. Without radiative cooling, the cold gas in the adiabatic model is heated up and disrupted by instabilities much faster compared to that in the radiative model.}
  \label{numden1}
\end{center}
\end{figure*}

Figure \ref{numden1} shows 2D slices at $Z=0$ of the gas number density. These panels correspond to our control adiabatic (top panel a) and radiative (bottom panel b) models with the standard separation distance, $\delta = 16$. The overall evolution of this pair of models, which is also relevant for the others, can be characterised by the following stages:

\begin{figure*}
\begin{center}
  \begin{tabular}{r r r}
  \multicolumn{1}{l}{(a) Cloud gas temperature} & \multicolumn{1}{l}{(b) Cloud gas density} & \multicolumn{1}{l}{(c) Dense gas mass fraction}  \\ 
       \hspace{-0.3cm}\resizebox{!}{44mm}{\includegraphics{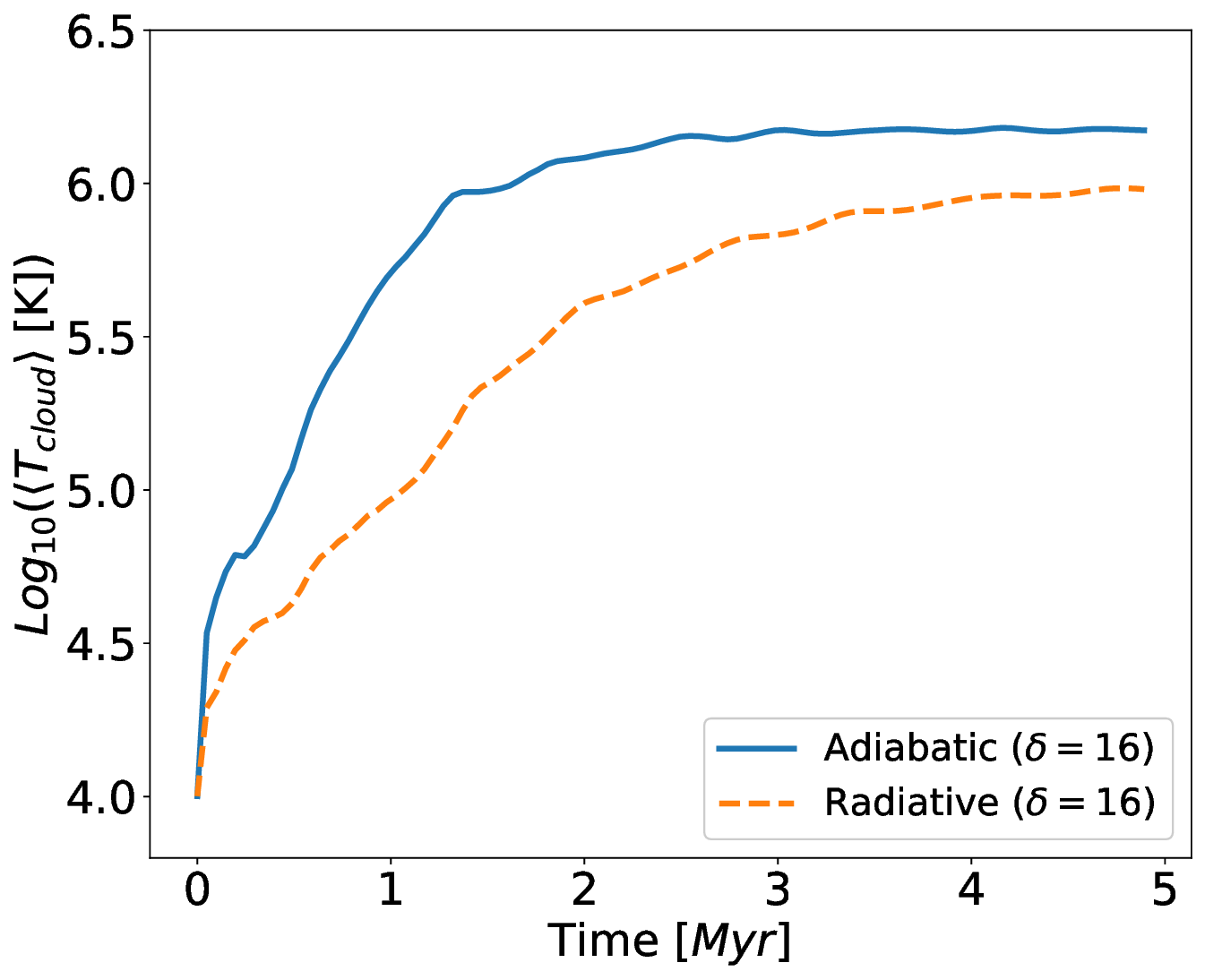}} &
       \hspace{-0.3cm}\resizebox{!}{44mm}{\includegraphics{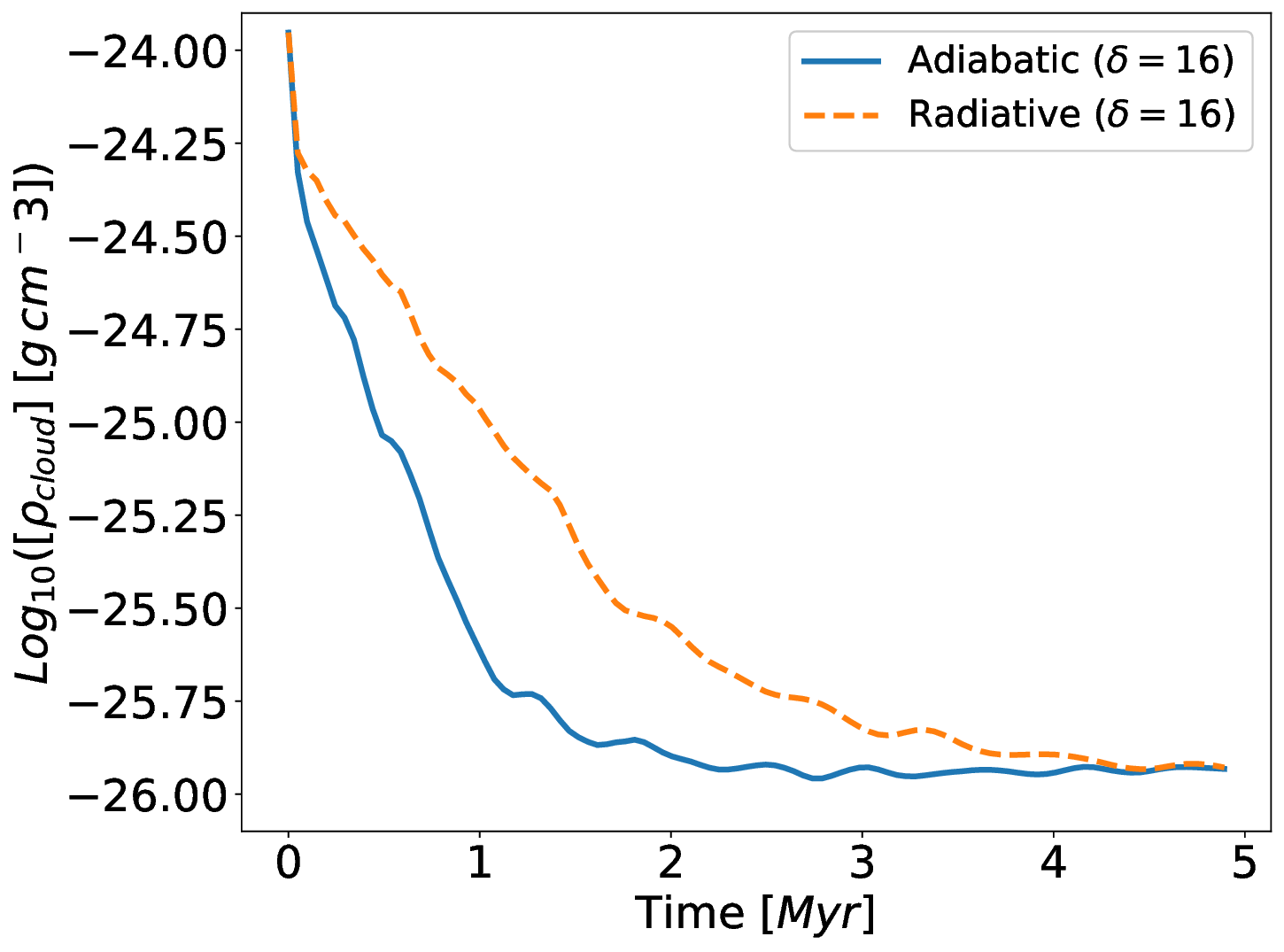}} &
       \hspace{-0.3cm}\resizebox{!}{44mm}{\includegraphics{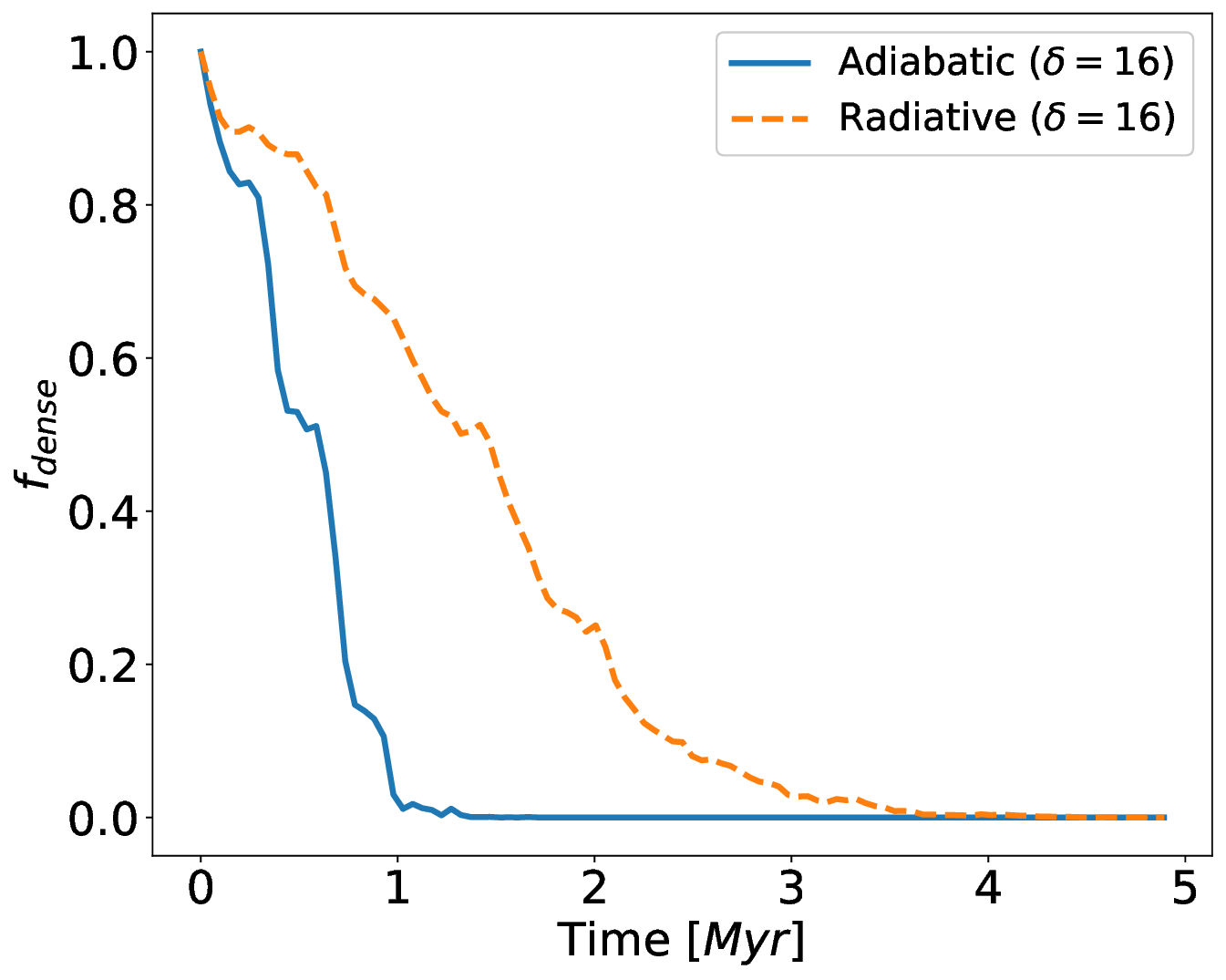}}\\
       \multicolumn{1}{l}{(d) Mixing fraction} & \multicolumn{1}{l}{ (e) Cold gas mass fraction} & \multicolumn{1}{l}{(f) Cloud gas velocity}  \\ 
       \hspace{-0.3cm}\resizebox{!}{44mm}{\includegraphics{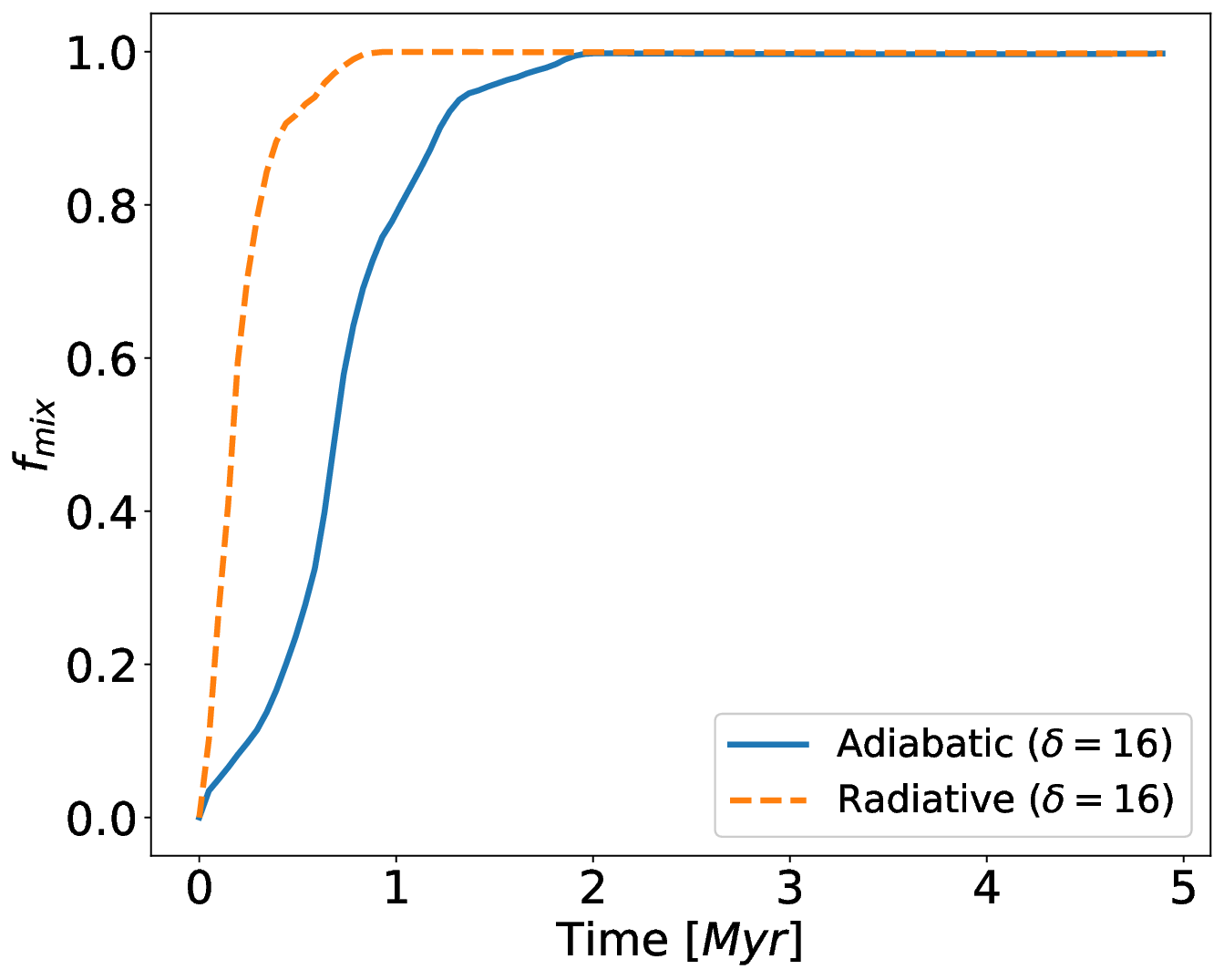}} & \hspace{-0.3cm}\resizebox{!}{44mm}{\includegraphics{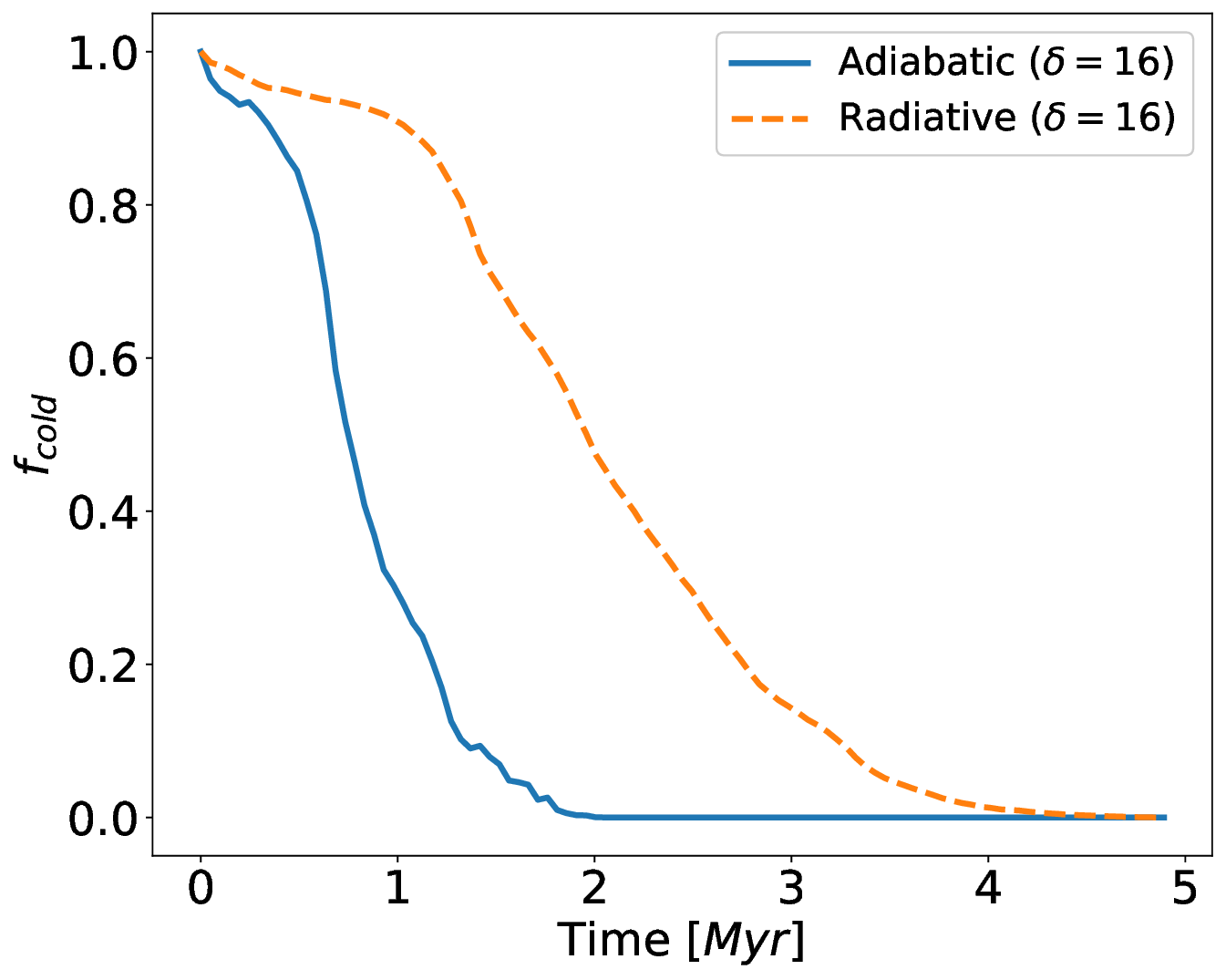}} & 
       \hspace{-0.3cm}\resizebox{!}{44mm}{\includegraphics{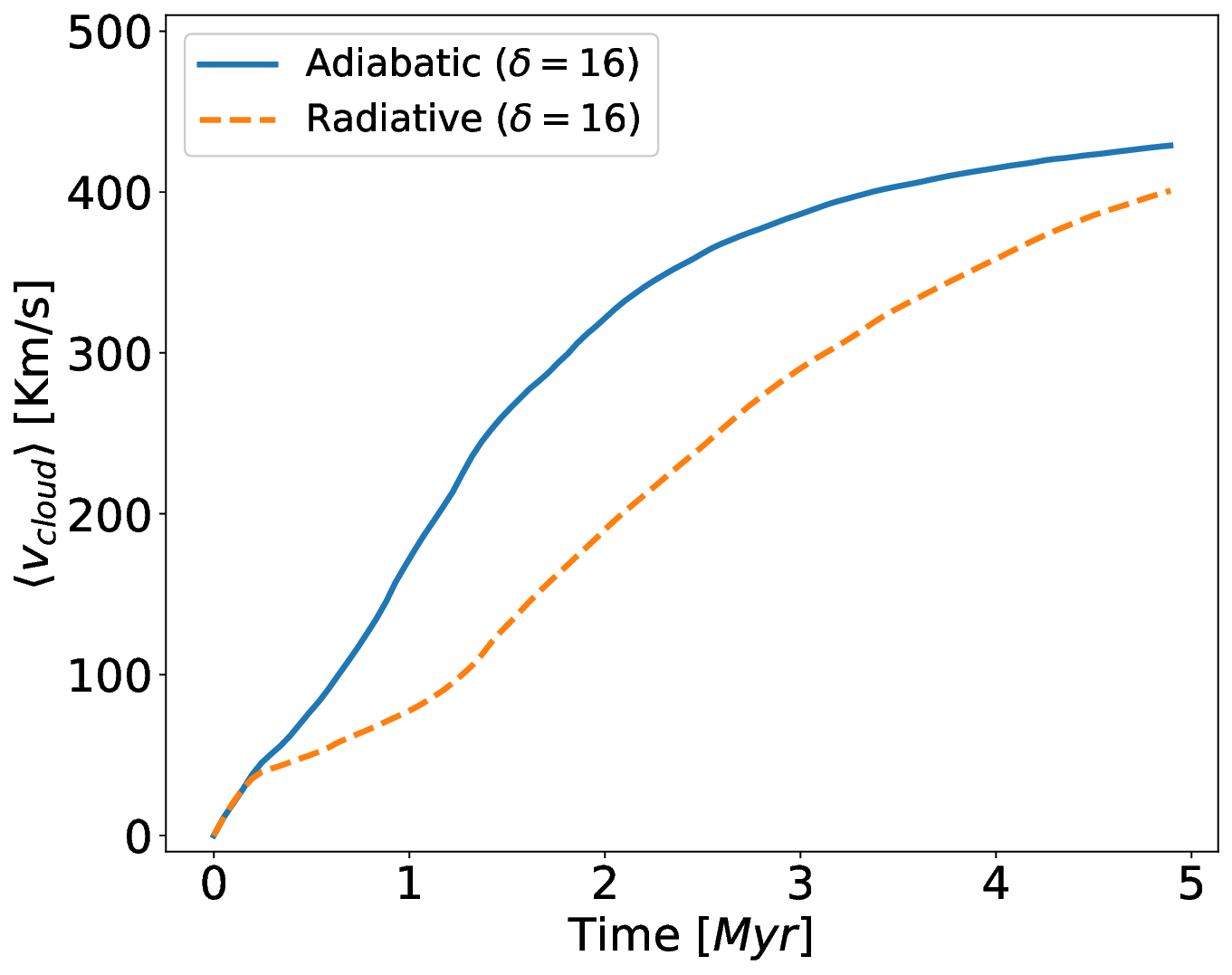}}\\
  \end{tabular}
  \caption{Time evolution of the mass-weighted average temperature (top left panel), mean density (top middle panel), dense gas mass fraction (top right panel), mixing fraction (bottom left panel), cold gas mass fraction (bottom middle panel), and mass-weighted average velocity (bottom right panel) of cloud material in the adiabatic and radiative $\delta = 16$ models. The radiative model extends the lifetime of dense gas compared to the adiabatic model by enabling the replenishment of dense gas along the flow, which maintains temperatures lower than the ambient temperature. Adiabatic clouds mix due to KH and RT instabilities, while radiative clouds mix more efficiently due to turbulence and cooling-induced pressure gradients led by condensation. Mixing aids the survival of cold gas in the radiative model, as mixed cloud material can condense back into a colder and denser phase. Adiabatic clouds accelerate through direct momentum transfer, while radiative clouds cannot accelerate through direct momentum transfer due to their high column densities.} 
  \label{evo2}
\end{center}
\end{figure*}

\begin{enumerate}
    \item \textbf{Shock formation:} In the first stage, the impact of the wind material on the front surfaces of the clouds produces internal shocks within them. The internal (refracted) shocks are accompanied by external (reflected) waves, which travel upstream. The reflected waves create bow shocks at the leading edges of the clouds, while the refracted shocks travel through the clouds. Due to pressure gradient forces and shock heating, which result from the initial impact, clouds start to accelerate and expand, stretching in the downstream direction (i.e. vertically along $Y$). As the clouds expand, they start to undergo hydrodynamic disruption, resulting in the formation of filamentary tails (akin to those seen in wind-single cloud models, e.g., \citealt{2017ApJ...845...69G,2019MNRAS.486.4526B}). Such tails are turbulent and comprise mixed gas, which exhibits densities and temperatures intermediate to that of the wind and the initial clouds.

    \item \textbf{Hydrodynamic shielding:} In the second stage, a mixture of wind gas and upstream cloud material continues to flow downstream, enveloping downstream clouds. The downstream motion of the wind triggers interactions between the filamentary tails of the upstream clouds with the front edges of the downstream clouds. The presence of mixed gas in between the clouds reduces the drag forces to which they are subjected and provides hydrodynamic shielding to the downstream clouds. This process plays a crucial role in preserving the structural integrity of downstream cloud cores because it reduces cloud acceleration and thus delays the growth of Rayleigh-Taylor (RT) instabilities (at least to a certain degree). Clouds then effectively protect each other from disruption associated with drag forces and RT instabilities. \cite{Forbes_2019} describes a similar process for adiabatic clouds in subsonic winds. Meanwhile, as the boundary conditions are periodic along $Y$, the turbulent tails of downstream clouds also interact with the upstream clouds. This interaction enhances the creation of a long gas stream of dense material (e.g. see the panels in middle column of Figure \ref{numden1} for $t=1.1\,\rm Myr$).
    
    
    \item \textbf{KH shredding and mixing:} In the third stage, the degree of mass loss and turbulent mixing become dominant. Clouds steadily lose mass from their surface layers via stripping caused by Kelvin-Helmholtz (KH) instabilities (\citealt{2016MNRAS.455.4274L}). The growth time of KH instabilities in hydrodynamical models depends mainly on the density contrast between both media, $\chi$, and the relative velocity at the boundary layer, $(v_{\rm wind} - v_{\rm cl})$ (see \citealt{chandrasekhar1961}). KH instabilities occur primarily at the sides of the clouds, where higher velocity shears exist, so hydrodynamic shielding has less impact on them than it has on RT instabilities. The swirling vortical motions associated with KH instabilities remove cloud material, which is carried downstream leading to the mixing of wind and cloud material. Shredding promotes acceleration as cloud cores become more diffuse owing to mass loss. The disruption process eventually forms long streamers of stripped gas that extend across the domain. The strength of the bow shocks also decreases over time as the clouds accelerate.
    
    \item \textbf{RT break-up and disruption:} In the fourth stage, the clouds have undergone substantial mass loss as a result of KH shredding. Additionally, their acceleration has reached a level that is conducive to the growth of long-wavelength RT instabilities. The RT instabilities favour the formation of low-density wind gas bubbles and high-density spikes of cloud material at the front of each cloud (see \citealt{Banda_16,2016MNRAS.457.4470P}). As the cross-sectional area increases, RT instabilities grow, which results in the disruption of the main cloud cores. Cores break up into smaller cloudlets. Some of these cloudlets eventually dissolve into the background medium, acquiring high velocities close to the wind speed. Ultimately, it is RT perturbations that cause the destruction of clouds. Low-density bubbles rapidly penetrate the denser layers of the clouds, causing disruption to the remaining cloud filaments. The remaining clouds then fragment into several cloudlets, which expand and mix further with the ambient gas. This process represents the final stage of cloud shredding and ultimately results in the complete destruction of the multi-cloud system.    
\end{enumerate}

It is important to remark that the break-up process can be accelerated or decelerated, depending on the inclusion or not of radiative cooling and on the initial separation distance, $\delta$, of the clouds. In particular, panel b in Figure \ref{numden1} shows that radiative cooling can prolong the lifetime of dense gas in the multicloud system, but does not prevent disruption.


\subsubsection{The role of radiative cooling}\label{re2}

Comparing our two control models presented in Figure \ref{numden1}, it is apparent that their evolution begins almost identically. A bow shock is formed as the wind interacts with the front surface of each cloud. In the adiabatic model, the shock produced from the interaction between the wind and the clouds results in the compression of the cloud and the subsequent injection of thermal energy. This energy is then converted into heat, which not only increases the temperature of the cloud (see Figure \ref{evo2}) but also contributes to the expansion and mixing of cloud and wind gas. Without an energy dissipation mechanism, dense gas is efficiently heated and disrupted by instabilities.\par

In contrast, radiative cooling removes the aforementioned thermal energy (injected by internal shocks) that would have been otherwise converted into heat. This prevents the pronounced increase in thermal pressure and temperature characteristic of adiabatic models. As a result, radiative cooling produces denser clouds where the cooled gas remains protected from instabilities, enveloped by a warm ($T \sim 10^{4.5}\,\rm K$) radiative layer of mixed, medium-density gas. As a result, the most significant effect of the inclusion of radiative cooling is that dense gas exhibits a longer lifetime.\par

The upper panels of Figure \ref{evo2} show the time evolution of the mass-weighted average temperature (see equation \ref{masswe}), the mean cloud density (see equation \ref{masswe}), and the dense gas mass fraction (see equation \ref{densefrac}), all of cloud material, in our control adiabatic and radiative models. In the absence of radiative cooling, the adiabatic wind-multicloud system maintains its dense gas mass only until $t\sim 1.2\,\rm Myr$ while the radiative model still maintains $\sim 60\%$ of dense material over the same time-scale. Eventually, at the end of the simulation time, both models lose all the dense material, but it is important to note that the radiative model prolongs the lifetime of dense gas at least three times compared to the adiabatic model. As cloud gas continues interacting with upstream clouds and the wind, it continues mixing, producing warm gas with temperatures $T< 10^5\,\rm K$. This denser gas is also subjected to erosion and shock heating, but strong cooling can overall slow down these processes. As a result, not all the gas in the radiative model is heated up to temperatures $T \gtrsim 10^{6}\,\rm K$ by the end of the simulation time.\par 



\begin{figure*}
\begin{center}
\begin{tabular}{c c}
\hspace{-0.9cm}
  \begin{tabular}{c c c c c}
       \multicolumn{4}{l}{a) Adiabatic ($\delta=16$)}\\ 
       \multicolumn{1}{l}{\hspace{-2mm}\hspace{+1.2mm}$t=0$} & \multicolumn{1}{c}{$49\,t_{\rm cross}=0.6\,\rm Myr$} & \multicolumn{1}{c}{$171.7\,t_{\rm cross}=2.1\,\rm Myr$} & \multicolumn{1}{c}{$310.6\,t_{\rm cross}=3.8\,\rm Myr$}  \\   
       \hspace{-0.3cm}\resizebox{!}{37mm}{\includegraphics{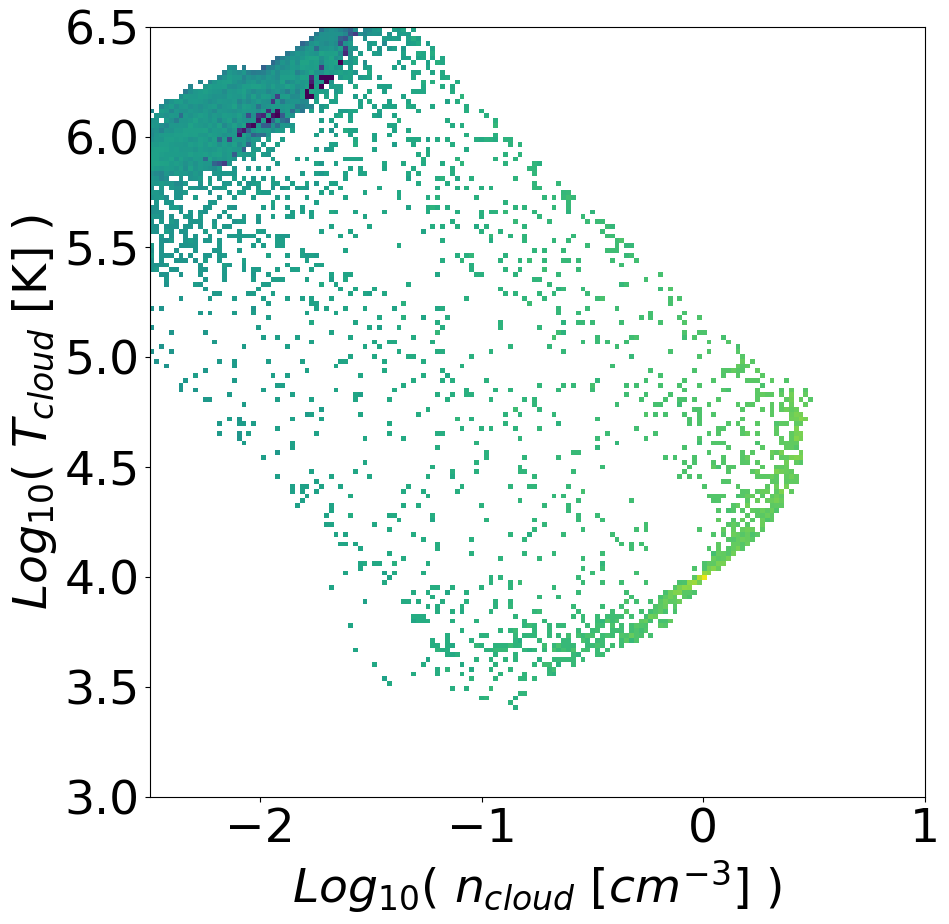}} & \hspace{-0.3cm}\resizebox{!}{37mm}{\includegraphics{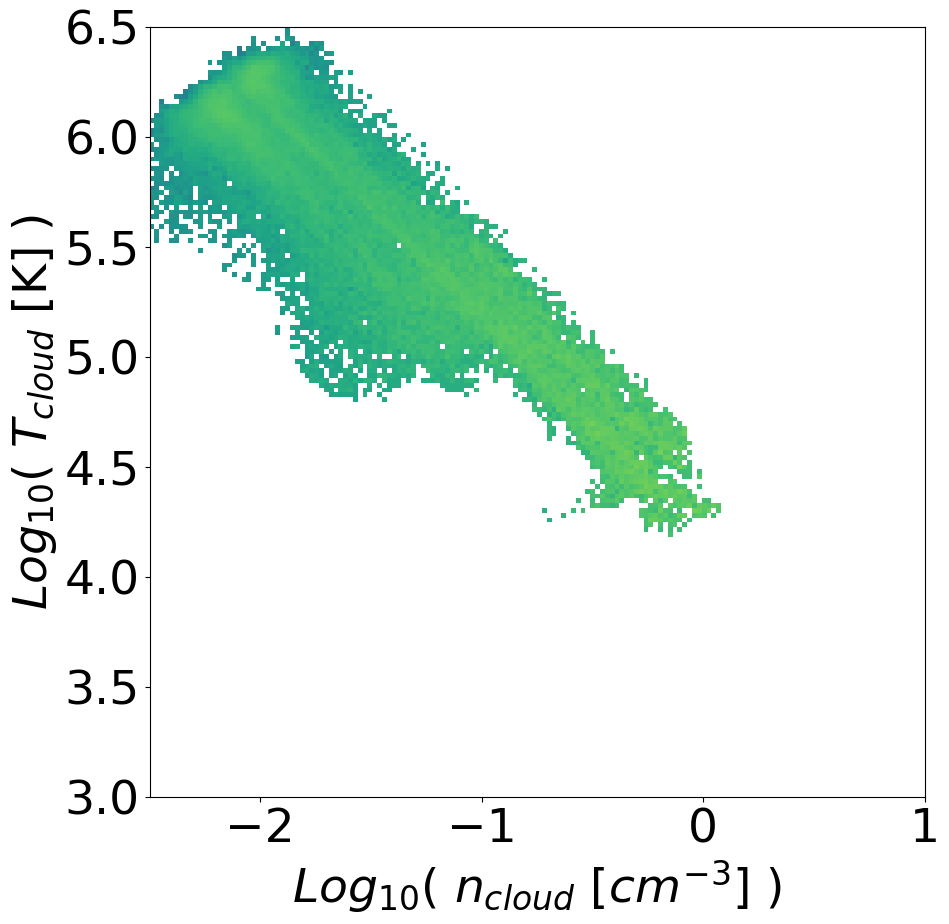}} & \hspace{-0.3cm}\resizebox{!}{37mm}{\includegraphics{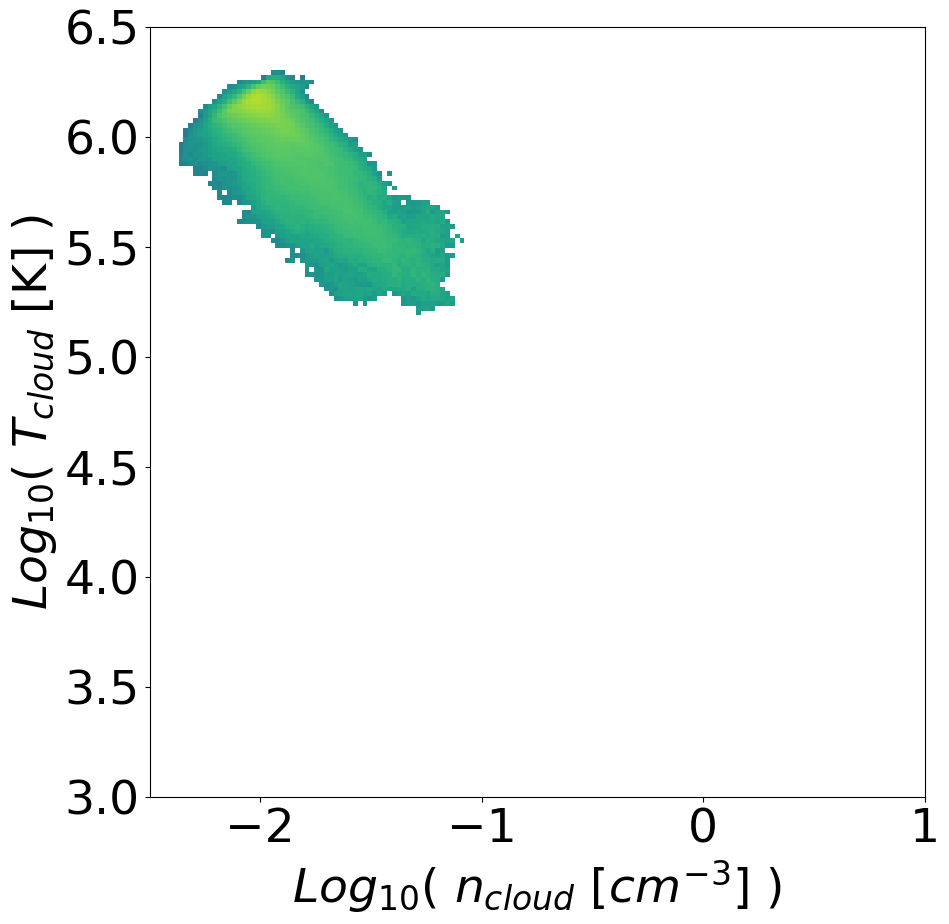}} & \hspace{-0.3cm}\resizebox{!}{37mm}{\includegraphics{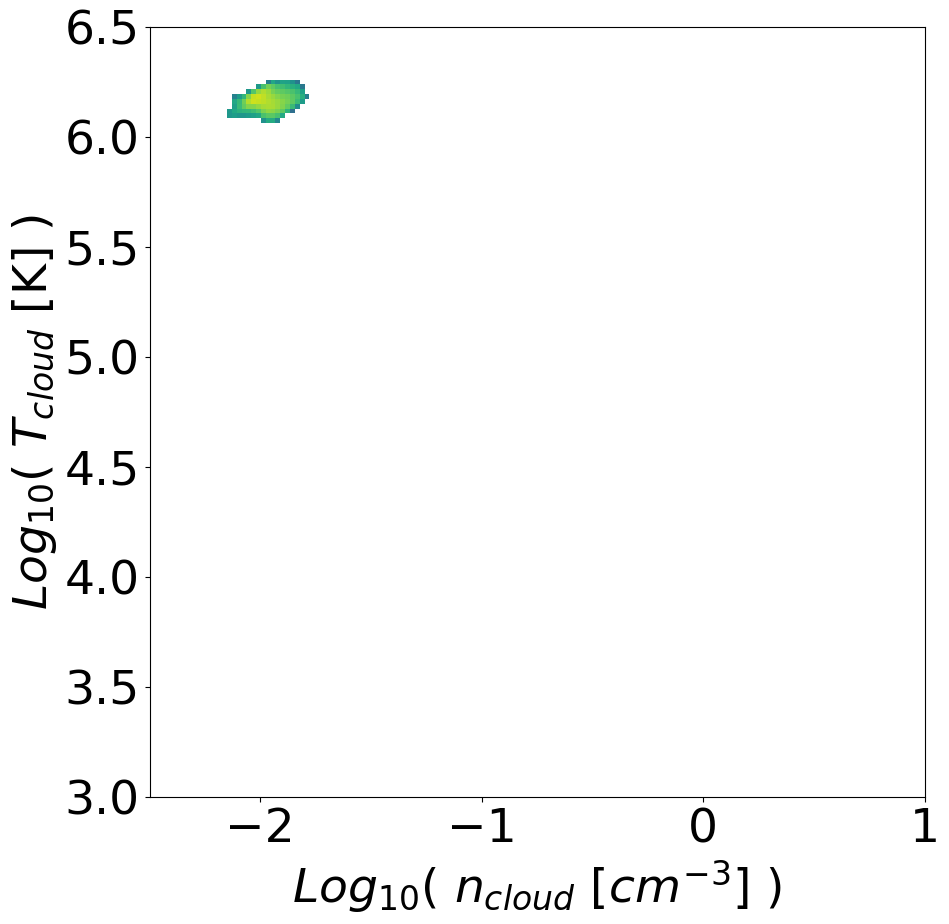}} \\
       \multicolumn{4}{l}{b) Radiative ($\delta=16$)}\\ 
       \multicolumn{1}{l}{\hspace{-2mm}\hspace{+1.2mm}$t=0$} & \multicolumn{1}{c}{$49\,t_{\rm cross}=0.6\,\rm Myr$} & \multicolumn{1}{c}{$171.7\,t_{\rm cross}=2.1\,\rm Myr$} & \multicolumn{1}{c}{$310.6\,t_{\rm cross}=3.8\,\rm Myr$}  \\         
       \hspace{-0.3cm}\resizebox{!}{37mm}{\includegraphics{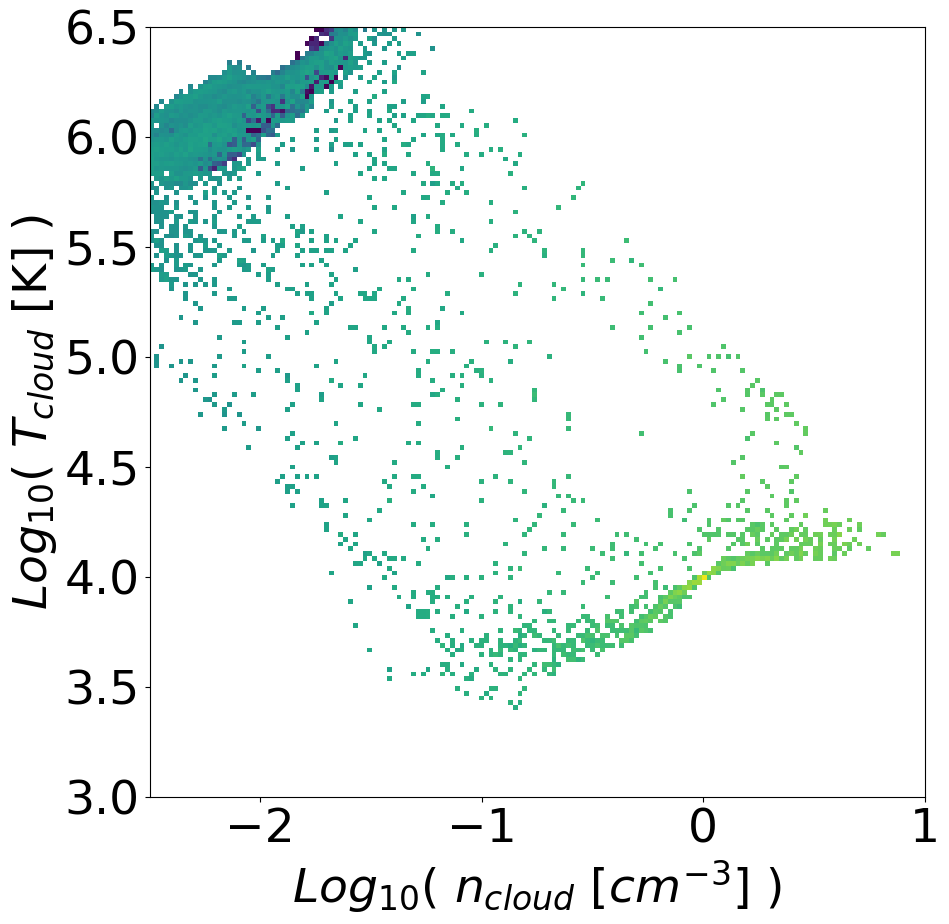}} & \hspace{-0.3cm}\resizebox{!}{37mm}{\includegraphics{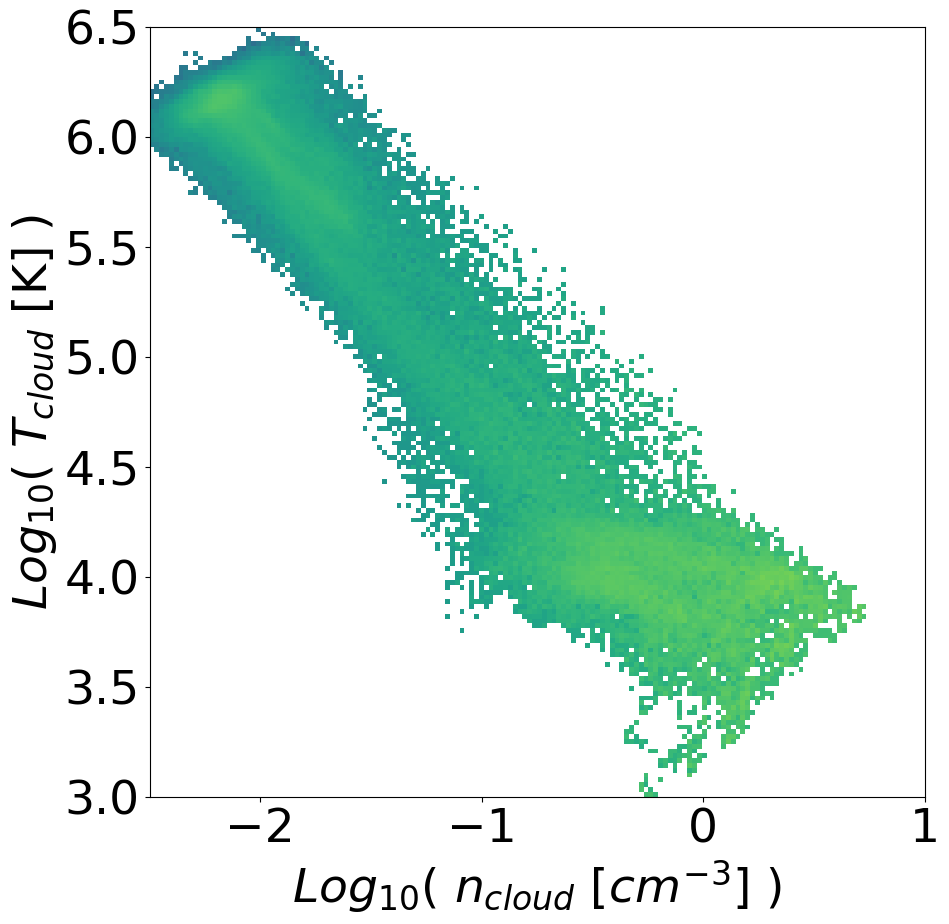}} & \hspace{-0.0cm}\resizebox{!}{37mm}{\includegraphics{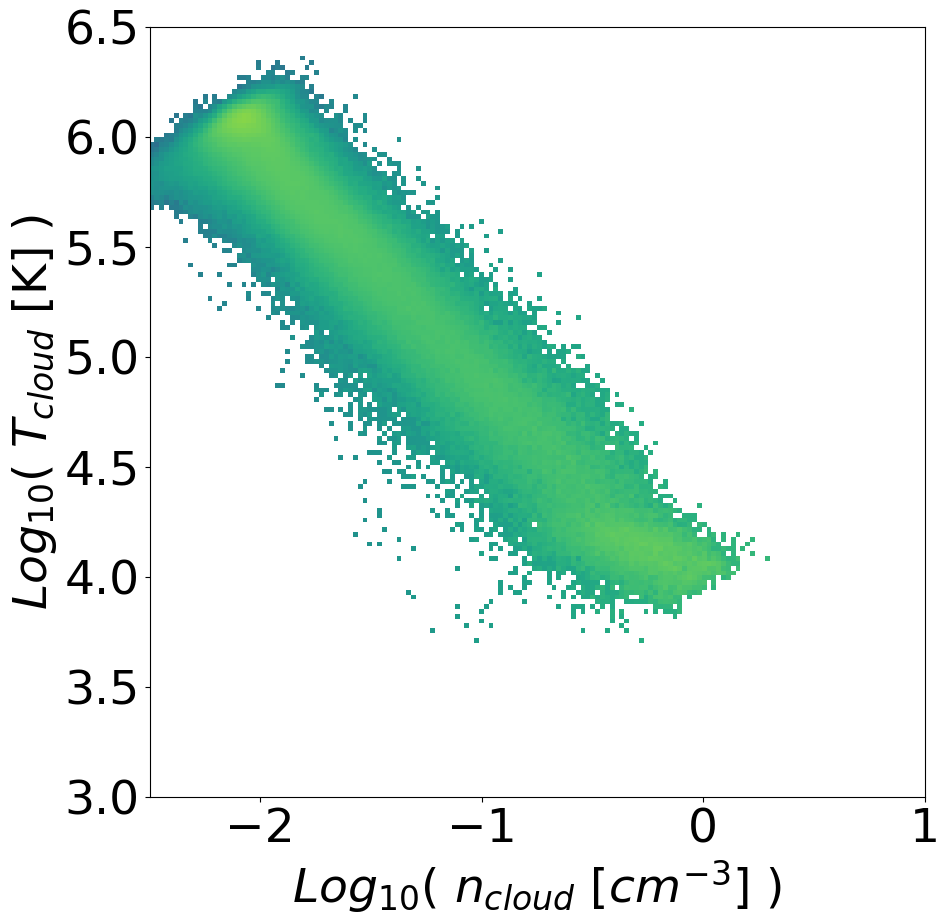}} & \hspace{-0.3cm}\resizebox{!}{37mm}{\includegraphics{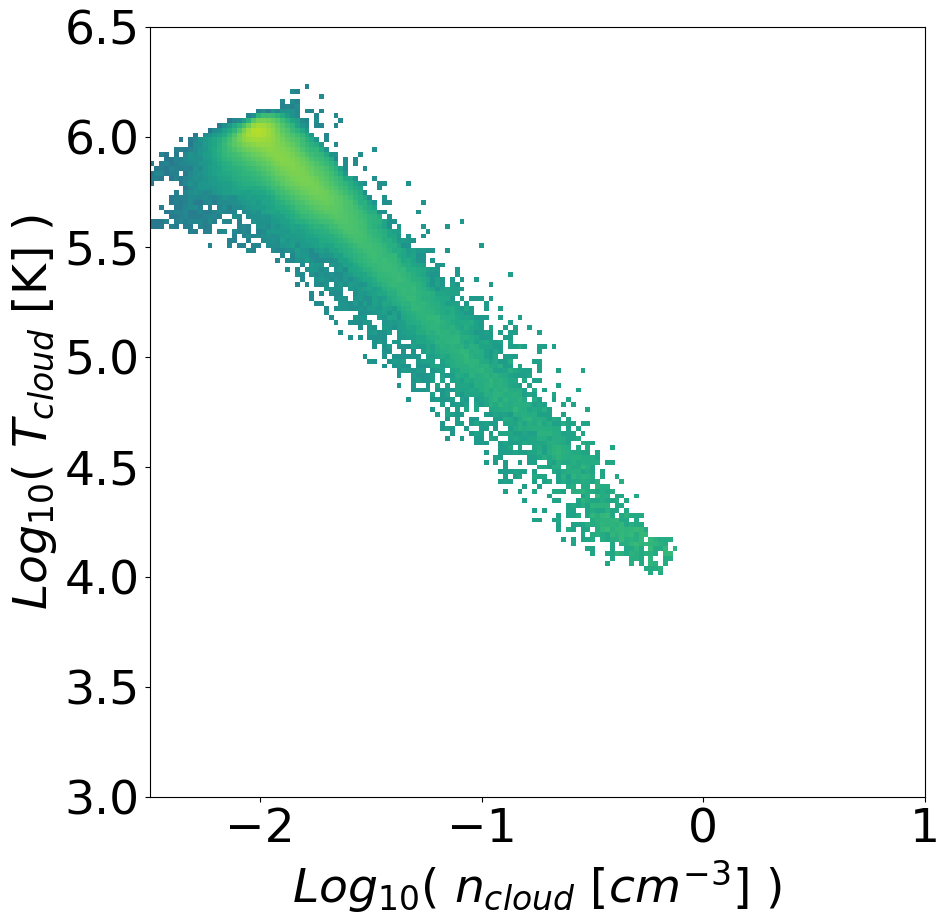}}\\
  \end{tabular}
  & 
  \hspace{-.4cm}\raisebox{-4.2cm}{\resizebox{9mm}{!}{\includegraphics{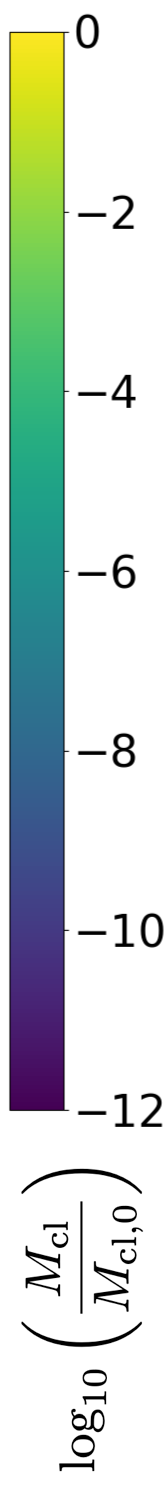}}}
\end{tabular}
  \caption{Mass-weighted phase diagrams displaying the two-dimensional distributions of temperature and density of cloud material in the adiabatic (top) and radiative (bottom) $\delta = 16$ model at five different times through the simulation (columns). Without mechanisms to release the excess energy injected by shocks, the cold gas in adiabatic clouds heats up and becomes unstable, leading to a shortage of dense cloud material at low temperatures. The radiative model exhibits less heating than the adiabatic case, resulting in a multiphase structure with a cooler, denser phase and a less dense, hotter phase.} 
  \label{hist1}
\end{center}
\end{figure*}

Figure \ref{hist1} shows the time evolution of the mass-weighted 2D phase diagrams of temperature versus number density of cloud material for our control adiabatic and radiative models with $\delta=16$. In the adiabatic model, the cloud material experiences rapid heating, with the majority of the material reaching temperatures $T > 10^4\,\rm K$ in only $t=0.6\,\rm Myr$. Due to the absence of mechanisms to release the extra energy injected via shocks, the cold gas is heated up, thus leaving little dense material at low temperatures. By $t_{\rm sim}=2.1\,\rm Myr$, most of the cloud mass corresponds to hot gas with temperatures $T \gtrsim 10^5\,\rm K$ which has also expanded transversely (i.e. along the X and Z directions, see Figure \ref{numden1}). The radiative model exhibits a different thermodynamic evolution. In the early stages of the simulation ($t=0.6\,\rm Myr$), the cloud gas pushed downstream by the wind is initially heated up, but then gradually cools down to temperatures $T\sim 10^{4}\,\rm K$. Cooling also leads to geometrically thinner tails of cloud material (see Figure \ref{numden1}). During most of the simulation time, two distinct gas phases are observed: a cooler and denser phase with $T\sim 10^{4}\,\rm K$ and $n\sim 1\,\rm cm^{-3}$, and a higher-temperature and lower-density phase with $T\sim 10^{6}\,\rm K$ and $n\sim 0.01\,\rm cm^{-3}$. The overall flow is therefore multi-phase, and resembles the thermodynamical structure seen in other galactic wind models (e.g. \citealt{Schneider_17,Banda_21}; \citealt*{2022ApJ...936..133C}).

\subsubsection{Gas Mixing and cloud dynamics}\label{renew}

The lower panels of Figure \ref{evo2} depict the mixing fraction between wind and cloud gas (equation \ref{mixfrac}), the cold gas mass fraction (equation \ref{coldgassfrac}), and the mass-weighted average velocity in cloud material (equation \ref{masswe}) for our control adiabatic and radiative $\delta = 16$ models. Both models exhibit a high degree of turbulence and mixing. In the adiabatic model, turbulent mixing and dispersion processes are associated with the generation of vortical motions via KH instabilities during the whole evolution and via RT instabilities at the late stages of the interaction (also seen in adiabatic wind-cloud simulations, e.g. \citealt{Banda_16,2017MNRAS.470.2427G}). Shear instabilities lead to mixing, erosion of low-density regions in the clouds, and the formation of geometrically thick tails downstream of each cloud.\par


On the other hand, in the radiative model, mixing and dispersion arise not only from KH and RT instabilities, but also from cooling-induced pressure gradients and turbulence driven by condensation (see \citealt{2019MNRAS.487..737J,2020ApJ...894L..24F} for further details on radiative mixing layers). As a result, radiative models exhibit a higher degree of mixing compared to adiabatic models (see panel d Figure \ref{evo2}). Condensation occurs from the early stages of evolution of radiative models, where warm, mixed gas with $T\sim 10^{4.5} - 10^{5.5}\,\rm K$ efficiently condenses into colder and denser material as the cooling rates peak at those temperatures. Condensation thus promotes the formation of a denser medium and geometrically thin tails behind the clouds (see Figure \ref{numden1}). Gas along such tails fragments into smaller gas clumps as time progresses, which results in the emergence of a stream of dense and turbulent clumpy gas. Similar multi-phase clumpy/misty flow structures have also been observed in other small-scale (\citealt{2020MNRAS.494L..27G,2022MNRAS.511..859G}) and large-scale (\citealt{2008ApJ...674..157C,2024MNRAS.527.9683T}) models of CGM clouds.\par




Radiative mixing also influences the cold gas mass fraction and the dynamics of cloud gas. Panel (e) of Figure \ref{evo2} shows that the radiative model preserves cold material (with $T< 10\,T_{\rm cl}$ for a significantly longer period compared to the adiabatic model. This occurs as a result of: 1) the delayed growth of KH instabilities thanks to the larger density contrasts caused by cooling-driven contraction, and 2) re-condensation of shocked cloud material that has already mixed and dispersed downstream of each cloud. On the other hand, Panel (f) of Figure \ref{evo2} shows that the cloud velocities in the adiabatic model are higher compared to the radiative one. Radiative models are subjected to efficient cooling of the warm, mixed gas, which leads to the formation of higher column densities. Denser gas occupies a smaller volume, so it is more difficult to accelerate via direct momentum transfer \citep[see][]{2015ApJ...805..158S,Banda_16}. Conversely, the expansion of adiabatic clouds results in larger cross sections and a higher rate of cloud acceleration. Towards the end of the simulations, however, we see that radiative clouds reduce the gap as gas that condenses back into the cold phase preserves some of the momentum that was present in the warm phase \citep[see discussions in][]{Schneider_20, Banda_21}.


\subsection{Hydrodynamic shielding}\label{Sec4.2}

The arrangement of clouds in a stream-like formation affects how they interact with the surrounding wind compared to isolated cloud formations. While isolated clouds are subjected to strong drag forces (\citealt{2019MNRAS.486.4526B}), clouds along gas streams have the capacity to hydrodynamically shield one another from the disruption caused by their interaction with the wind gas (see \citealt{Aluzas_12,Banda_20}). As noted in the previous section, the morphology of downstream clouds is not only influenced by the background wind, but also by the upstream clouds. Two processes are crucial for hydrodynamic shielding in multi-cloud systems: the first one involves the reduction of drag forces on individual clouds (which lessens acceleration and delays RT instabilities), and the second one is the interaction between clouds as a result of the displacement induced by the supersonic wind (which eventually leads to interactions between upstream cloud tails and downstream cloud cores). Generally, then, cloud material subjected to hydrodynamic shielding can persist for longer periods due to its ability to survive dynamic shredding and drag.\par

The filaments created from the stripping of the outer layers of the clouds, as a result of instabilities, serve as a protective cover for the downstream cloud cores once they reach them. As the wind continues to flow downstream, it also interacts with downward clouds which means that part of the gas that already directly interacted with the preceding cloud, starts to interact with downstream clouds. This means that the evolution of the individual clouds along the gas stream is not independent, but relies on their interaction with: 1) the hot wind, 2) the other clouds further upstream, and 3) mixed gas that has already interacted with upstream cold material. The higher the separation distance ($\delta$) within the clouds, the less probable it is for hydrodynamic shielding to act; and vice versa. However, this largely depends, as we show henceforth, on whether the wind is adiabatic or radiative, and whether it is subsonic or supersonic. For adiabatic models, there is a threshold value for the cloud separation distance (\citealt{Forbes_2019}), $\delta_{\rm shield}$, so that when $\delta<\delta_{\rm shield}$, hydrodynamic shielding occurs. However, we find that the value of $\delta_{\rm shield}$ varies between subsonic and supersonic cases (see below).\par


\begin{figure*}
\begin{center}
  \begin{tabular}{c c c c c c c c c c c}
       \multicolumn{5}{l}{(a) Adiabatic ($\delta=2$)} & \multicolumn{5}{l}{(d) Adiabatic ($\delta=16$)}\\
       \multicolumn{1}{c}{{\scriptsize$t=0$}} & \multicolumn{1}{c}{{\scriptsize$49.1\,t_{\rm cross}$}} & \multicolumn{1}{c}{{\scriptsize$155.3\,t_{\rm cross}$}} & \multicolumn{1}{c}{{\scriptsize$253.5\,t_{\rm cross}$}}  & \multicolumn{1}{c}{{\scriptsize$351.6\,t_{\rm cross}$}}& \multicolumn{1}{c}{{\scriptsize$t=0$}} & \multicolumn{1}{c}{{\scriptsize$49.1\,t_{\rm cross}$}} & \multicolumn{1}{c}{{\scriptsize$155.3\,t_{\rm cross}$}} & \multicolumn{1}{c}{{\scriptsize$253.5\,t_{\rm cross}$}}  & \multicolumn{1}{c}{{\scriptsize$351.6\,t_{\rm cross}$}}\\   
       \hspace{-0.3cm}\resizebox{!}{51mm}{\includegraphics{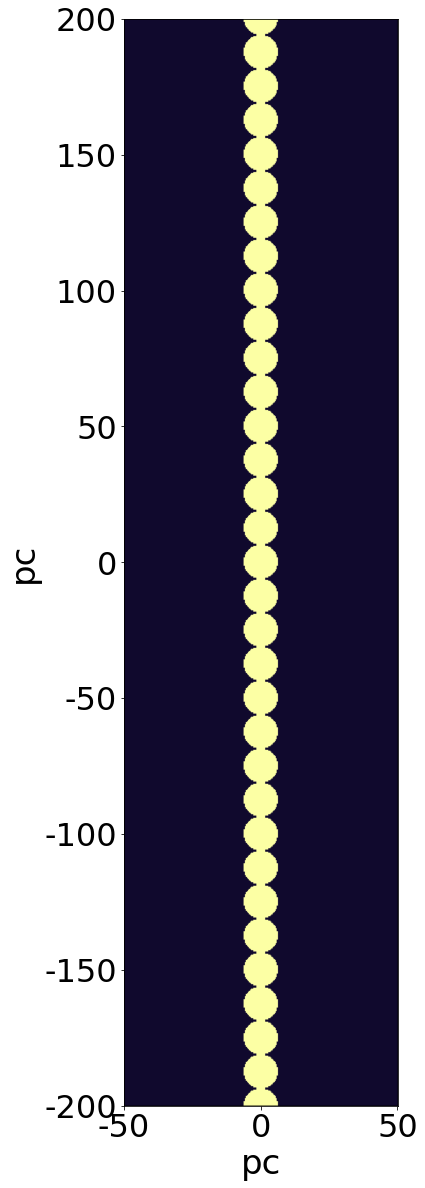}} & \hspace{-0.5cm}\resizebox{!}{51mm}{\includegraphics{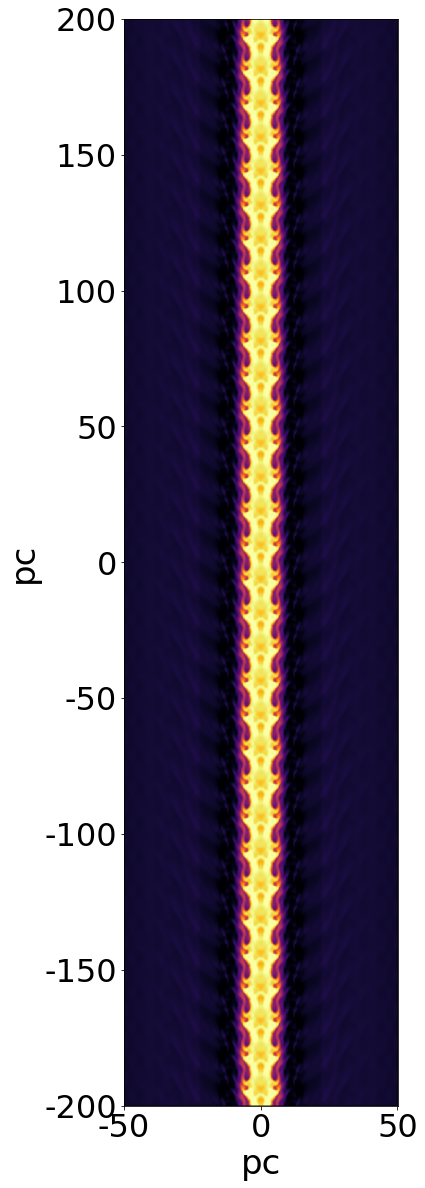}} & \hspace{-0.5cm}\resizebox{!}{51mm}{\includegraphics{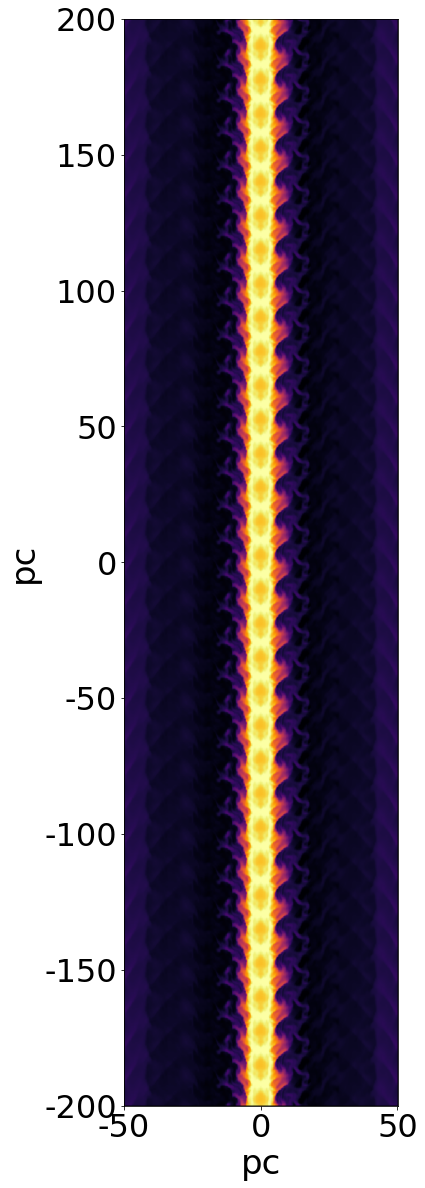}} & \hspace{-0.5cm}\resizebox{!}{51mm}{\includegraphics{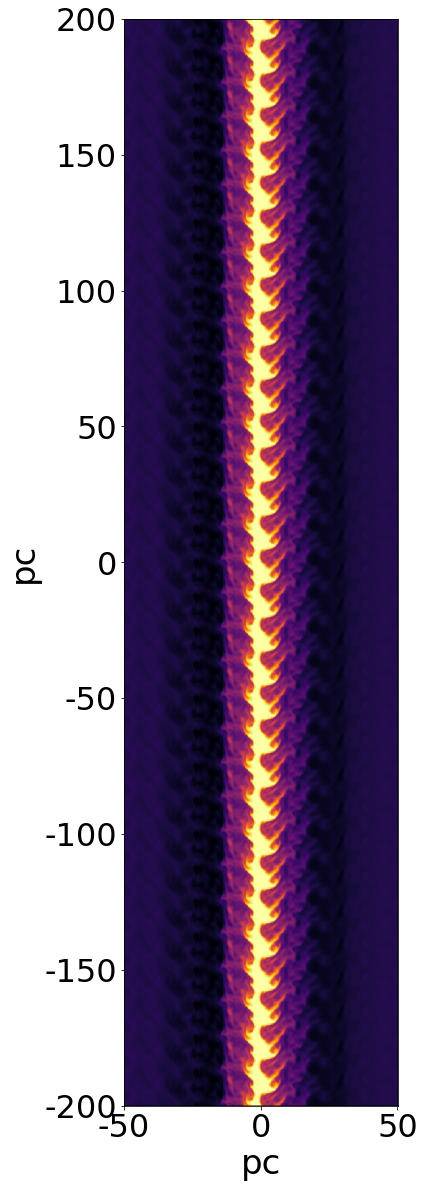}} &
       \hspace{-0.5cm}\resizebox{!}{51mm}{\includegraphics{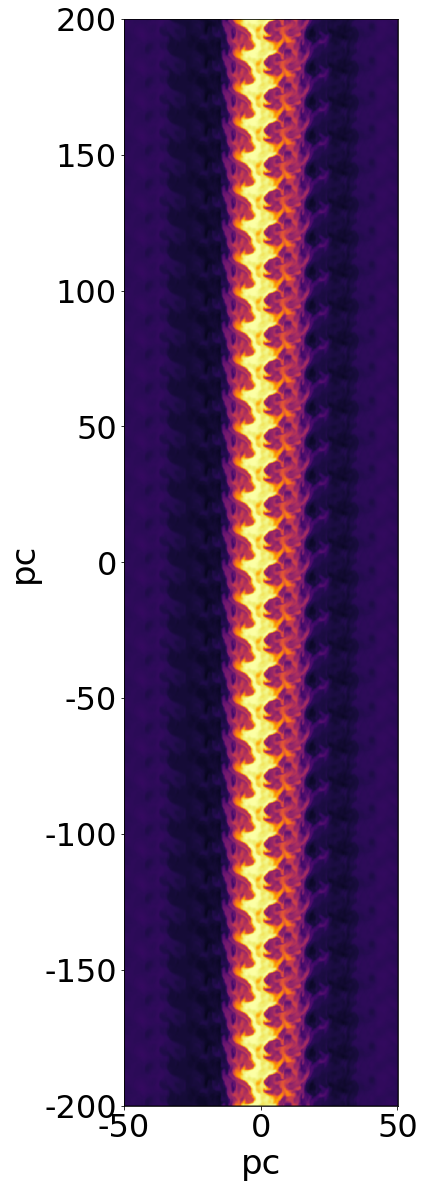}} &
       \hspace{0.1cm}
       \hspace{-0.5cm}\resizebox{!}{51mm}{\includegraphics{2Dslices/newadi16numden16_000.png}} & \hspace{-0.5cm}\resizebox{!}{51mm}{\includegraphics{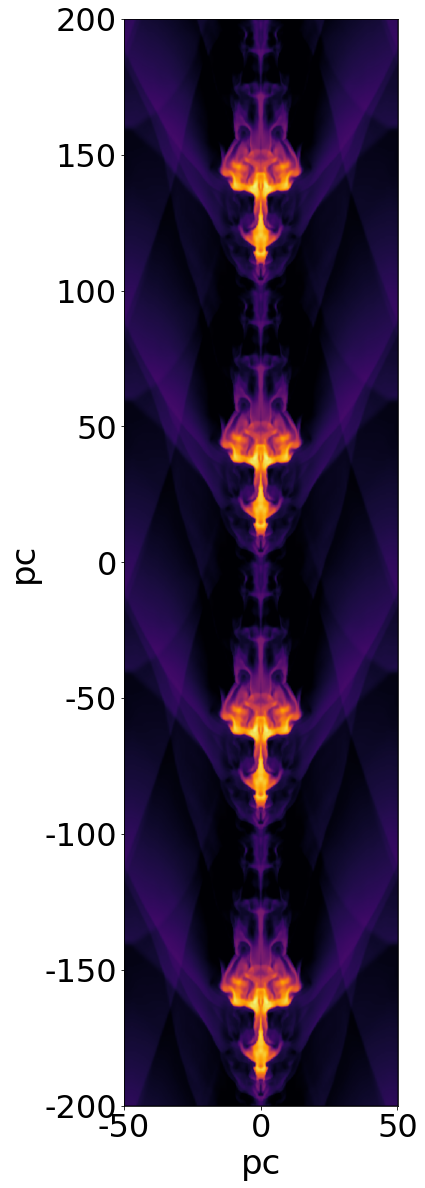}} & \hspace{-0.5cm}\resizebox{!}{51mm}{\includegraphics{2Dslices/newadi16numden16_040.png}} & \hspace{-0.5cm}\resizebox{!}{51mm}{\includegraphics{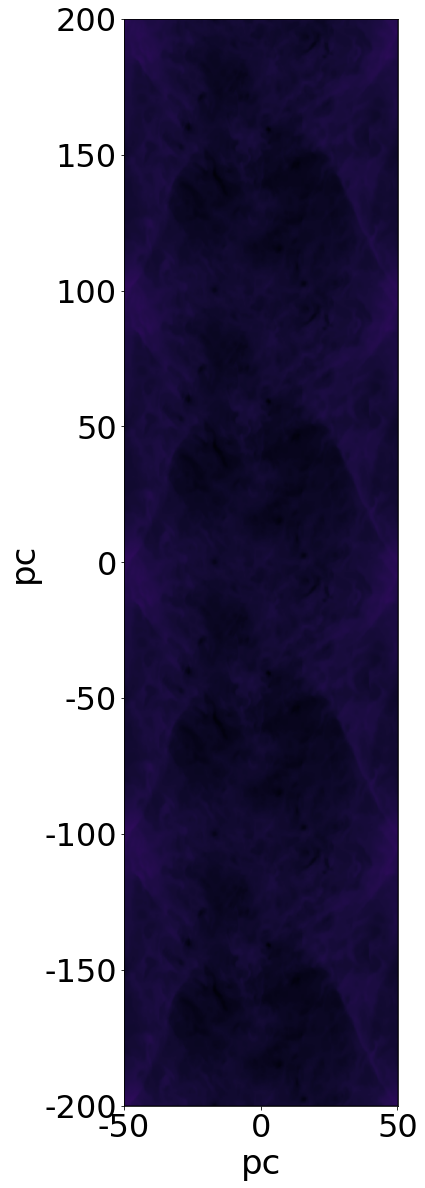}} &
       \hspace{-0.5cm}\resizebox{!}{51mm}{\includegraphics{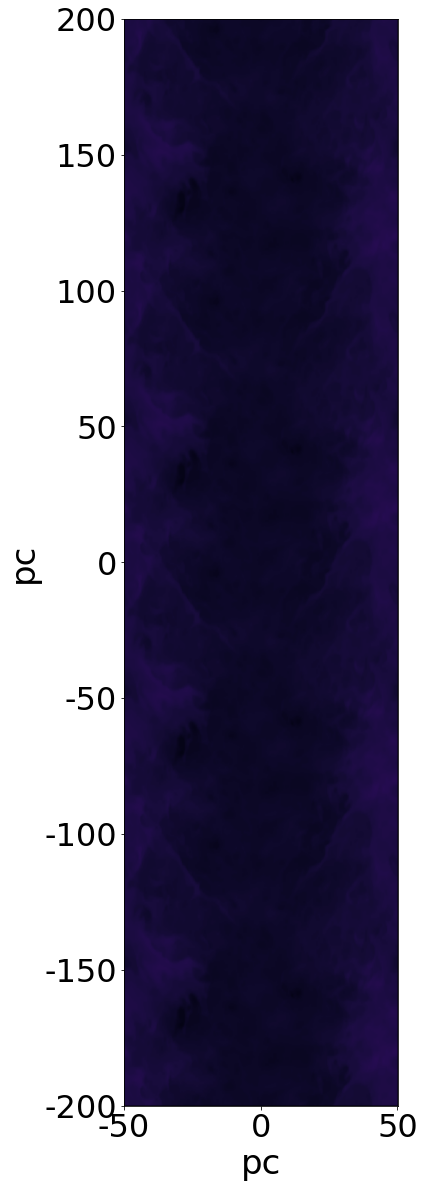}}
    \\
       \multicolumn{5}{l}{(b) Adiabatic ($\delta=4$)} & \multicolumn{5}{l}{(e) Adiabatic ($\delta=32$)}\\
       \multicolumn{1}{c}{{\scriptsize$t=0$}} & \multicolumn{1}{c}{{\scriptsize$49.1\,t_{\rm cross}$}} & \multicolumn{1}{c}{{\scriptsize$155.3\,t_{\rm cross}$}} & \multicolumn{1}{c}{{\scriptsize$253.5\,t_{\rm cross}$}}  & \multicolumn{1}{c}{{\scriptsize$351.6\,t_{\rm cross}$}}& \multicolumn{1}{c}{{\scriptsize$t=0$}} & \multicolumn{1}{c}{{\scriptsize$49.1\,t_{\rm cross}$}} & \multicolumn{1}{c}{{\scriptsize$155.3\,t_{\rm cross}$}} & \multicolumn{1}{c}{{\scriptsize$253.5\,t_{\rm cross}$}}  & \multicolumn{1}{c}{{\scriptsize$351.6\,t_{\rm cross}$}}\\   
    \hspace{-0.3cm}\resizebox{!}{51mm}{\includegraphics{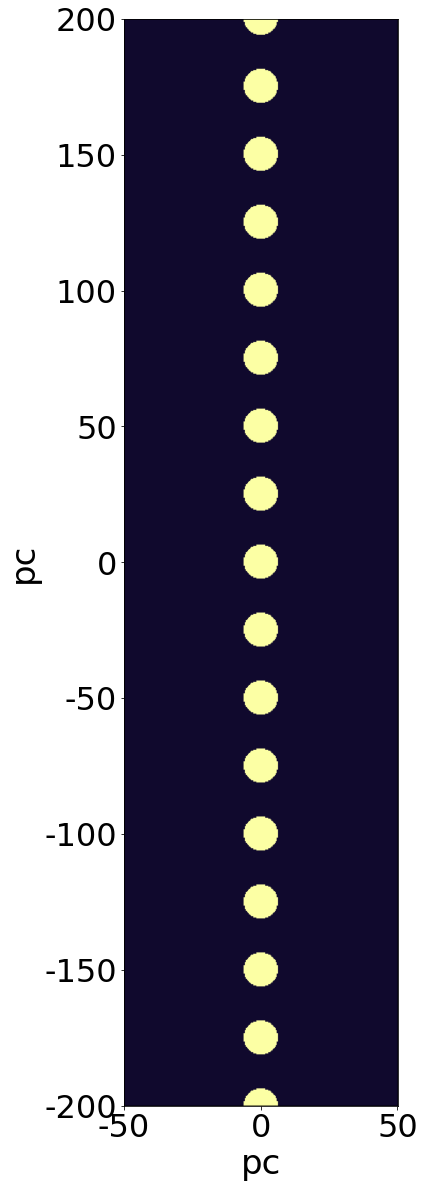}} & \hspace{-0.5cm}\resizebox{!}{51mm}{\includegraphics{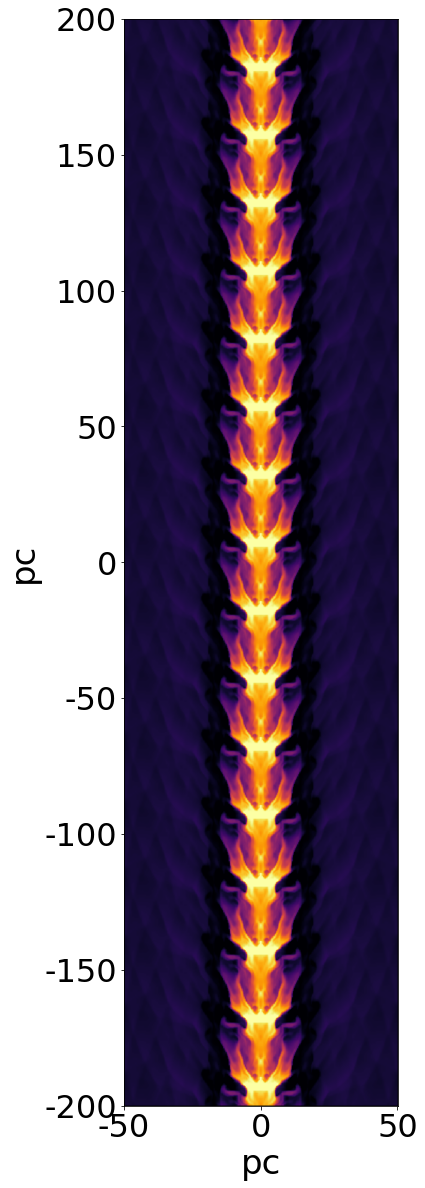}} & \hspace{-0.5cm}\resizebox{!}{51mm}{\includegraphics{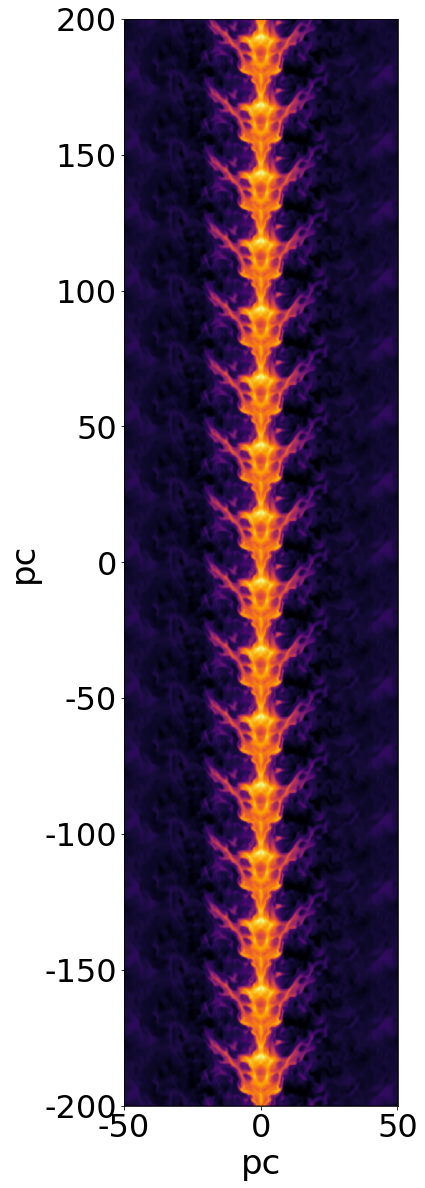}} & \hspace{-0.5cm}\resizebox{!}{51mm}{\includegraphics{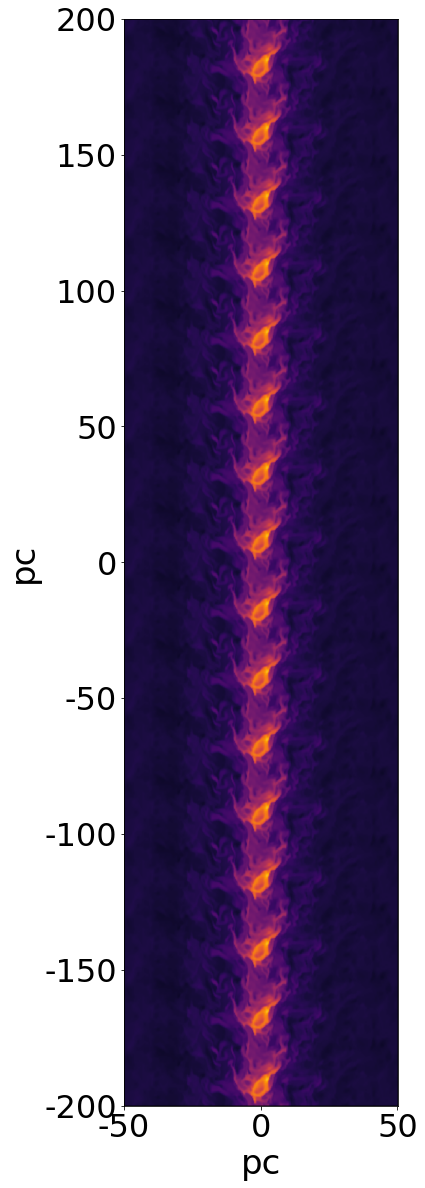}} &
    \hspace{-0.5cm}\resizebox{!}{51mm}{\includegraphics{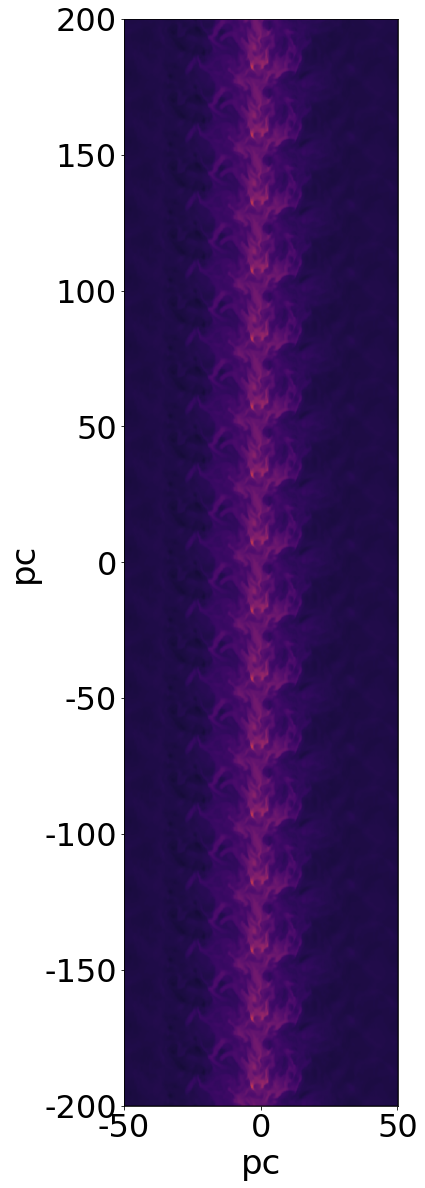}} &
    \hspace{0.1cm}
    \hspace{-0.5cm}\resizebox{!}{51mm}{\includegraphics{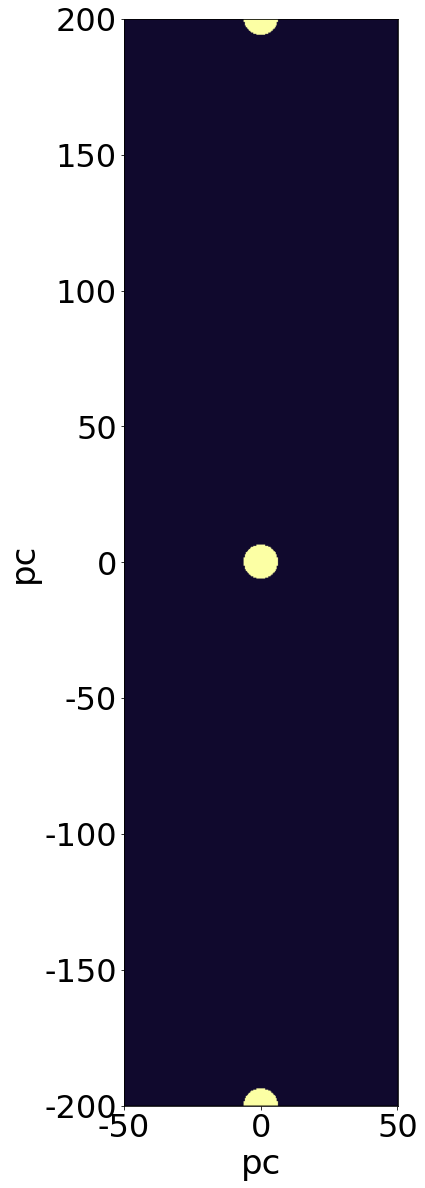}} & \hspace{-0.5cm}\resizebox{!}{51mm}{\includegraphics{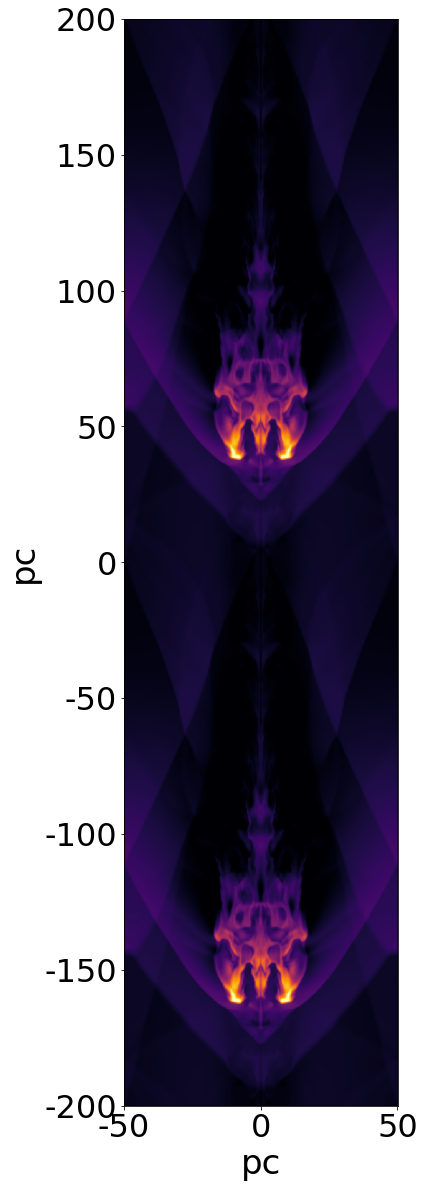}} & \hspace{-0.5cm}\resizebox{!}{51mm}{\includegraphics{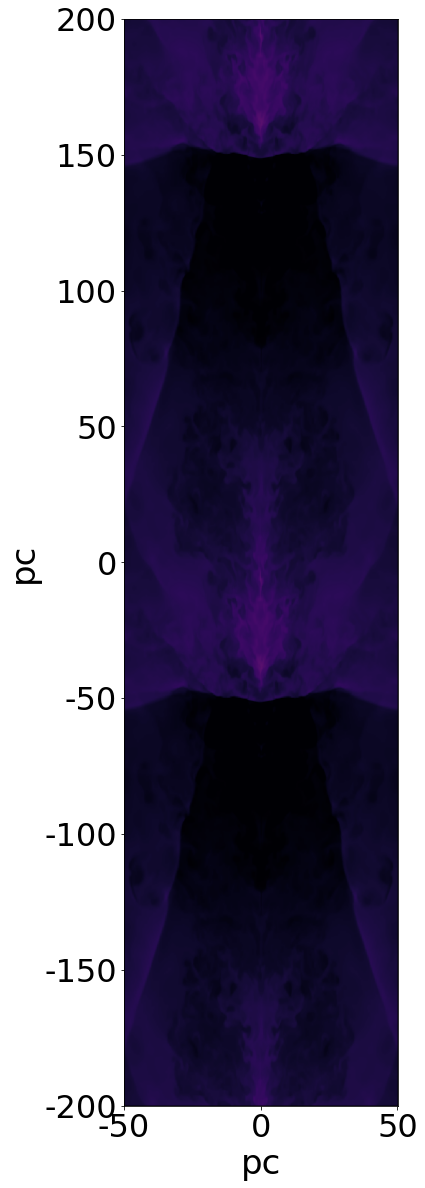}} & \hspace{-0.5cm}\resizebox{!}{51mm}{\includegraphics{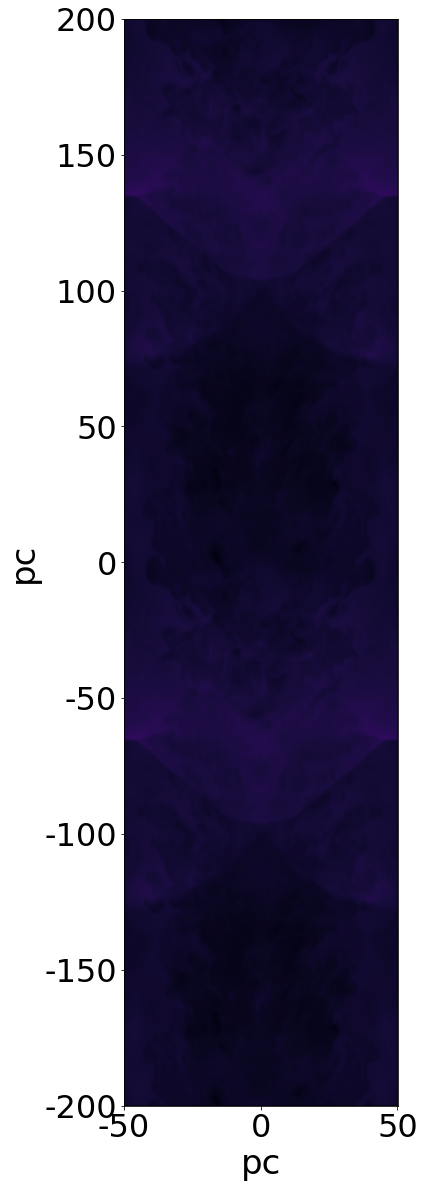}} &
    \hspace{-0.5cm}\resizebox{!}{51mm}{\includegraphics{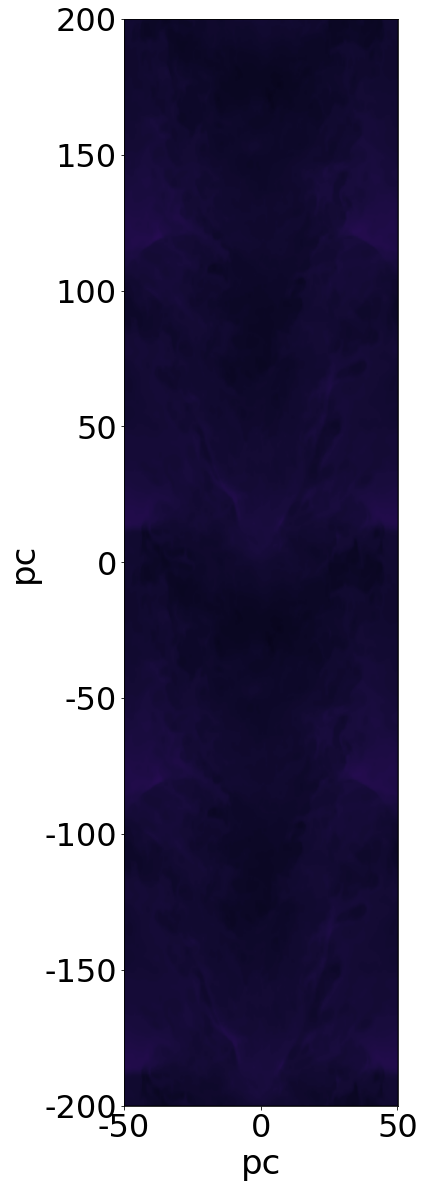}} 
    \\
    \multicolumn{5}{l}{(c) Adiabatic ($\delta=8$)} & \multicolumn{5}{l}{(f) Adiabatic ($\delta=64$)}\\
       \multicolumn{1}{c}{{\scriptsize$t=0$}} & \multicolumn{1}{c}{{\scriptsize$49.1\,t_{\rm cross}$}} & \multicolumn{1}{c}{{\scriptsize$155.3\,t_{\rm cross}$}} & \multicolumn{1}{c}{{\scriptsize$253.5\,t_{\rm cross}$}}  & \multicolumn{1}{c}{{\scriptsize$351.6\,t_{\rm cross}$}}& \multicolumn{1}{c}{{\scriptsize$t=0$}} & \multicolumn{1}{c}{{\scriptsize$49.1\,t_{\rm cross}$}} & \multicolumn{1}{c}{{\scriptsize$155.3\,t_{\rm cross}$}} & \multicolumn{1}{c}{{\scriptsize$253.5\,t_{\rm cross}$}}  & \multicolumn{1}{c}{{\scriptsize$351.6\,t_{\rm cross}$}}\\   
    \hspace{-0.3cm}\resizebox{!}{51mm}{\includegraphics{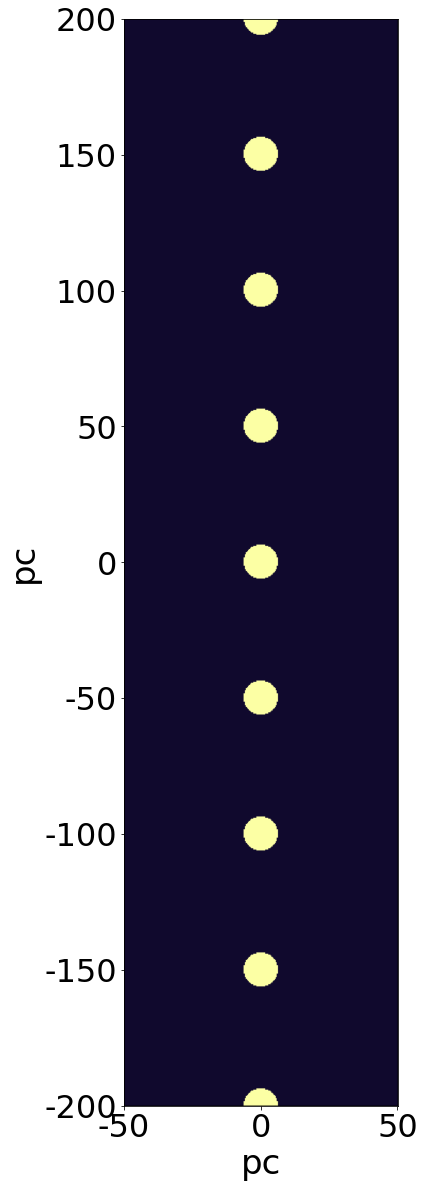}} & \hspace{-0.5cm}\resizebox{!}{51mm}{\includegraphics{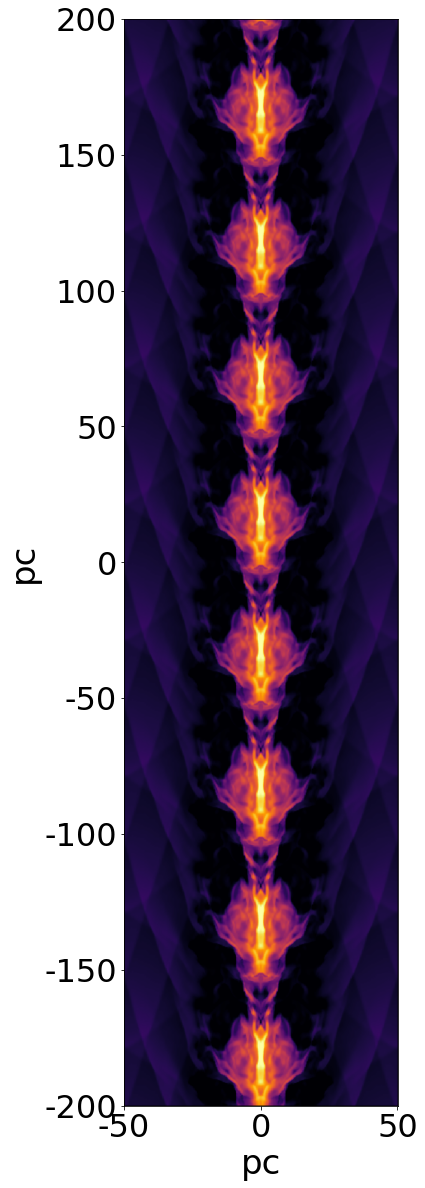}} & \hspace{-0.5cm}\resizebox{!}{51mm}{\includegraphics{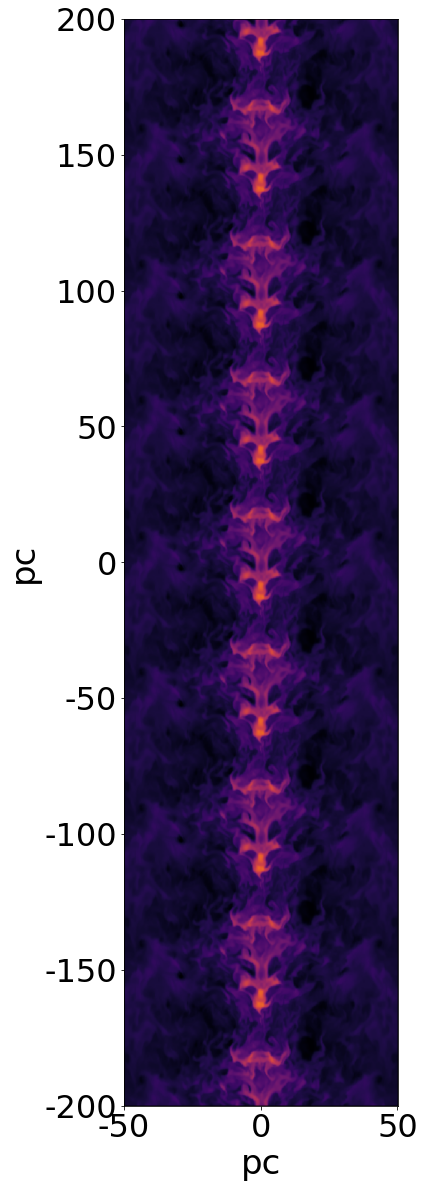}} & \hspace{-0.5cm}\resizebox{!}{51mm}{\includegraphics{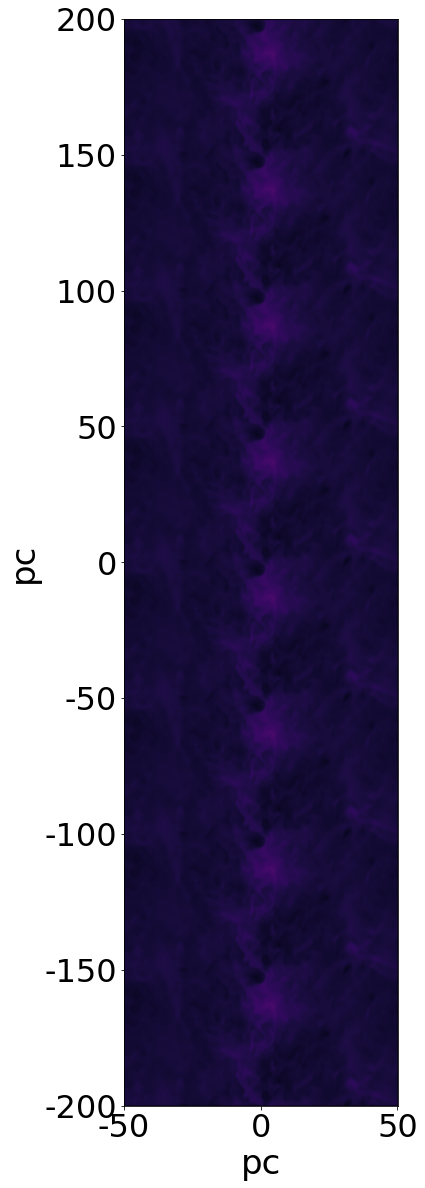}} &
    \hspace{-0.5cm}\resizebox{!}{51mm}{\includegraphics{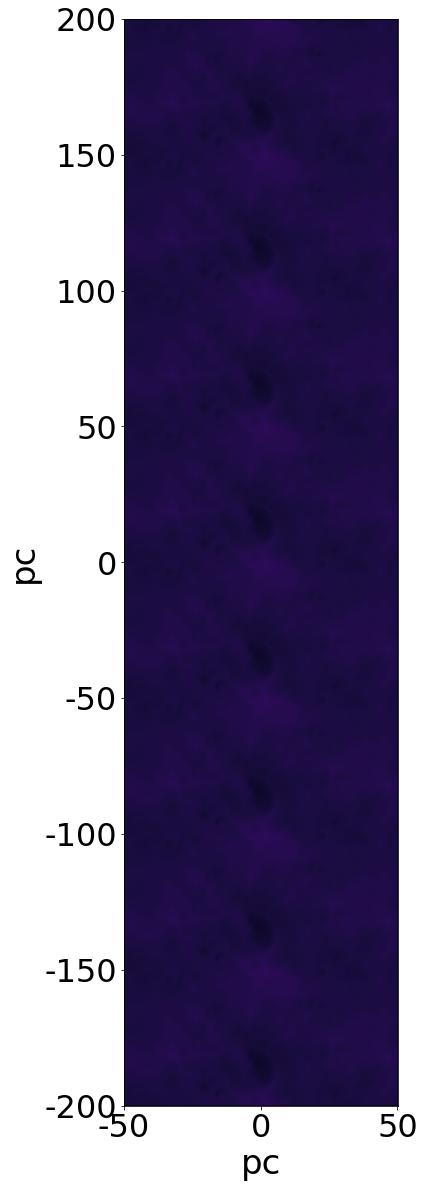}} &
    \hspace{0.1cm}
    \hspace{-0.5cm}\resizebox{!}{51mm}{\includegraphics{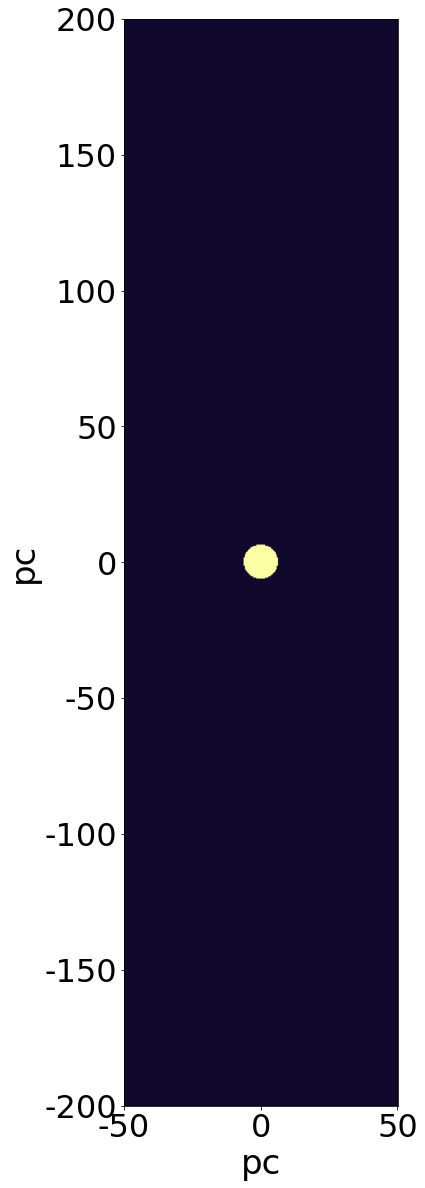}} & \hspace{-0.5cm}\resizebox{!}{51mm}{\includegraphics{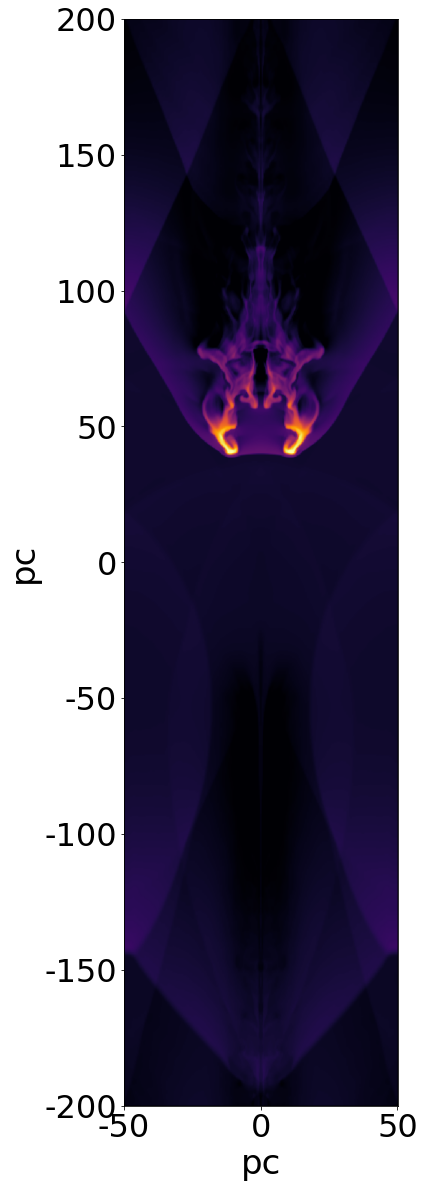}} & \hspace{-0.5cm}\resizebox{!}{51mm}{\includegraphics{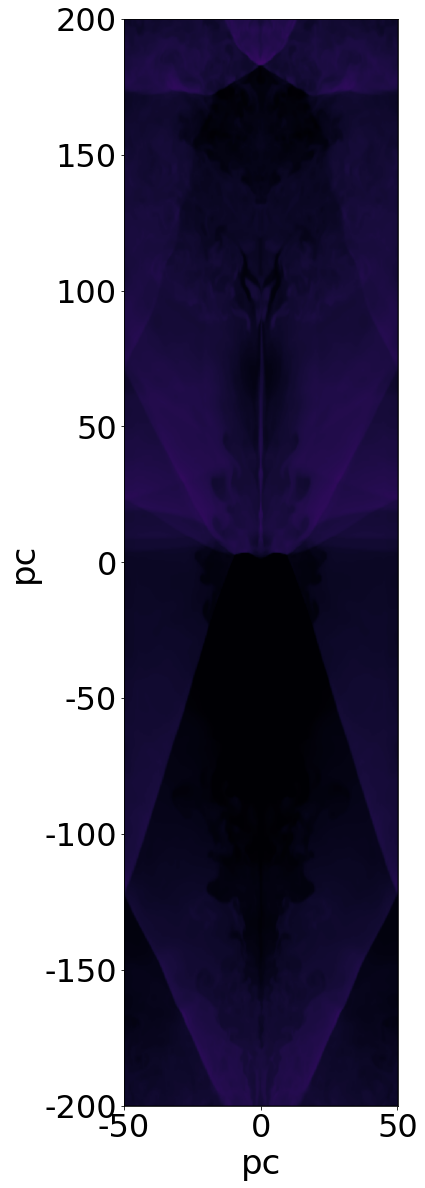}} & \hspace{-0.5cm}\resizebox{!}{51mm}{\includegraphics{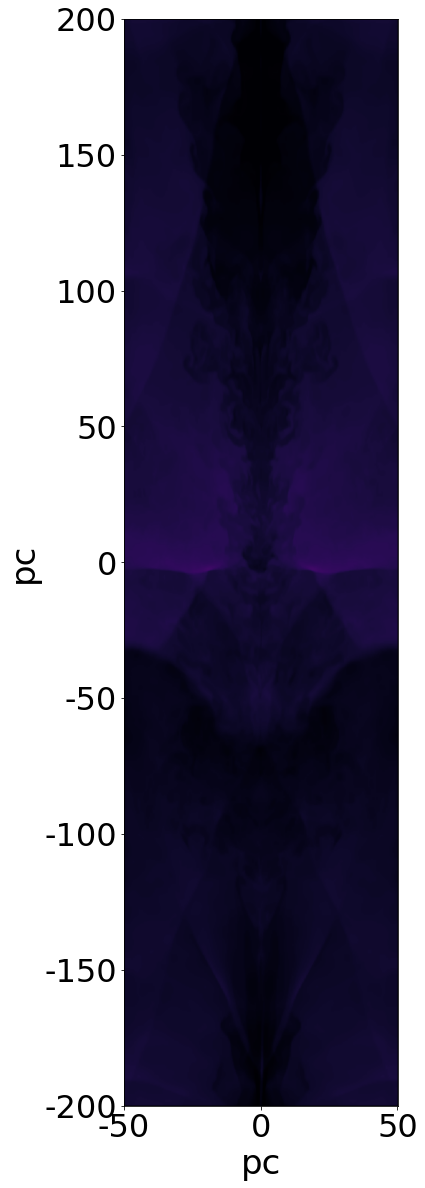}} &
    \hspace{-0.5cm}\resizebox{!}{51mm}{\includegraphics{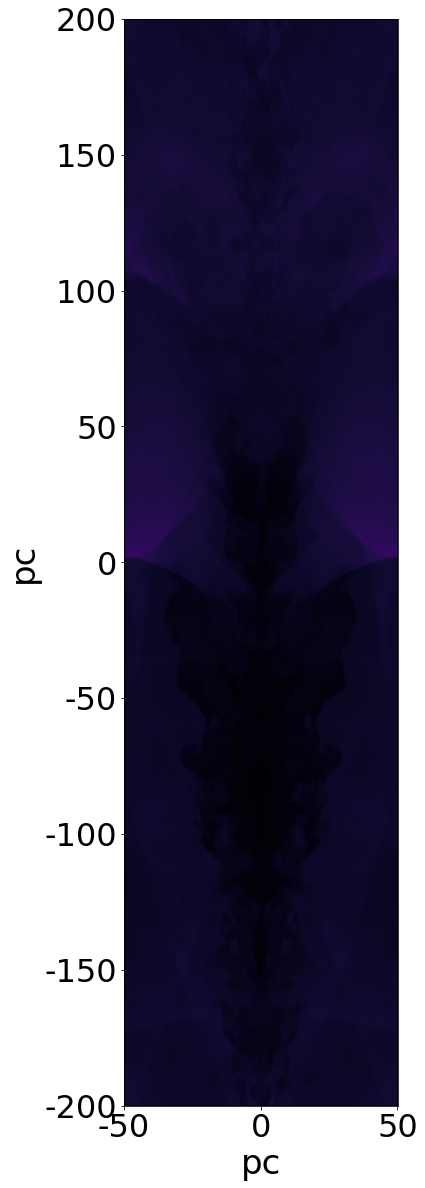}} &
  \end{tabular}
  \begin{tabular}{c@{\hspace{-.035cm}}c}
    \resizebox{95mm}{!}{\includegraphics{2Dslices/ARndbar.png}}
  \end{tabular}
  \caption{2D slices at $Z=0$ of the gas number density in adiabatic models with different cloud separation distances of $\delta = 2$ (upper left panel), $\delta = 4$ (middle left panel), $\delta = 8$ (lower left panel), $\delta = 16$ (upper right panel), $\delta = 32$ (middle right panel), and $\delta = 64$ (lower right panel) at five different times through the simulation (shown in columns). Hydrodynamic drag forces cause clouds separated by large distances to disintegrate and mix completely with the surrounding medium at a rapid pace. In contrast, a group of closely spaced clouds have the capacity to shield each other from drag forces and the erosive effects of dynamical instabilities.}
  \label{numden2}
\end{center}
\end{figure*}

\begin{figure*}
\begin{center}
  \begin{tabular}{c c c c c c c c c c c}
       \multicolumn{5}{l}{(a) Radiative ($\delta=2$)} & \multicolumn{5}{l}{(d) Radiative ($\delta=16$)}\\
       \multicolumn{1}{c}{{\scriptsize$t=0$}} & \multicolumn{1}{c}{{\scriptsize$49.1\,t_{\rm cross}$}} & \multicolumn{1}{c}{{\scriptsize$155.3\,t_{\rm cross}$}} & \multicolumn{1}{c}{{\scriptsize$253.5\,t_{\rm cross}$}}  & \multicolumn{1}{c}{{\scriptsize$351.6\,t_{\rm cross}$}}& \multicolumn{1}{c}{{\scriptsize$t=0$}} & \multicolumn{1}{c}{{\scriptsize$49.1\,t_{\rm cross}$}} & \multicolumn{1}{c}{{\scriptsize$155.3\,t_{\rm cross}$}} & \multicolumn{1}{c}{{\scriptsize$253.5\,t_{\rm cross}$}}  & \multicolumn{1}{c}{{\scriptsize$351.6\,t_{\rm cross}$}}\\    
       \hspace{-0.3cm}\resizebox{!}{51mm}{\includegraphics{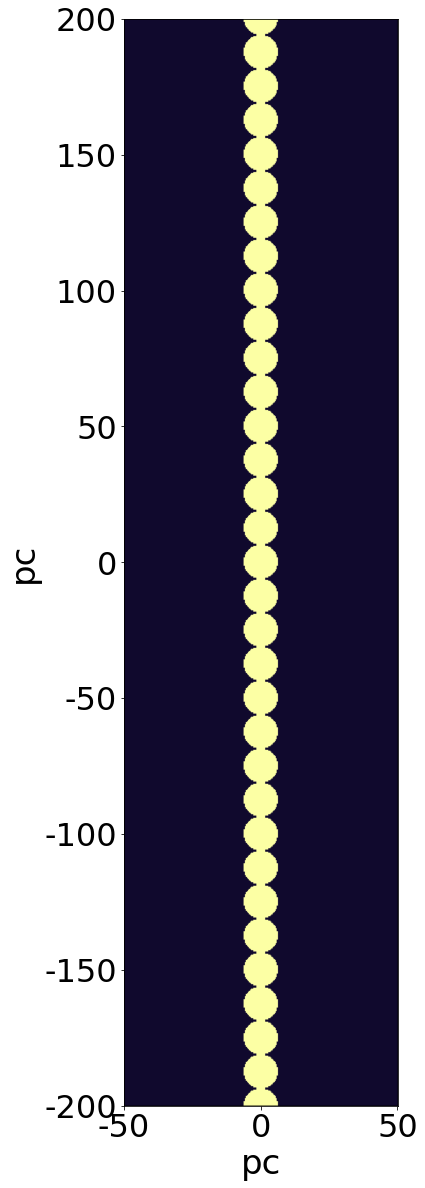}} & \hspace{-0.5cm}\resizebox{!}{51mm}{\includegraphics{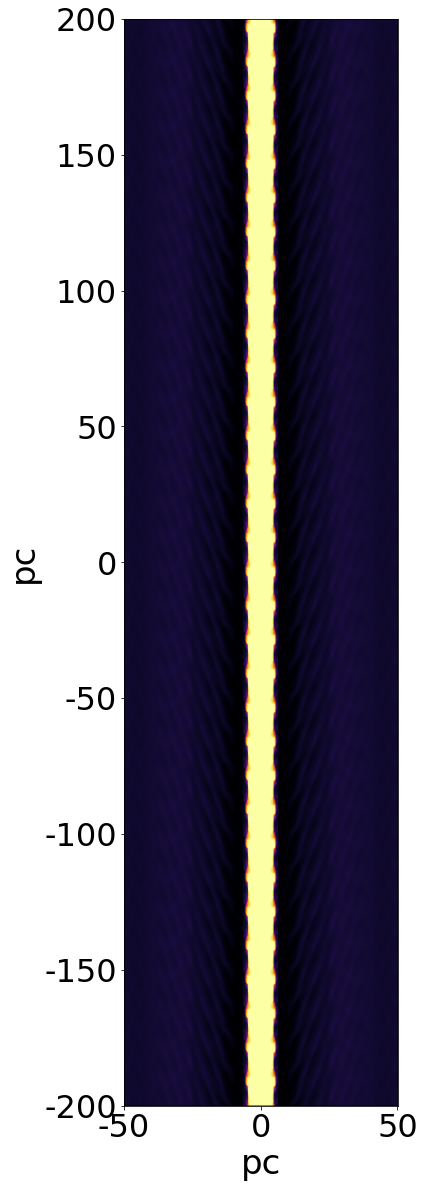}} & \hspace{-0.5cm}\resizebox{!}{51mm}{\includegraphics{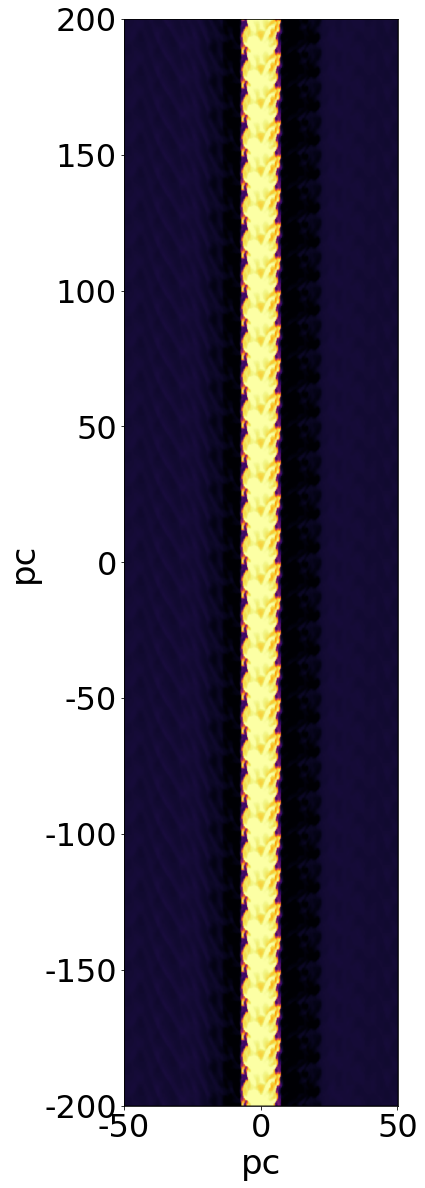}} & \hspace{-0.5cm}\resizebox{!}{51mm}{\includegraphics{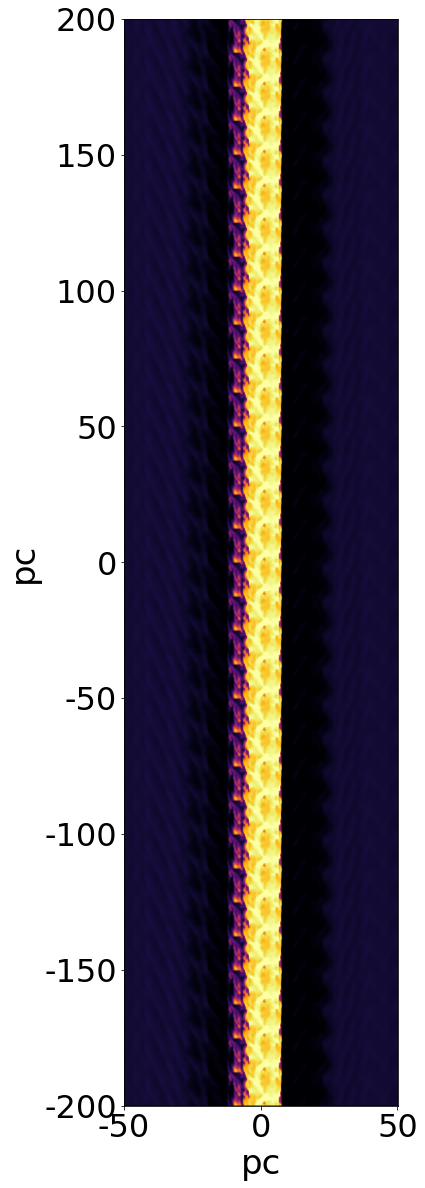}} &
       \hspace{-0.5cm}\resizebox{!}{51mm}{\includegraphics{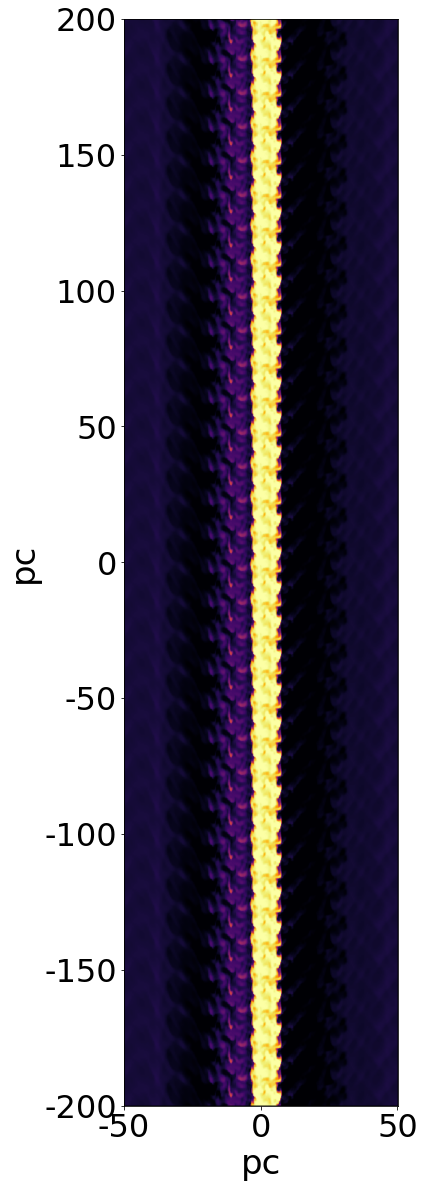}} &
       \hspace{0.1cm}
       \hspace{-0.5cm}\resizebox{!}{51mm}{\includegraphics{2Dslices/newradwithout16numden16_000.png}} & \hspace{-0.5cm}\resizebox{!}{51mm}{\includegraphics{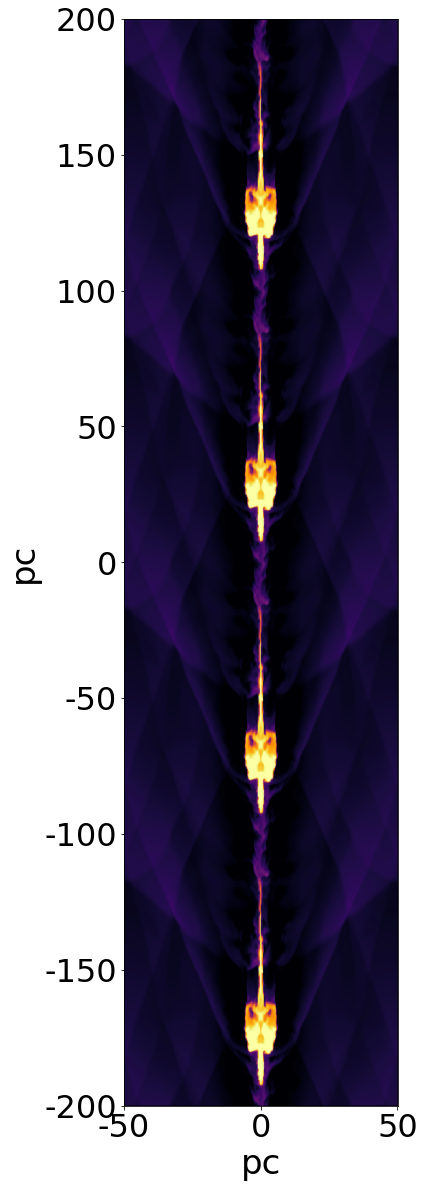}} & \hspace{-0.5cm}\resizebox{!}{51mm}{\includegraphics{2Dslices/newradwithout16numden16_040.png}} & \hspace{-0.5cm}\resizebox{!}{51mm}{\includegraphics{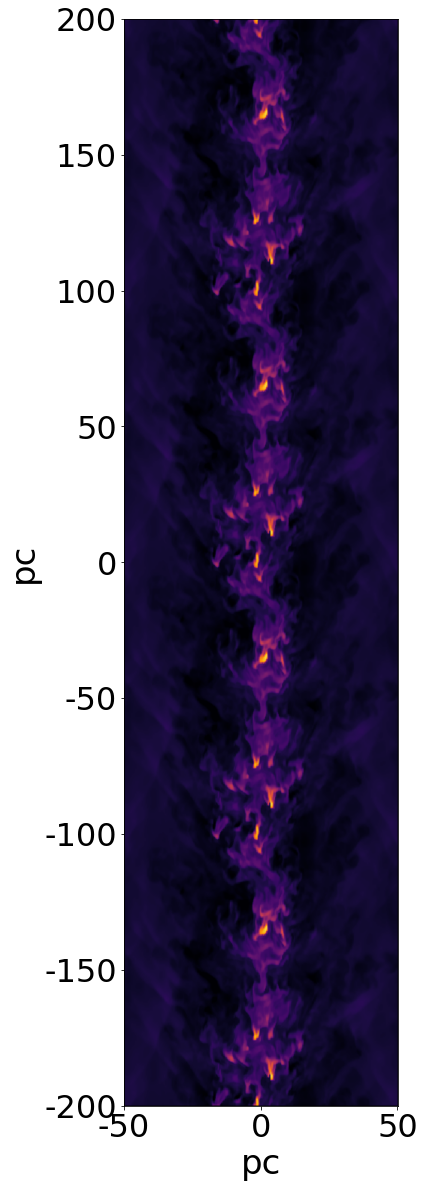}} &
       \hspace{-0.5cm}\resizebox{!}{51mm}{\includegraphics{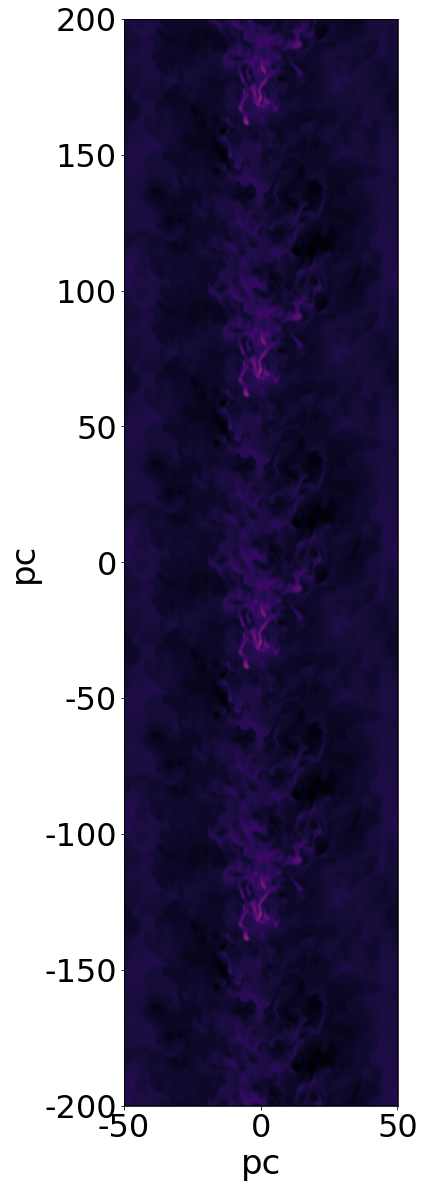}}
    \\
       \multicolumn{5}{l}{(b) Radiative ($\delta=4$)} & \multicolumn{5}{l}{(e) Radiative ($\delta=32$)}\\
       \multicolumn{1}{c}{{\scriptsize$t=0$}} & \multicolumn{1}{c}{{\scriptsize$49.1\,t_{\rm cross}$}} & \multicolumn{1}{c}{{\scriptsize$155.3\,t_{\rm cross}$}} & \multicolumn{1}{c}{{\scriptsize$253.5\,t_{\rm cross}$}}  & \multicolumn{1}{c}{{\scriptsize$351.6\,t_{\rm cross}$}}& \multicolumn{1}{c}{{\scriptsize$t=0$}} & \multicolumn{1}{c}{{\scriptsize$49.1\,t_{\rm cross}$}} & \multicolumn{1}{c}{{\scriptsize$155.3\,t_{\rm cross}$}} & \multicolumn{1}{c}{{\scriptsize$253.5\,t_{\rm cross}$}}  & \multicolumn{1}{c}{{\scriptsize$351.6\,t_{\rm cross}$}}\\   
    \hspace{-0.3cm}\resizebox{!}{51mm}{\includegraphics{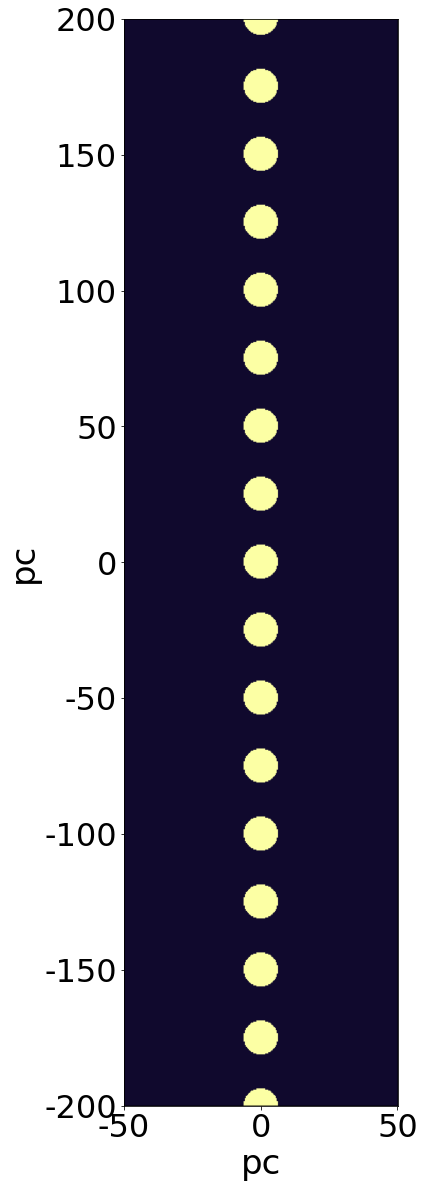}} & \hspace{-0.5cm}\resizebox{!}{51mm}{\includegraphics{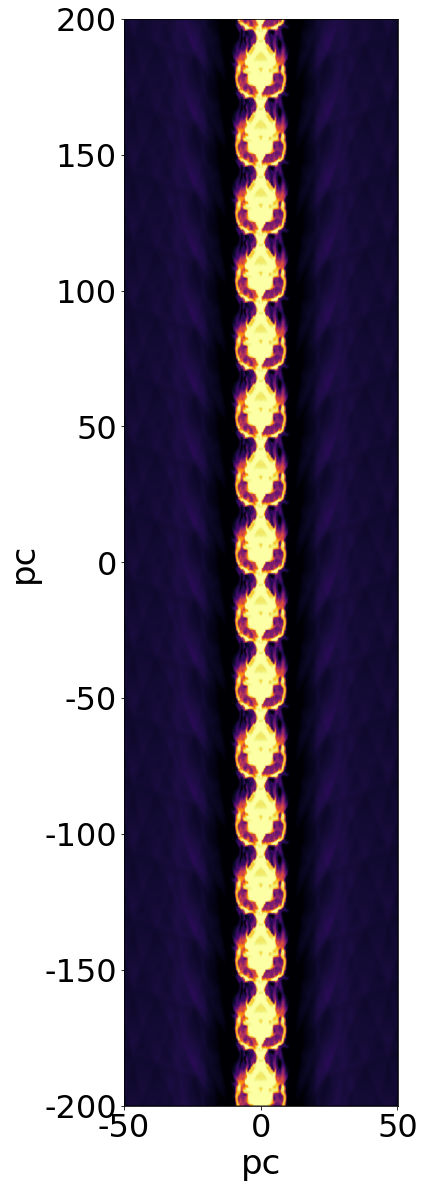}} & \hspace{-0.5cm}\resizebox{!}{51mm}{\includegraphics{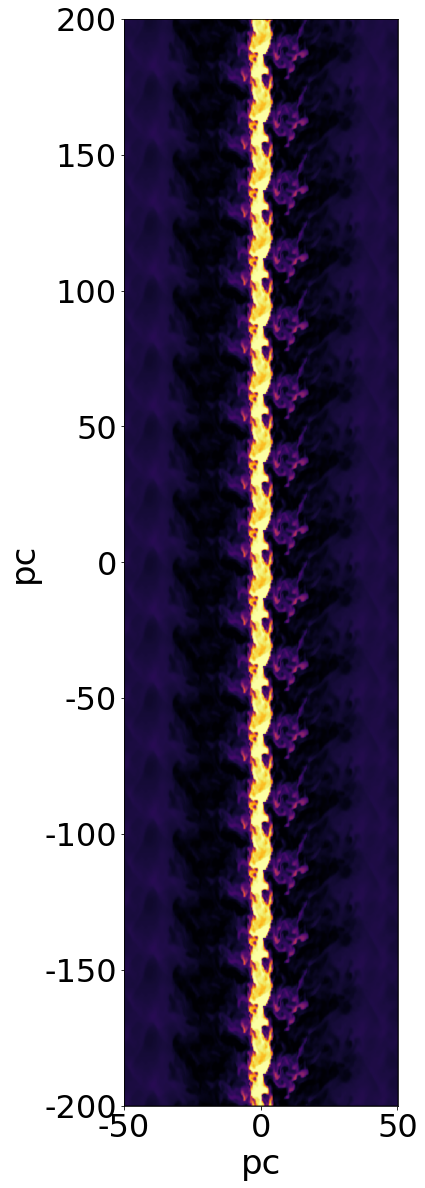}} & \hspace{-0.5cm}\resizebox{!}{51mm}{\includegraphics{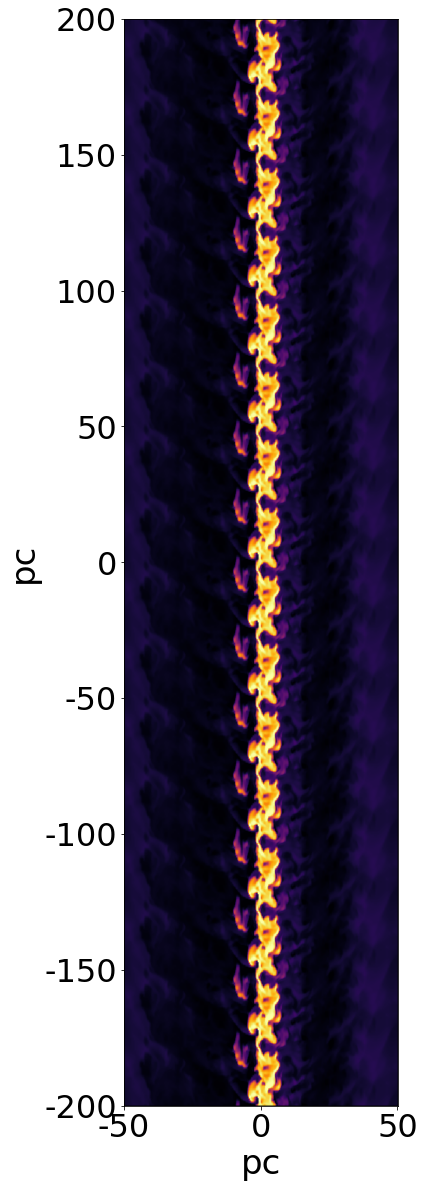}} &
    \hspace{-0.5cm}\resizebox{!}{51mm}{\includegraphics{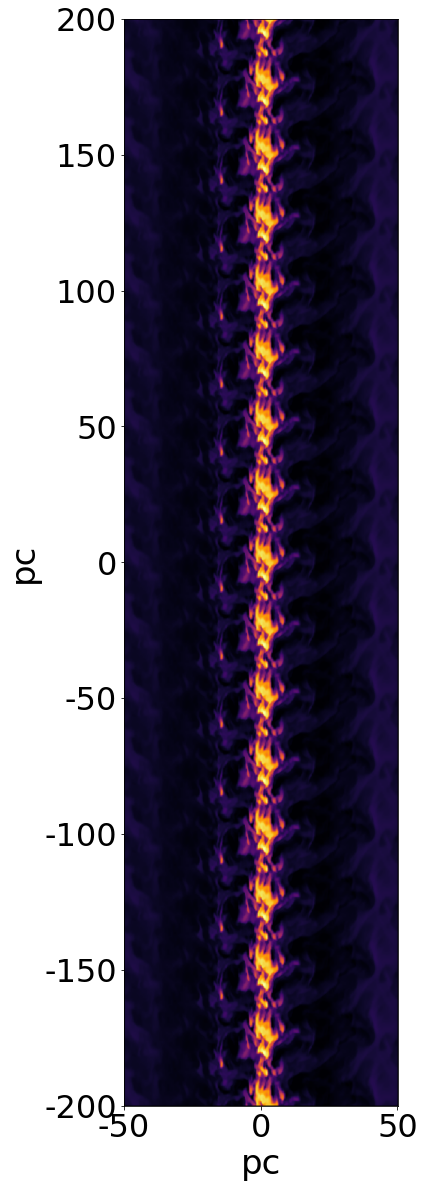}} &
    \hspace{0.1cm}
    \hspace{-0.5cm}\resizebox{!}{51mm}{\includegraphics{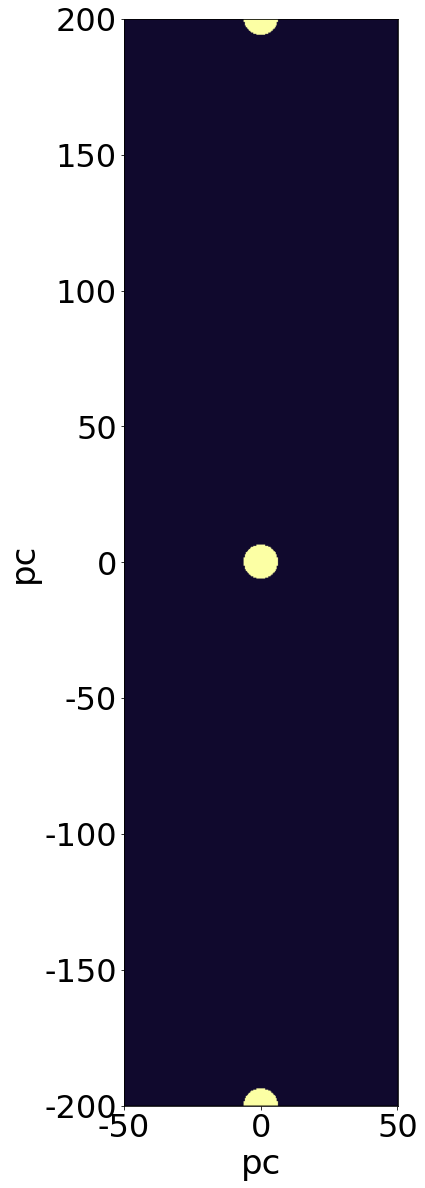}} & \hspace{-0.5cm}\resizebox{!}{51mm}{\includegraphics{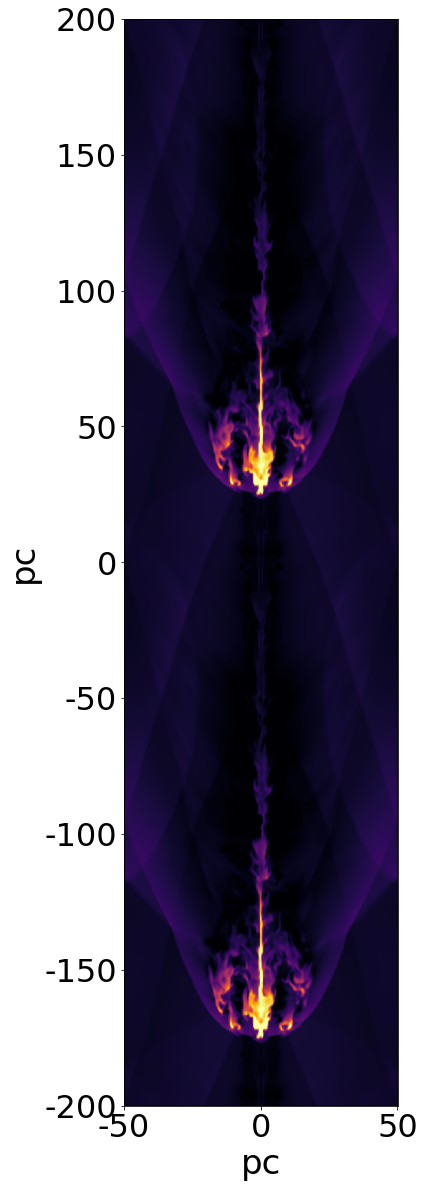}} & \hspace{-0.5cm}\resizebox{!}{51mm}{\includegraphics{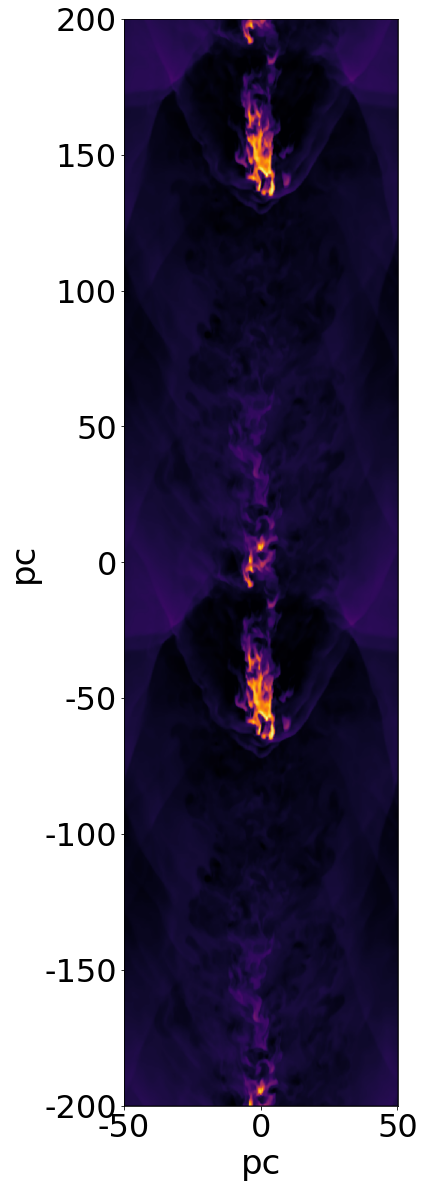}} & \hspace{-0.5cm}\resizebox{!}{51mm}{\includegraphics{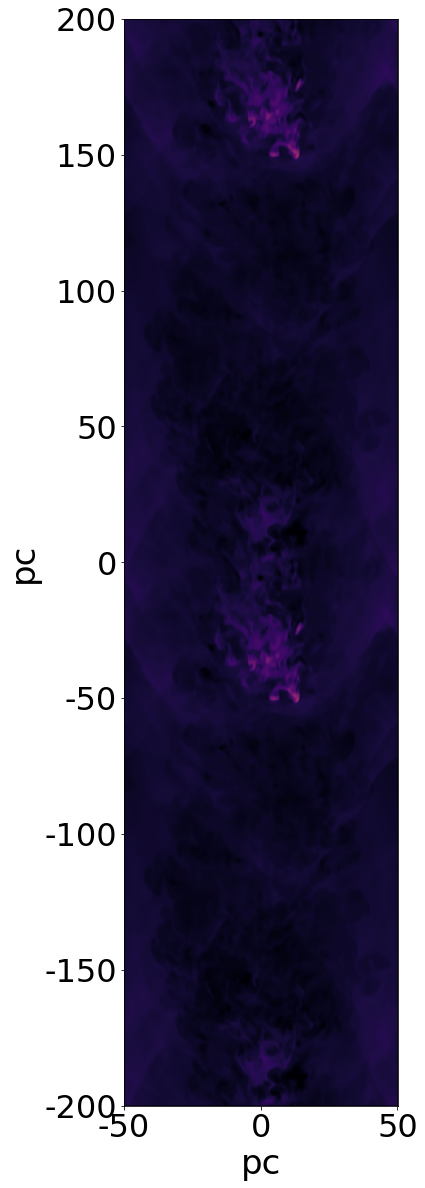}} &
    \hspace{-0.5cm}\resizebox{!}{51mm}{\includegraphics{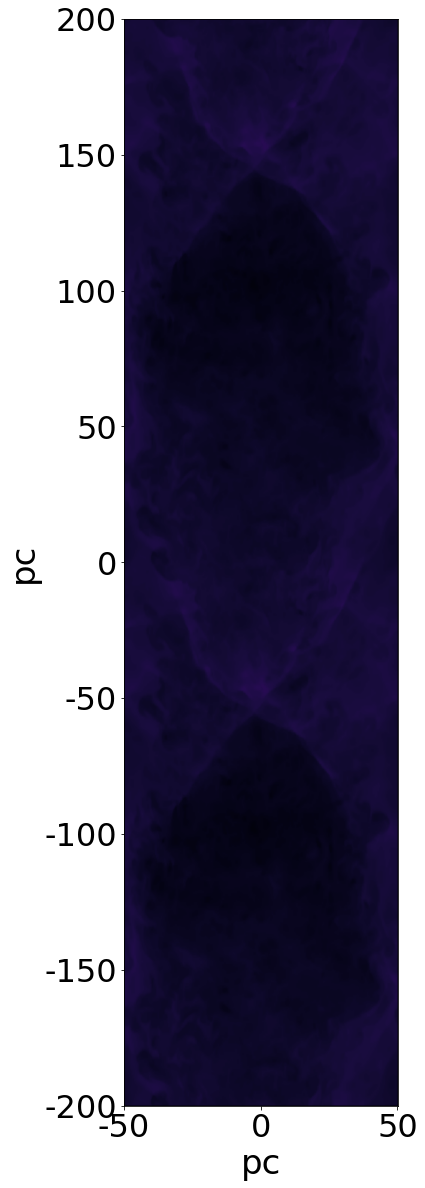}}
    \\
       \multicolumn{5}{l}{(c) Radiative ($\delta=8$)} & \multicolumn{5}{l}{(f) Radiative ($\delta=64$)}\\
       \multicolumn{1}{c}{{\scriptsize$t=0$}} & \multicolumn{1}{c}{{\scriptsize$49.1\,t_{\rm cross}$}} & \multicolumn{1}{c}{{\scriptsize$155.3\,t_{\rm cross}$}} & \multicolumn{1}{c}{{\scriptsize$253.5\,t_{\rm cross}$}}  & \multicolumn{1}{c}{{\scriptsize$351.6\,t_{\rm cross}$}}& \multicolumn{1}{c}{{\scriptsize$t=0$}} & \multicolumn{1}{c}{{\scriptsize$49.1\,t_{\rm cross}$}} & \multicolumn{1}{c}{{\scriptsize$155.3\,t_{\rm cross}$}} & \multicolumn{1}{c}{{\scriptsize$253.5\,t_{\rm cross}$}}  & \multicolumn{1}{c}{{\scriptsize$351.6\,t_{\rm cross}$}}\\   
    \hspace{-0.3cm}\resizebox{!}{51mm}{\includegraphics{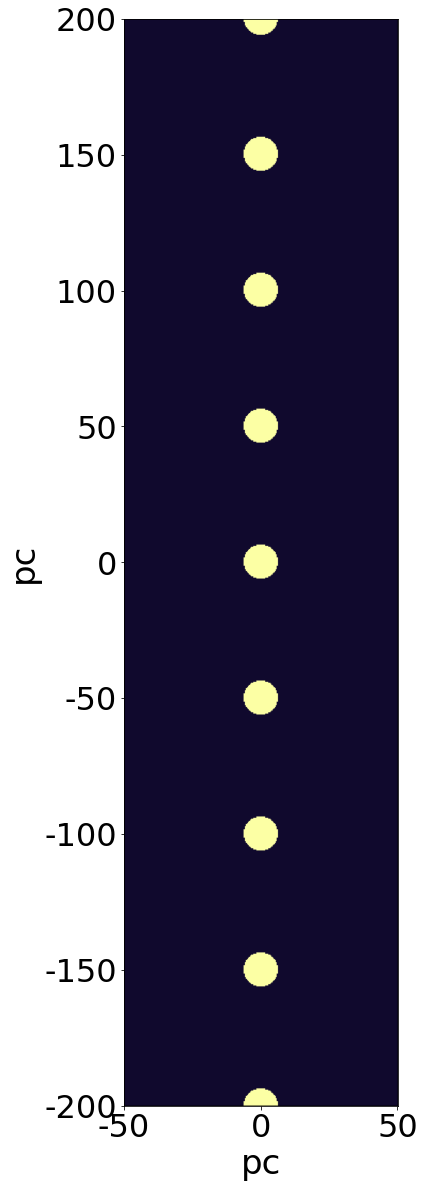}} & \hspace{-0.5cm}\resizebox{!}{51mm}{\includegraphics{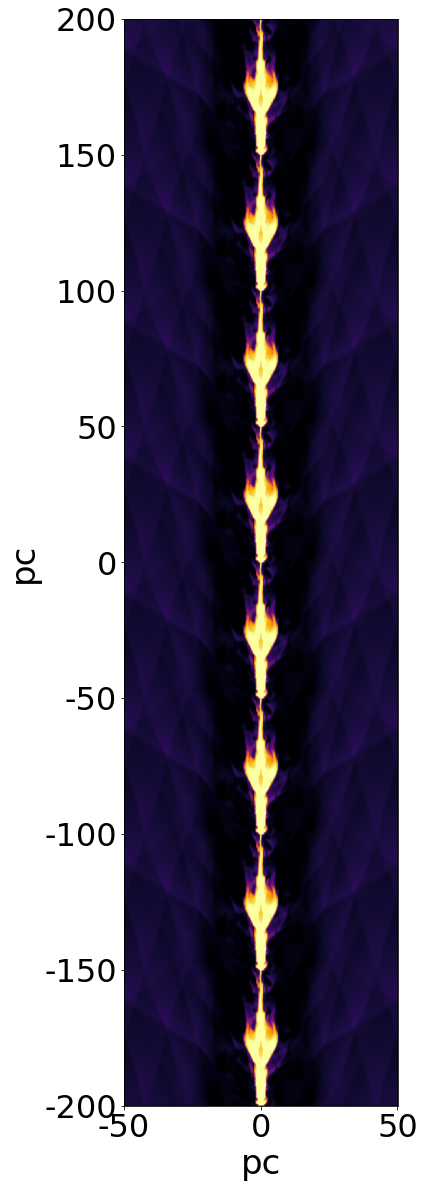}} & \hspace{-0.5cm}\resizebox{!}{51mm}{\includegraphics{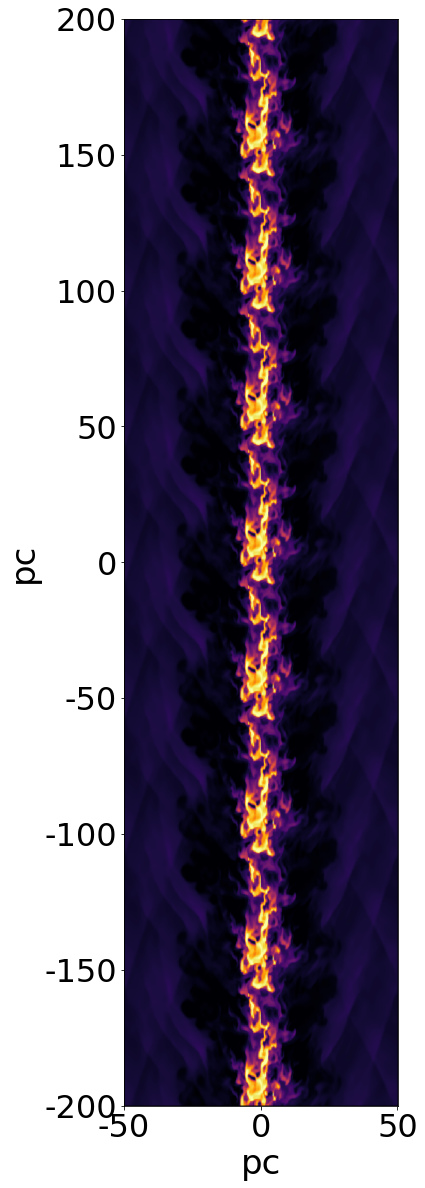}} & \hspace{-0.5cm}\resizebox{!}{51mm}{\includegraphics{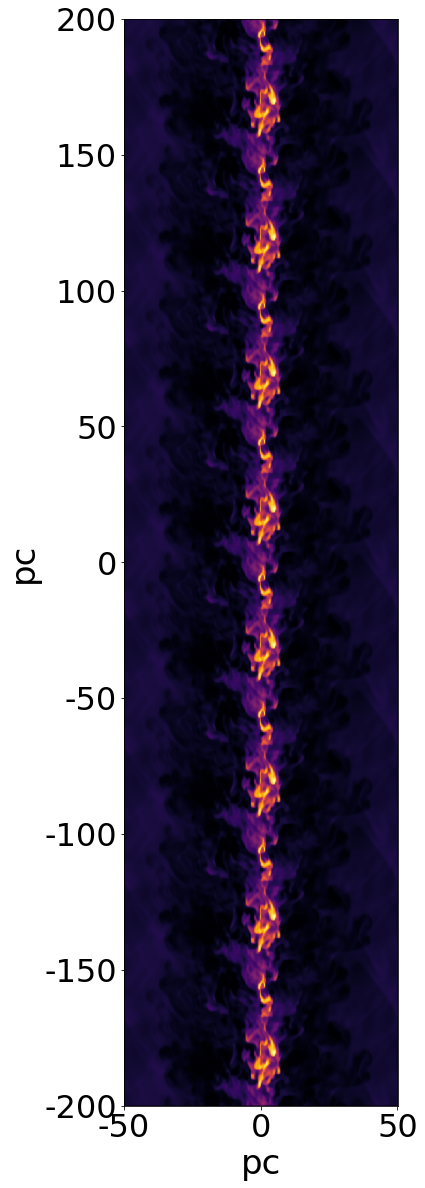}} &
    \hspace{-0.5cm}\resizebox{!}{51mm}{\includegraphics{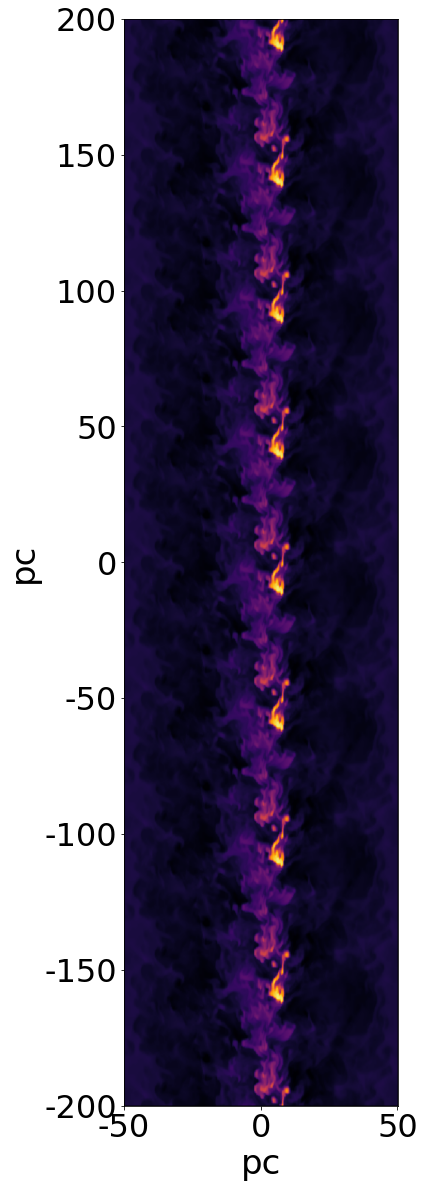}} &
    \hspace{0.1cm}
    \hspace{-0.5cm}\resizebox{!}{51mm}{\includegraphics{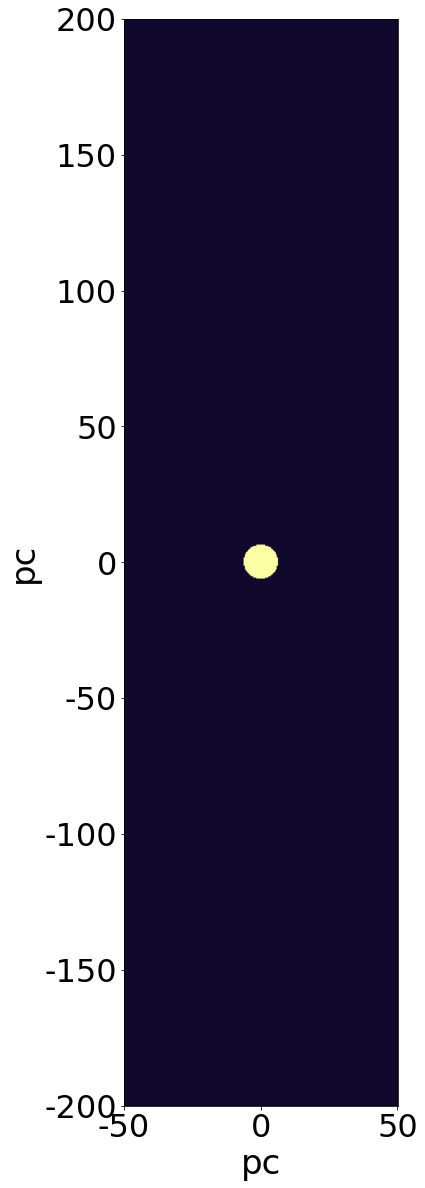}} & \hspace{-0.5cm}\resizebox{!}{51mm}{\includegraphics{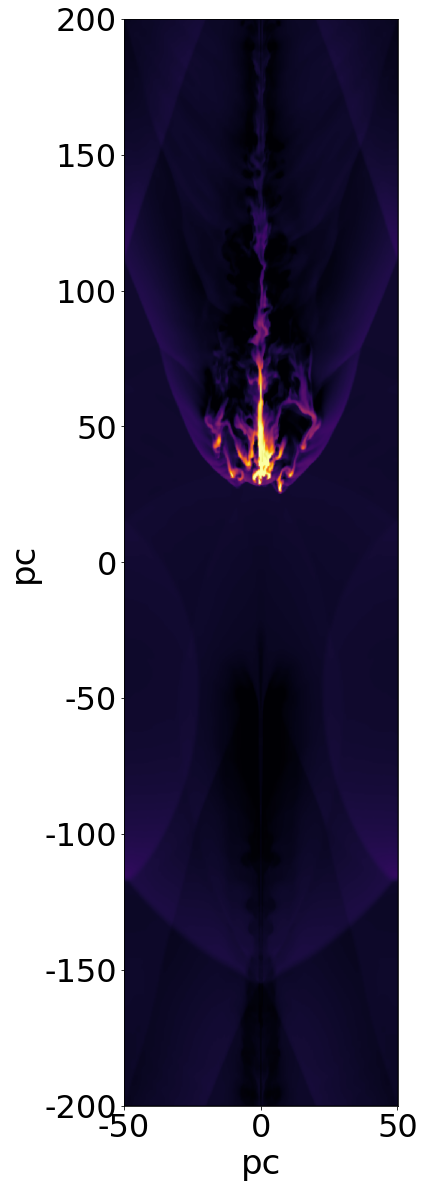}} & \hspace{-0.5cm}\resizebox{!}{51mm}{\includegraphics{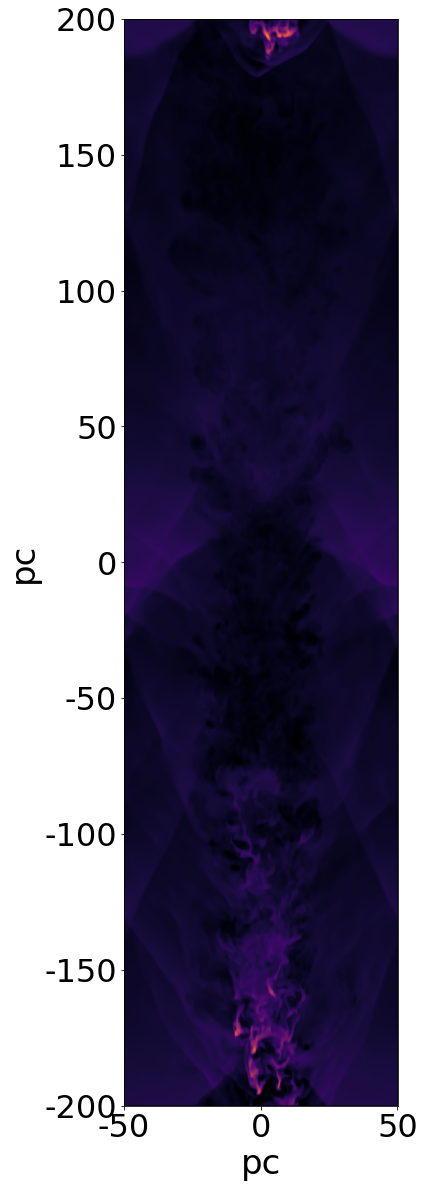}} & \hspace{-0.5cm}\resizebox{!}{51mm}{\includegraphics{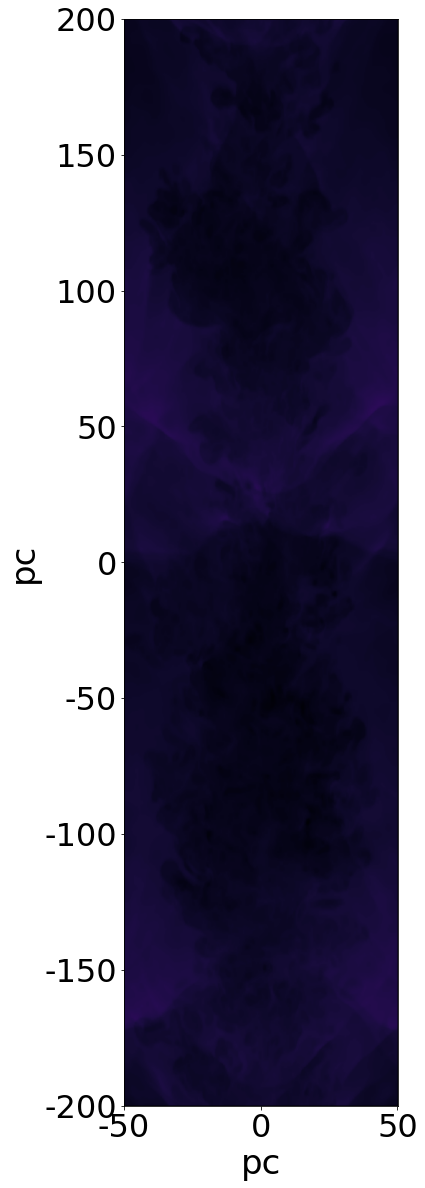}} &
    \hspace{-0.5cm}\resizebox{!}{51mm}{\includegraphics{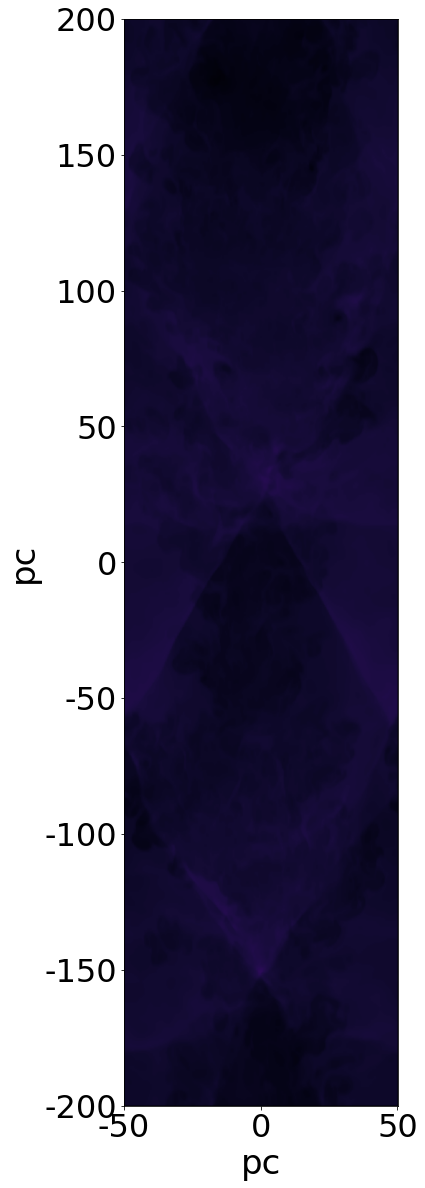}} 
  \end{tabular}
  \begin{tabular}{c@{\hspace{-.035cm}}c}
 \resizebox{95mm}{!}{\includegraphics{2Dslices/ARndbar.png}}
  \end{tabular}
  \caption{2D slices at $Z=0$ of the gas number density in radiative models with different cloud separation distances $\delta = 2$ (top left panel), $\delta = 4$ (middle left panel), $\delta = 8$ (bottom left panel), $\delta = 16$ (top right panel), $\delta = 32$ (middle right panel), and $\delta = 64$ (bottom right panel) at five different times through the simulation (columns). Cooling can form a multiphase medium characterised by continuous condensation, which allows for the preservation of a large fraction of dense gas. Interactions between clouds shield the cold material in downstream clouds from drag forces from the background medium by interposing a layer of disrupted re-condensed material. This reduces the effective drag forces acting on downstream clouds. Decreasing the inter-cloud separation distance results in a much smoother transition. This is because the hydrodynamic shielding effect is more pronounced when we consider radiative cooling at smaller separation distances, which helps to mitigate the disruptive effects of the drag force.} 
  \label{numden3}
\end{center}
\end{figure*}

\subsubsection{The role of the inter-cloud separation distance ($\delta$)}\label{hs1}

\begin{figure*}
\begin{center}
  \begin{tabular}{r r r} 
  \multicolumn{3}{l}{(a) Abiabatic models}\\
    \multicolumn{1}{l}{Cloud gas temperature} & \multicolumn{1}{l}{Cloud gas density} & \multicolumn{1}{l}{Dense gas mass fraction}  \\ 
       \hspace{-0.3cm}\resizebox{!}{44mm}{\includegraphics{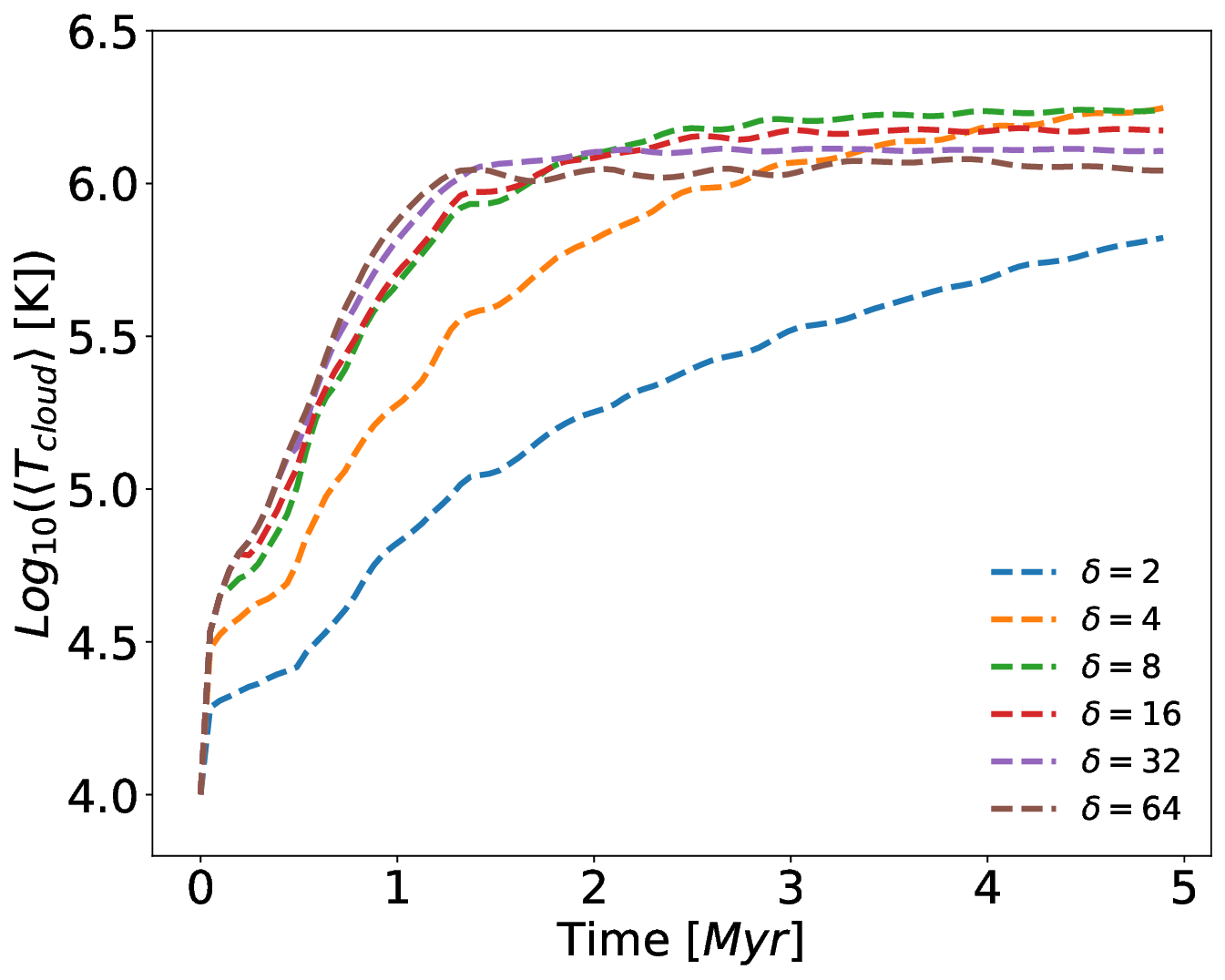}} & \hspace{-0.3cm}\resizebox{!}{44mm}{\includegraphics{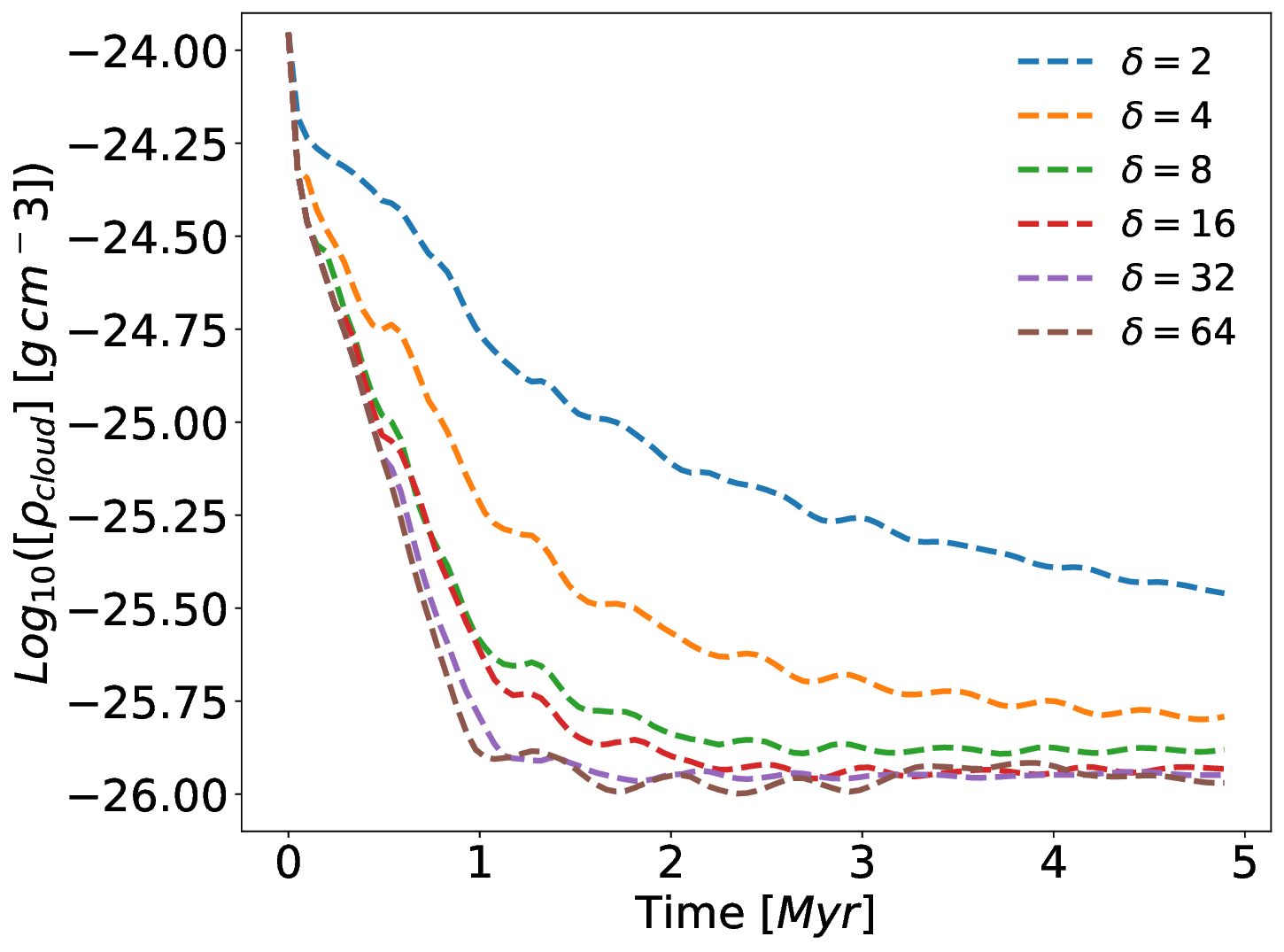}} & \hspace{-0.3cm}\resizebox{!}{44mm}{\includegraphics{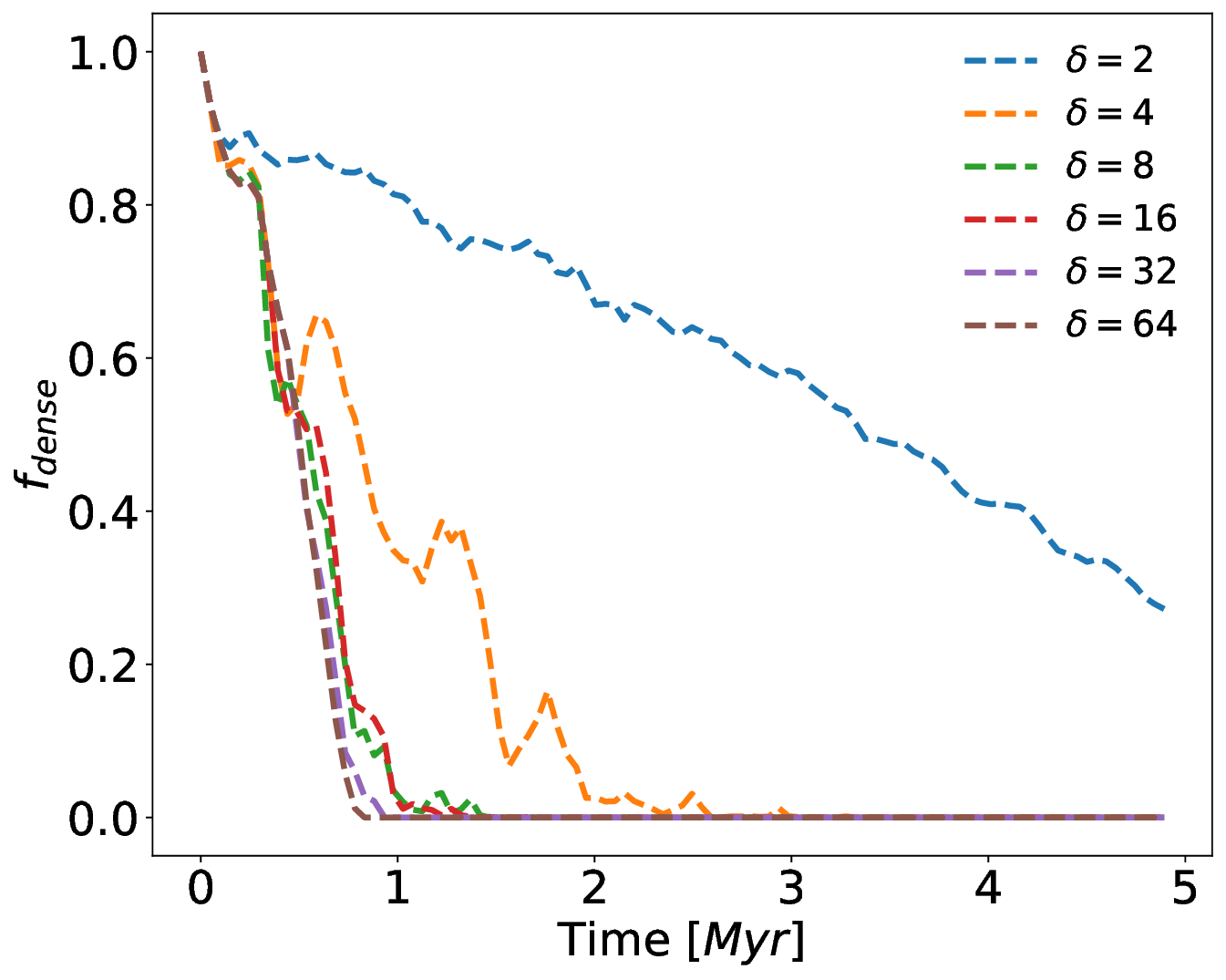}}
       \\
  \multicolumn{3}{l}{(b) Radiative models}\\
    \multicolumn{1}{l}{Cloud gas temperature} & \multicolumn{1}{l}{Cloud gas density} & \multicolumn{1}{l}{Dense gas mass fraction}  \\ 
       \hspace{-0.3cm}\resizebox{!}{44mm}{\includegraphics{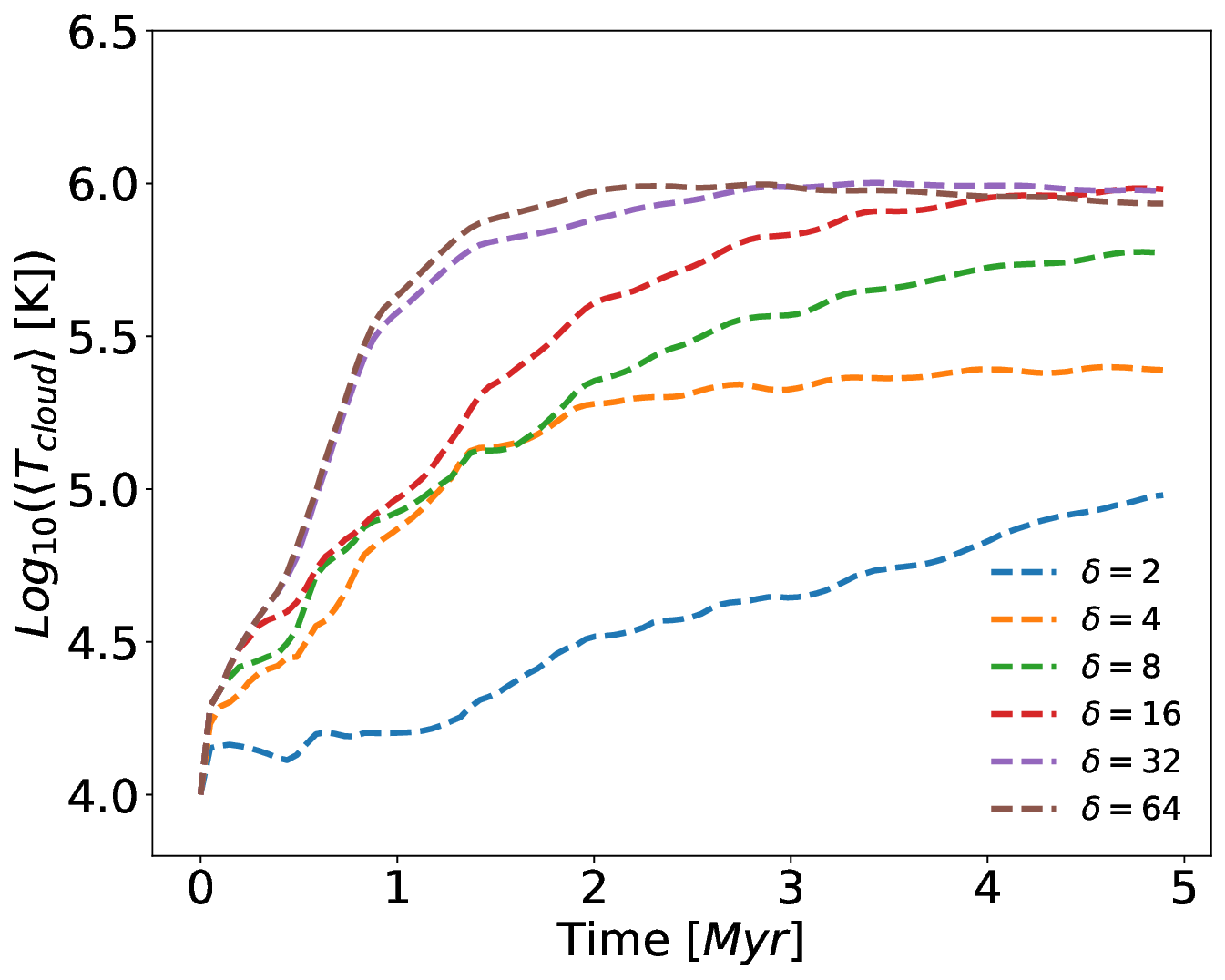}} & \hspace{-0.3cm}\resizebox{!}{44mm}{\includegraphics{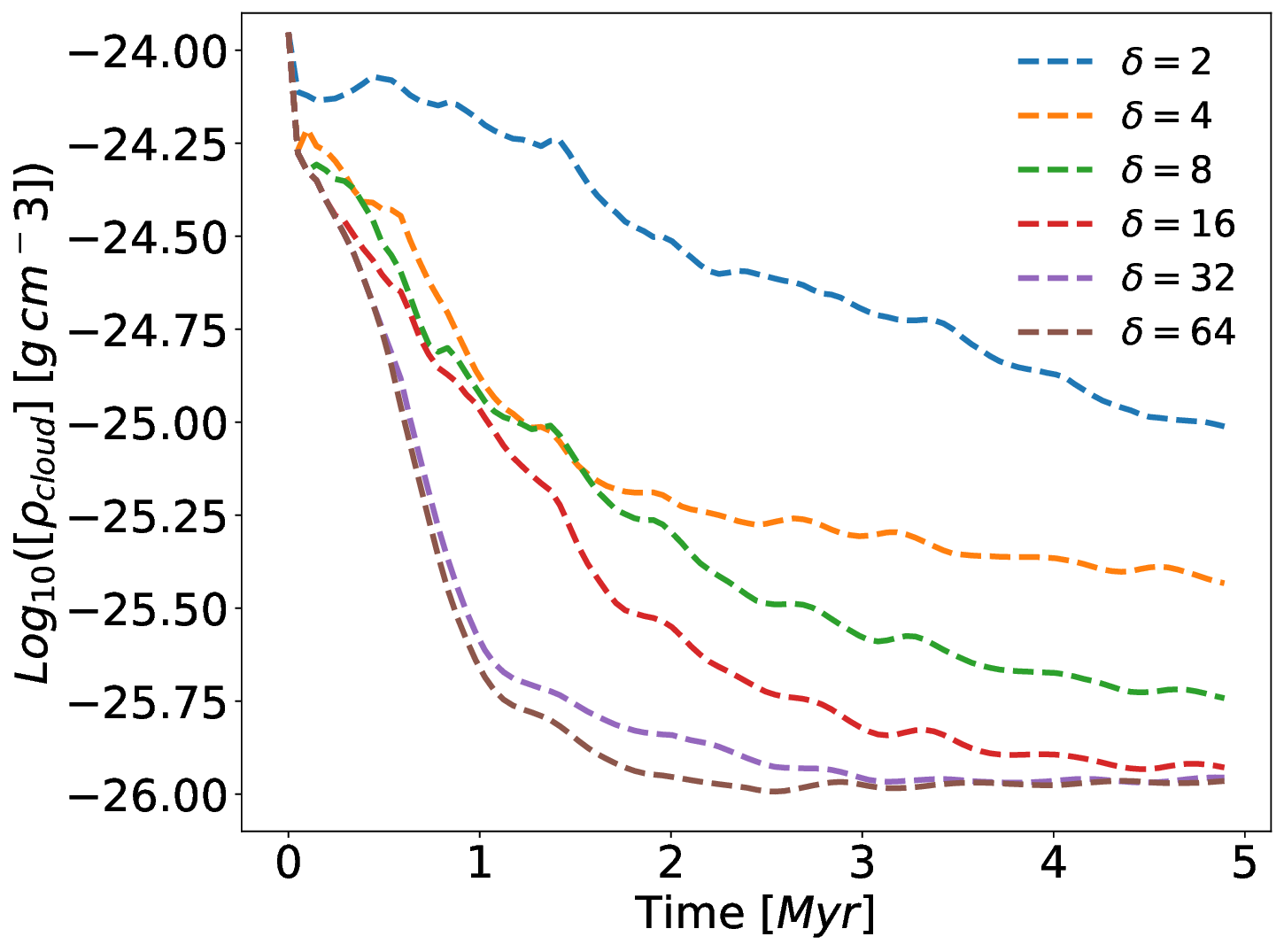}} & \hspace{-0.3cm}\resizebox{!}{44mm}{\includegraphics{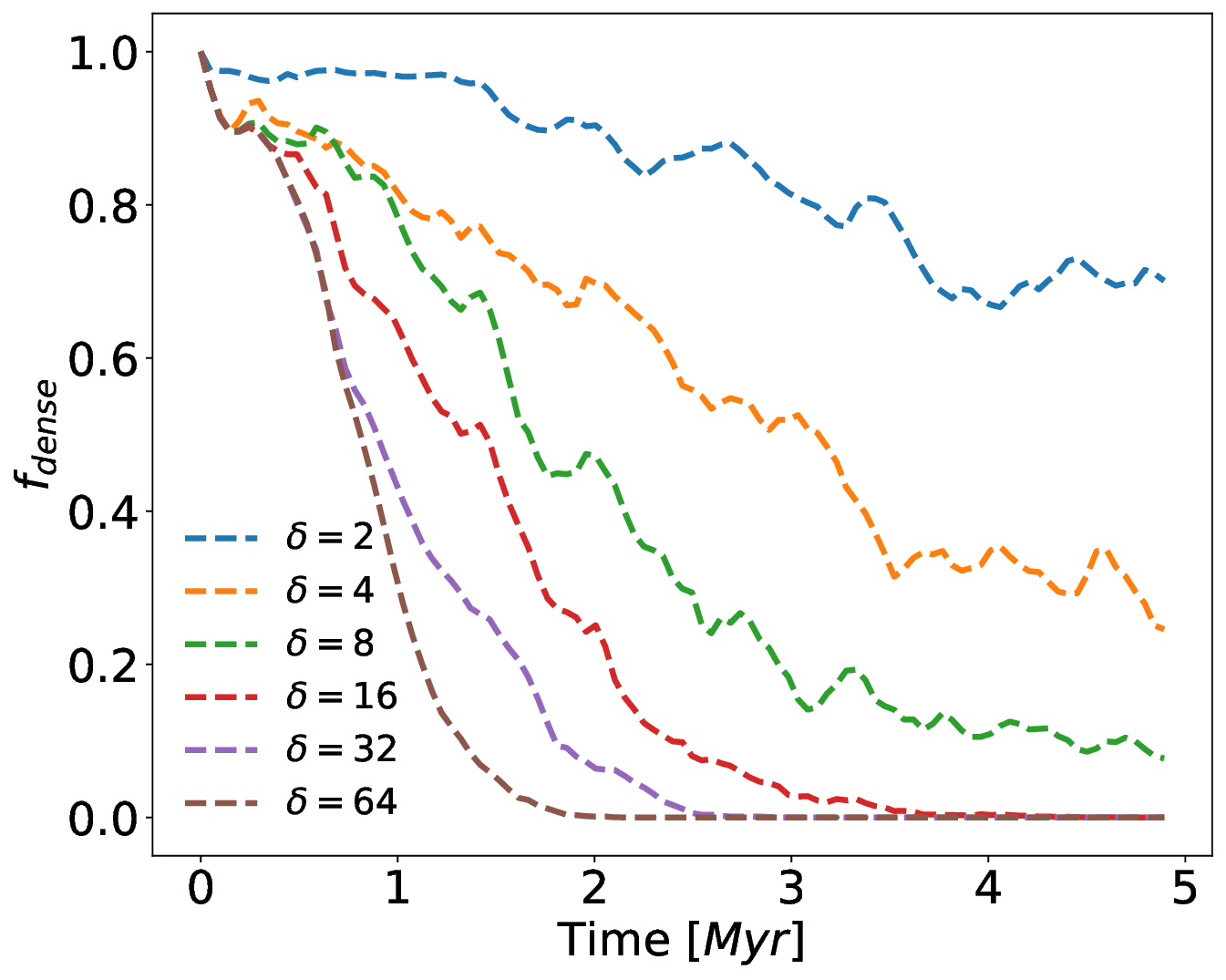}}
       \\
  \end{tabular}
  \caption{Time evolution of the mass-weighted average temperature (left column), mean density (middle column), and dense gas mass fraction (right column) for different values of the parameter $\delta$ in the adiabatic (top) and radiative (bottom) models. Smaller cloud spacing enhances the protection of cold material from shredding and erosion, preventing a drastic temperature increase. Shielding in adiabatic models requires smaller $\delta_{\rm shield}$ than radiative models, so condensation is key for gas survival. 
  } 
  \label{evo3}
\end{center}
\end{figure*}

Figures \ref{numden2} and \ref{numden3} show 2D slices at $Z=0$ of the gas number density for several cloud separation distances, ranging from $\delta = 2$ to $\delta = 64$, for adiabatic and radiative models, respectively. In Figure \ref{evo3}, we report the evolution of various parameters, such as the mass-weighted average temperature, the mean density, and the dense gas mass fraction of cloud material for the same models (adiabatic in the top row and radiative in the bottom row). All these maps and curves follow the same trends discussed in Section \ref{Sec4.1}, i.e. cloud gas in adiabatic models reaches higher temperatures, loses more mass, and retains less dense gas than their radiative counterparts. In addition, these figures reveal extra effects: (i) a very clear dependence on $\delta$ as the trend indicates that the closer the clouds are placed, the best the protection against shredding and erosion of cold material, which in turn prevents a drastic increase in temperature, (ii) the threshold separation distance for hydrodynamic shielding to occur is smaller in our supersonic models than in the subsonic and transonic models reported in \citep{Forbes_2019}, and (iii) hydrodynamic shielding operates differently in adiabatic and radiative models.\par

Regarding point (i), the effects of placing a group of cold clouds in a hot background wind vary widely with $\delta$. Clouds, separated by large distances (i.e. with high $\delta$ values), are unable to withstand the hydrodynamic drag forces, and disintegrate and mix with the surrounding medium at a very fast pace (in agreement with results from adiabatic single-cloud models by e.g. \citealt{Nakamura_2006,2016MNRAS.457.4470P}). In contrast, groups of clouds, separated by systematically smaller distances are able to reduce drag and preserve increasingly larger fractions of dense gas. Thus, reducing the separation distance between clouds results in a substantial increase in hydrodynamic shielding. When the separation distance is set to $\delta = 64$, the cloud can be considered as quasi-isolated because hydrodynamic shielding is absent. Therefore, there is a significant difference between quasi-isolated clouds that are rapidly disrupted and those that provide mutual shielding.

\subsubsection{Hydrodynamic shielding in subsonic vs. supersonic cases}
\label{hs2}

Regarding point (ii), closely-spaced clouds can effectively act as a cylindrical-like stream of cold material, and whether hydrodynamic shielding occurs or not depends on how fast information from upstream clouds travels downstream. Following \citep{Forbes_2019}, we set that time to the cloud-crushing time, obtaining the following expression for the threshold $\delta$ value in adiabatic subsonic models (see Figure \ref{numden2}):
\begin{equation}\label{eq:sub_delta}
    \delta^{\rm sub}_{\rm shield} \approx 2+\sqrt{\chi}\left(1+\frac{1}{{\cal M}_{\rm wind}}\right),
\end{equation}

\noindent where the extra term of $2$ in front results from considering the actual separation distance between the clouds, which is ($\delta-2\,r_{\rm cl}$).




In adiabatic supersonic cases (like our models in Figure \ref{numden3}), the flow exceeds the wind sound speed, so applying the same definition as above results in a much smaller threshold separation distance of:
\begin{equation}\label{eq:sup_delta}
    \delta^{\rm sup}_{\rm shield} \approx 2+\sqrt{\chi}\left(\frac{1}{{\cal M}_{\rm wind}}\right).
\end{equation}

This is reasonable because in supersonic flows downstream clouds are ``unaware'' of the destruction of upstream clouds and information arrives with shocked gas in the wind flow, which then results in a more drastic constraint for hydrodynamic shielding to emerge. Our simulations then show that in adiabatic models the threshold separation distance of supersonic cases becomes much smaller than in subsonic cases (cf. Figure 2 in \citealt{Forbes_2019}, where a smoother transition can be seen for increasing $\delta$ values). The resulting dependence of the threshold value, $\delta_{\rm shield}$, on the wind Mach number is shown in Figure \ref{plot:shield} for various cases.\par

For our wind Mach number, using equation \ref{eq:sup_delta} we obtain $\delta^{\rm sup}_{\rm shield}\approx 4.9$, which is in agreement with the results presented in Figure \ref{evo3}, particularly with $f_{\rm dense}$ (top right panel in Figure \ref{evo3}). This panel shows strong hydrodynamic shielding only for $\delta \lesssim 4$ (much smaller than what the subsonic equation \ref{eq:sub_delta} predicts, i.e., $\delta^{\rm sub}_{\rm shield}\approx 14.9$), and very weak or no shielding for larger distances. It is clear that hydrodynamic shielding in adiabatic supersonic cases requires very small separation distances between the clouds as they can be disrupted faster than what hydrodynamic shielding requires to develop. In other words, adiabatic supersonic models require clouds to be in almost direct contact for hydrodynamic shielding to operate, otherwise it is absent.


\subsubsection{Hydrodynamic shielding in adiabatic vs. radiative models}
\label{hs3}

Regarding point (iii), we find that separation $\delta$ values have a higher impact on radiative clouds than they do on adiabatic clouds, pointing to differences in how hydrodynamic shielding operates in adiabatic and radiative models. This is due to the combination of two effects, namely the damping of shear instabilities and the re-condensation of cold gas, which are both present in radiative models. Firstly, dynamical instabilities are not as effective in the radiative cases owing to the higher densities, so dense material is preserved in narrower stream-like structures. Comparing adiabatic and radiative models with the same $\delta$, we find that radiative models consistently exhibit a longer lifespan of dense cloud material for all $\delta$ values. As an example, when the separation value is $\delta=2$ (see Figure \ref{evo3}), the multicloud system preserves a dense-gas mass fraction of $\sim 30 \%$ in the adiabatic run and $>70\%$ in the radiative run by the end of the simulation period ($t\sim 5\,\rm Myr)$.\par

Secondly, in radiative cases cooling promotes the condensation of warm mixed material into the cold phase, contributing to the preservation of gas with temperatures $T\lesssim 10^{5}\,\rm K$. Since the cooling time of mixed gas is small compared to other dynamical times, the mixed phase has sufficient time to cool and populate the inter-cloud region in between the clouds. Re-condensed gas effectively decreases the distance between dense gas along the stream, thus offering protection to downstream dense cores by reducing drag forces. Re-condensation via radiative cooling is then essential for hydrodynamic shielding to emerge in supersonic cases. To estimate $\delta_{\rm shield}$ in radiative supersonic scenarios, a more relevant length scale to consider is the cooling length of cold and mixed gas. This length scale tells us how far cloud gas travels before it re-condenses. Since the cooling length is very small compared to the cloud radius for our simulations, cloud gas is rapidly replenished along the stream, and hydrodynamic shielding can operate much more effectively than in adiabatic supersonic counterparts. Considering $(\delta\,\ell_{\rm cool}-2\,r_{\rm cl})$ as the relevant travel distance results in the following condition for $\delta_{\rm shield}$ in radiative cases:
\begin{equation}\label{eq:sup_rad_delta}
    \delta^{\rm sup, rad}_{\rm shield} \approx \frac{r_{\rm cl}}{\ell_{\rm cool}}\left[2+\sqrt{\chi}\left(\frac{1}{{\cal M}_{\rm wind}}\right)\right].
\end{equation}

The equation above implies that for a fixed ${\cal M}_{\rm wind}$, the critical separation distance between clouds depends on how weak or strong cooling is (as quantified by the cooling length, $\ell_{\rm cool}$). For the CGM-motivated parameters we chose for our models, $\ell_{\rm cool}=0.15\,r_{\rm cl}$, so radiative cooling is strong. As a result, $\delta^{\rm sup,rad}_{\rm shield} \approx 33.8$ and the net effect of cooling is to facilitate hydrodynamic shielding at much larger distances than in adiabatic models for the same initial cloud separation distance. Figures \ref{evo3} and \ref{evo4} (see $f_{\rm dense}$ and $f_{\rm cold}$) confirm this prediction as the boundary between shielding and no shielding appears to be at $\delta=32$. Figure \ref{plot:shield} shows how $\delta_{\rm shield}$ across ${\cal M}_{\rm wind}$ for our radiative models (solid, dark blue line) and for models with other cooling lengths (dashed, grey lines). For stronger cooling, clouds can be farther apart and still experience hydrodynamical shielding, while for weaker cooling they would need to be more closely spaced. As expected, for weak cooling ($\ell_{\rm cool}\sim r_{\rm cl}$), equation \ref{eq:sup_rad_delta} reverts back to the adiatic case (see equation \ref{eq:sup_delta}).


\subsection{Shielding effects on cloud survival and dynamics}\label{hs4_m}

\subsubsection{Cloud mixing and survival of cold gas}\label{hs4}

The left-hand side column of Figure \ref{evo4} shows the mixing fractions in both adiabatic (top) and radiative (bottom) models. We find that: (i) mixing is significantly enhanced as we increase the value of $\delta$, and (ii) for a fixed separation distance $\delta$, mixing fractions in adiabatic cases are always lower than in their radiative counterparts. Regarding (i), the models with larger $\delta = 16, 32, 64$ show similar mixing profiles, indicating that when clouds are farther apart, there is greater disruption by turbulence induced by KH and RT instabilities. Conversely, when clouds are closer together, i.e. $\delta = 2, 4, 8$, the degree of mixing is significantly reduced due to hydrodynamic shielding, which protects the clouds from these instabilities. Regarding (ii), unlike adiabatic clouds, radiative clouds undergo mixing not only via dynamical instabilities and the turbulence they induce, but also via cooling-induced pressure gradients driven by condensation of gas in the warm mixed phase. As a result, mixing is not as fast and efficient in adiabatic models as in radiative models. This is in agreement with the results presented in \cite{Banda_21} (cf. with their Figure 7).\par

Mixing also has a direct influence on the cold, dense gas content in the flow. The middle column of Figure \ref{evo4} shows the cold gas mass fraction ($f_{\rm cold}$), which exhibit the same effects as before with clouds in closer configurations retaining more cold material than clouds with large $\delta$, and radiative clouds showing higher amounts of cold gas compared to their adiabatic peers. For instance, in the $\delta = 2$ case, the adiabatic model retains $\sim 40\%$ of its cold gas material by the end of the simulation time ($t\sim 5$ Myr), while higher $\delta$ models lose all their cold gas and the radiative model is able to retain more than twice this amount, holding over $\sim 90\%$ of the initial cold cloud material. Thus, hydrodynamic shielding and the aforementioned re-condensation mechanism prove to be particularly powerful in sustaining cold gas throughout the entire evolution. In line with our analysis in Section \ref{hs3}, the re-condensation of warm mixed gas triggered in radiative models also favours a smoother transition in cold gas content among simulations with different $\delta$ values, compared to adiabatic models where a very sharp transition occurs for $\sim \delta^{\rm sup}_{\rm shield}\approx 4.9$.  

\begin{figure}
\begin{center}
  \begin{tabular}{c} 
       \hspace{-0.6cm}\resizebox{!}{82mm}{\includegraphics{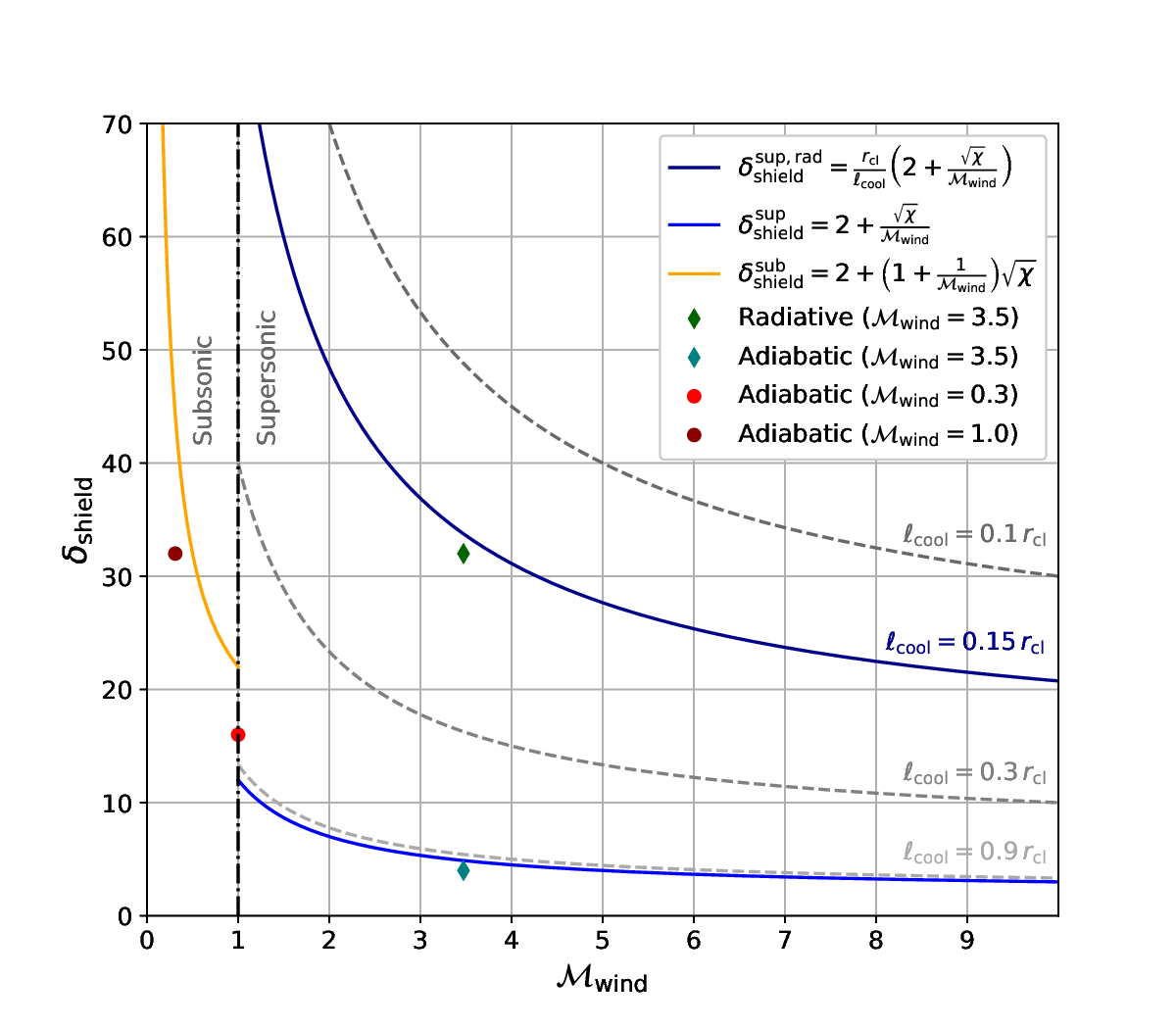}}\\
  \end{tabular}
  \caption{Dependence of $\delta_{\rm shield}$ on the wind Mach number, ${\cal M}_{\rm wind}$, for different models (solid lines): adiabatic subsonic in yellow (see \citealt{Forbes_2019}), adiabatic supersonic in light blue (see Section \ref{hs2}), and radiative supersonic in dark blue (see Section \ref{hs3}). For comparison, the simulations results are shown with markers and the predictions for other cooling regimes are shown with dashed lines.} 
  \label{plot:shield}
\end{center}
\end{figure}






\begin{figure*}
\begin{center}
  \begin{tabular}{c c c} 
  \multicolumn{3}{l}{(a) Abiabatic models}\\
         \multicolumn{1}{l}{Mixing fraction} & \multicolumn{1}{l}{Cold gas mass fraction} & \multicolumn{1}{l}{Cloud gas velocity}  \\ 
       \hspace{-0.3cm}\resizebox{!}{44mm}{\includegraphics{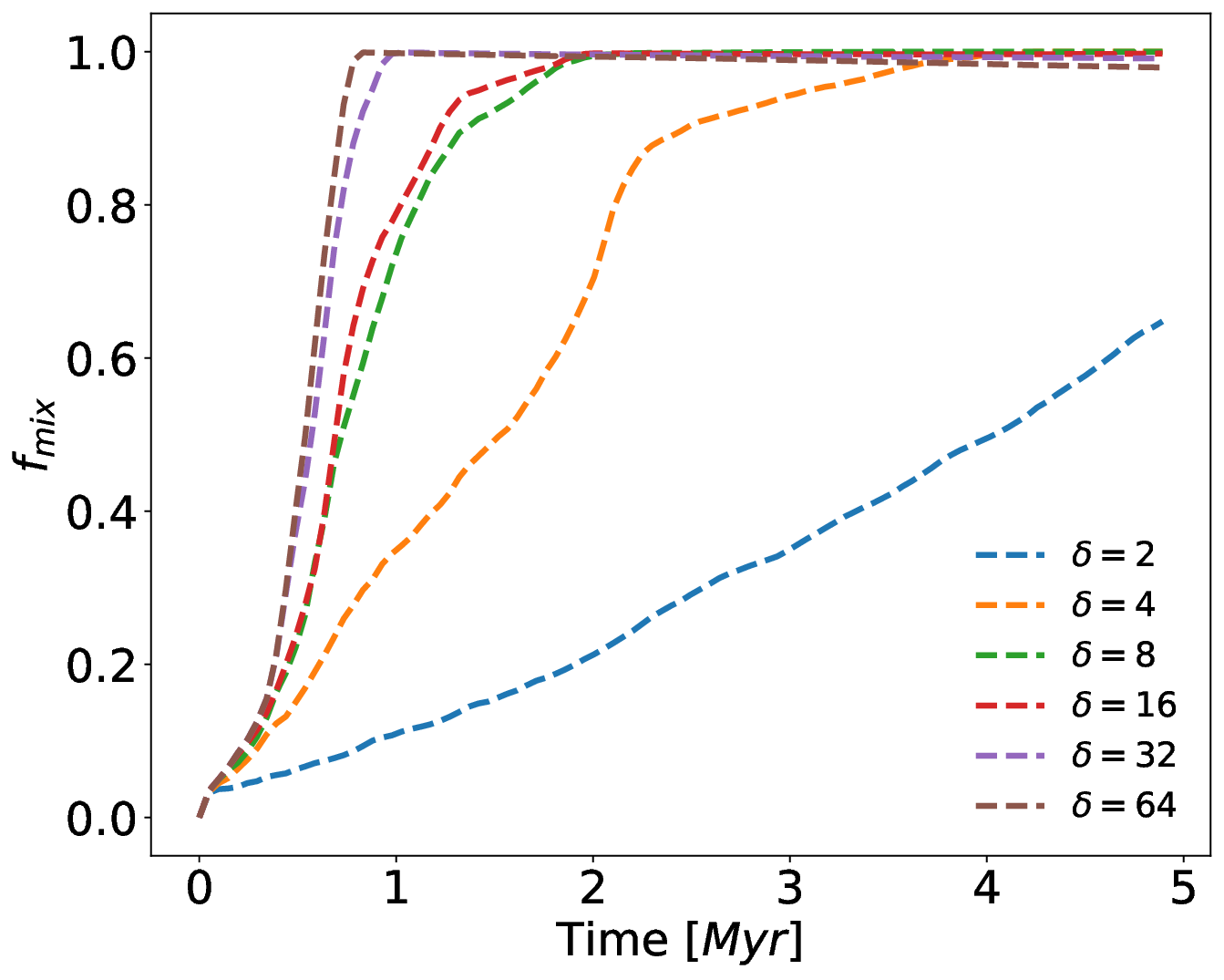}} & \hspace{-0.3cm}\resizebox{!}{44mm}{\includegraphics{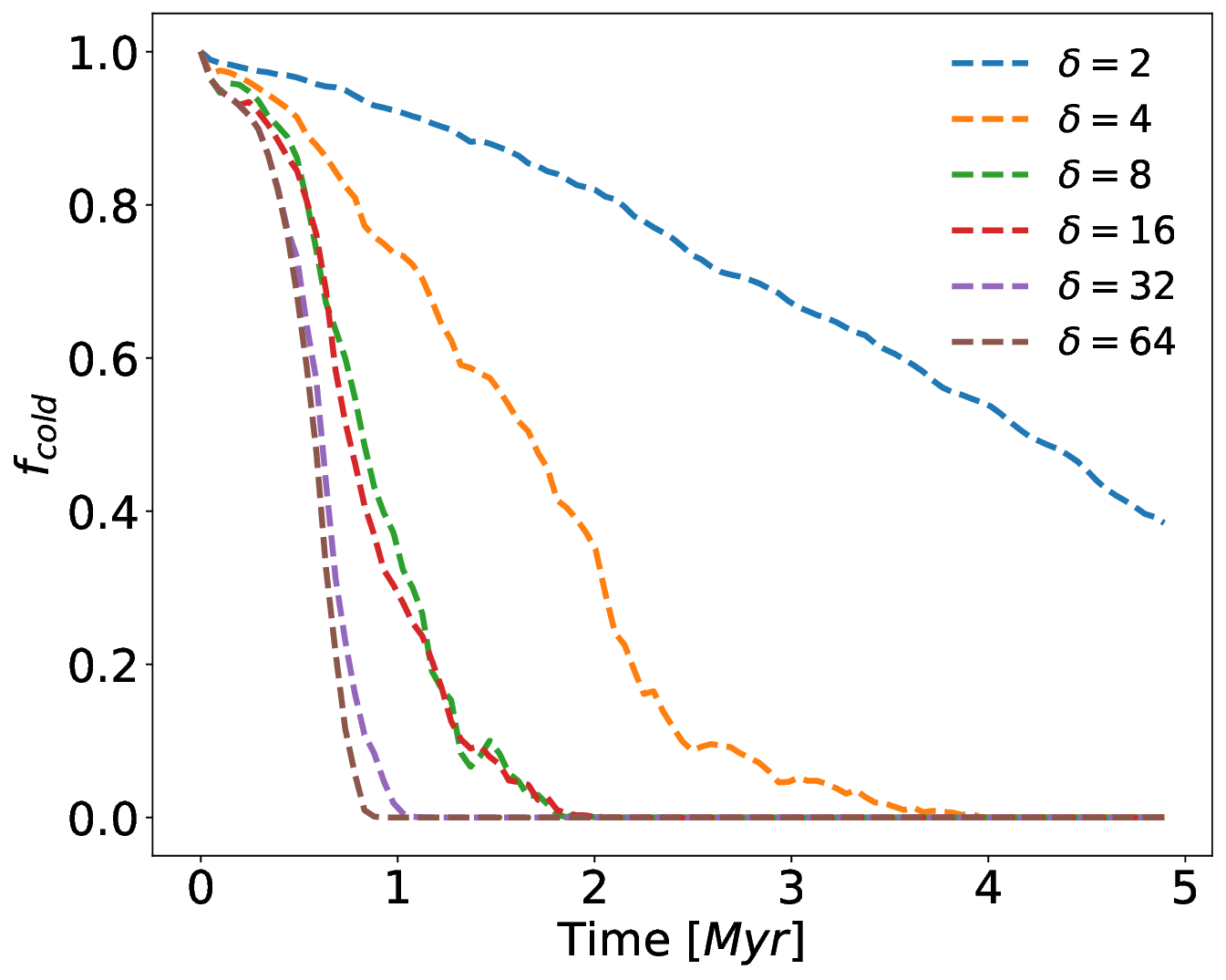}} & \hspace{-0.3cm}\resizebox{!}{44mm}{\includegraphics{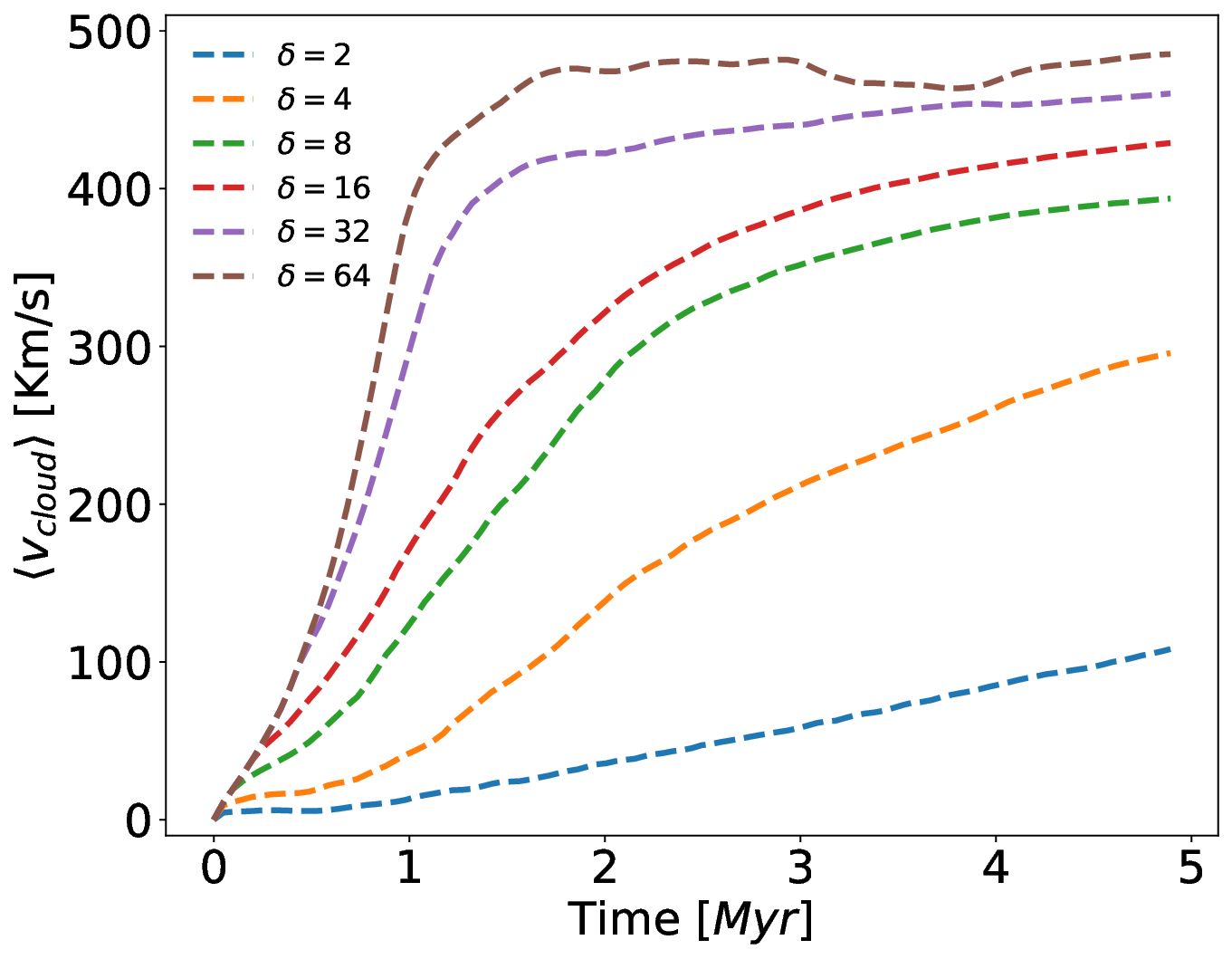}}
       \\
       \multicolumn{3}{l}{(b) Radiative models}\\
        \multicolumn{1}{l}{Mixing fraction} & \multicolumn{1}{l}{Cold gas mass fraction} & \multicolumn{1}{l}{Cloud gas velocity}  \\ 
       \hspace{-0.3cm}\resizebox{!}{44mm}{\includegraphics{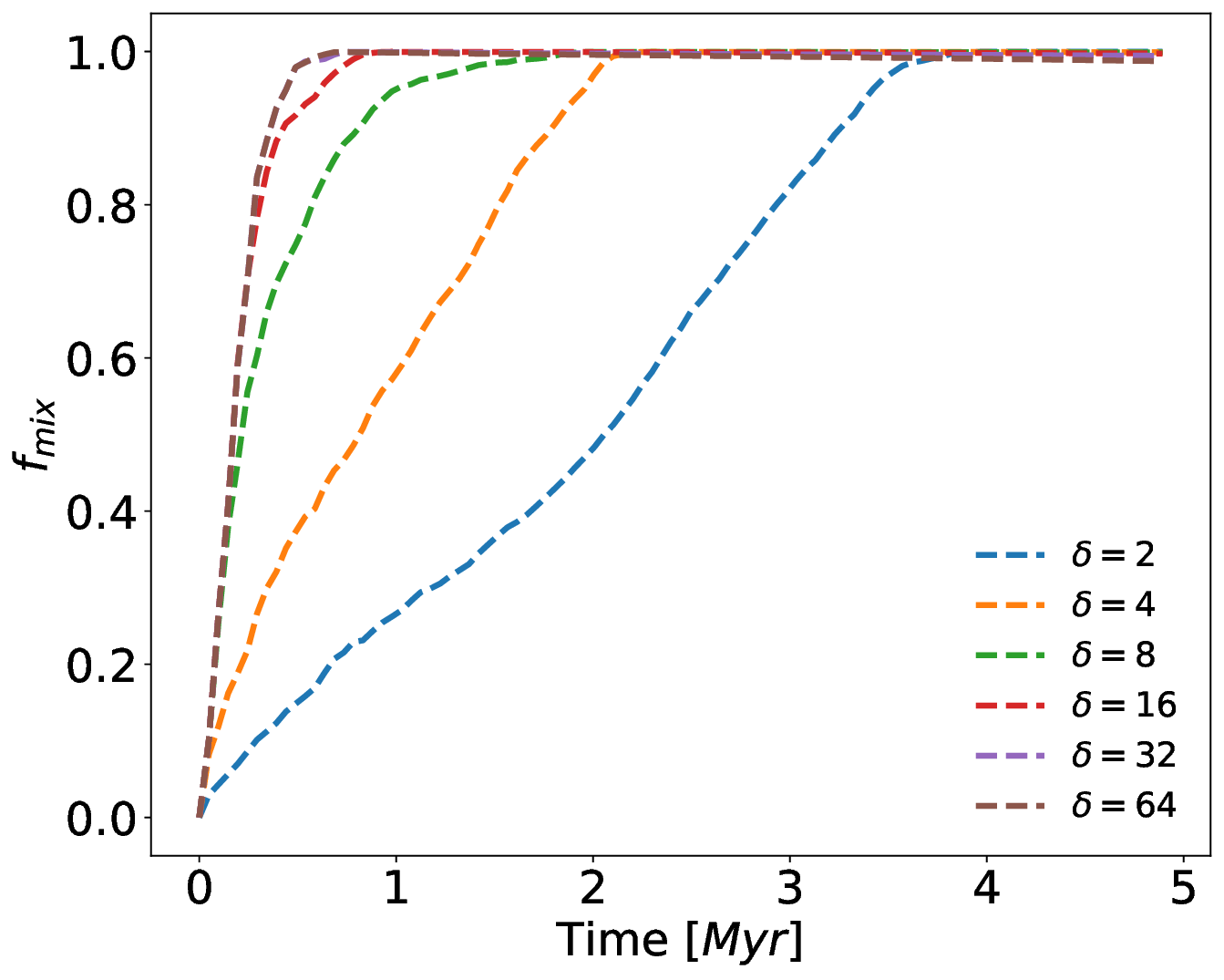}} & \hspace{-0.3cm}\resizebox{!}{44mm}{\includegraphics{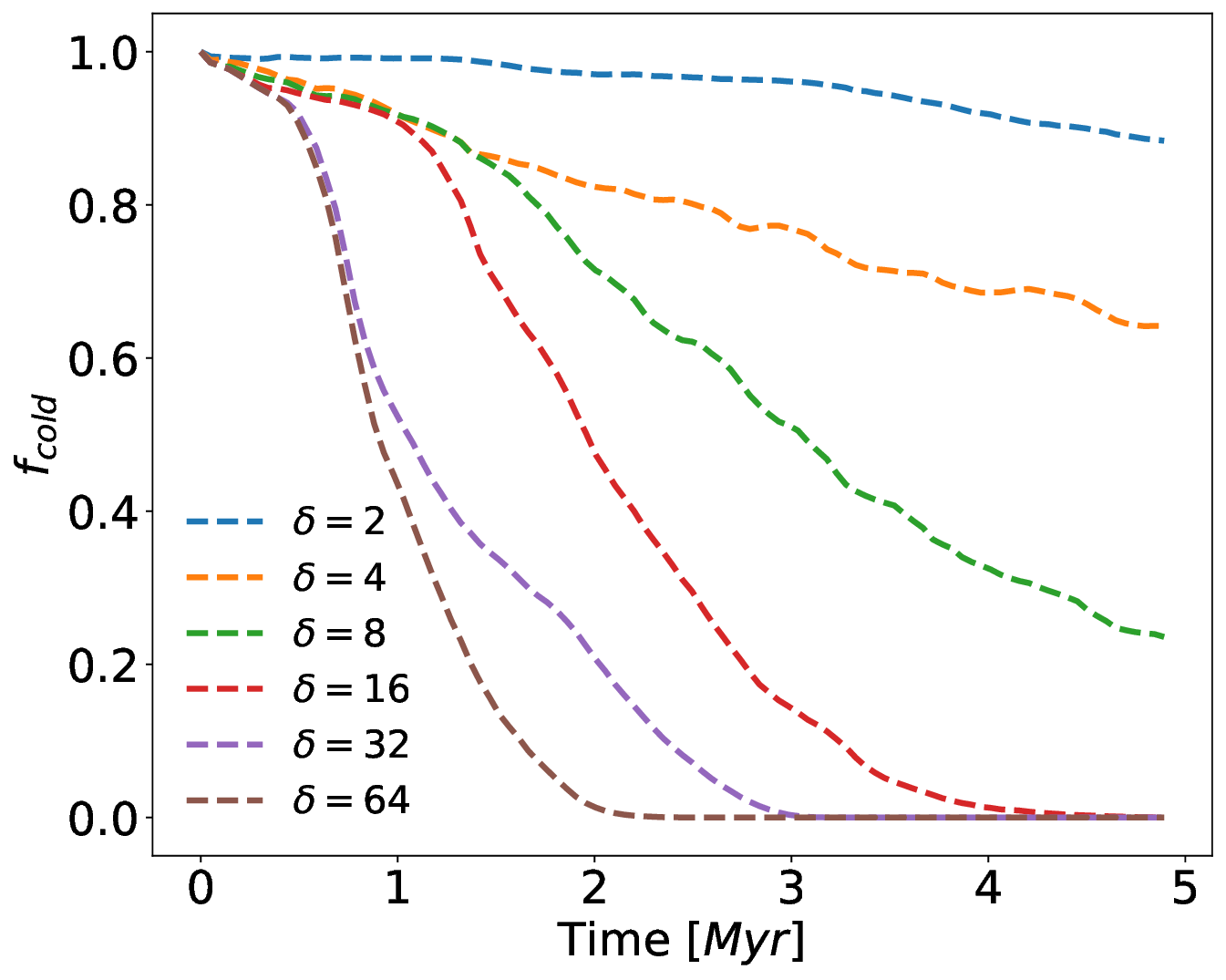}} & \hspace{-0.3cm}\resizebox{!}{44mm}{\includegraphics{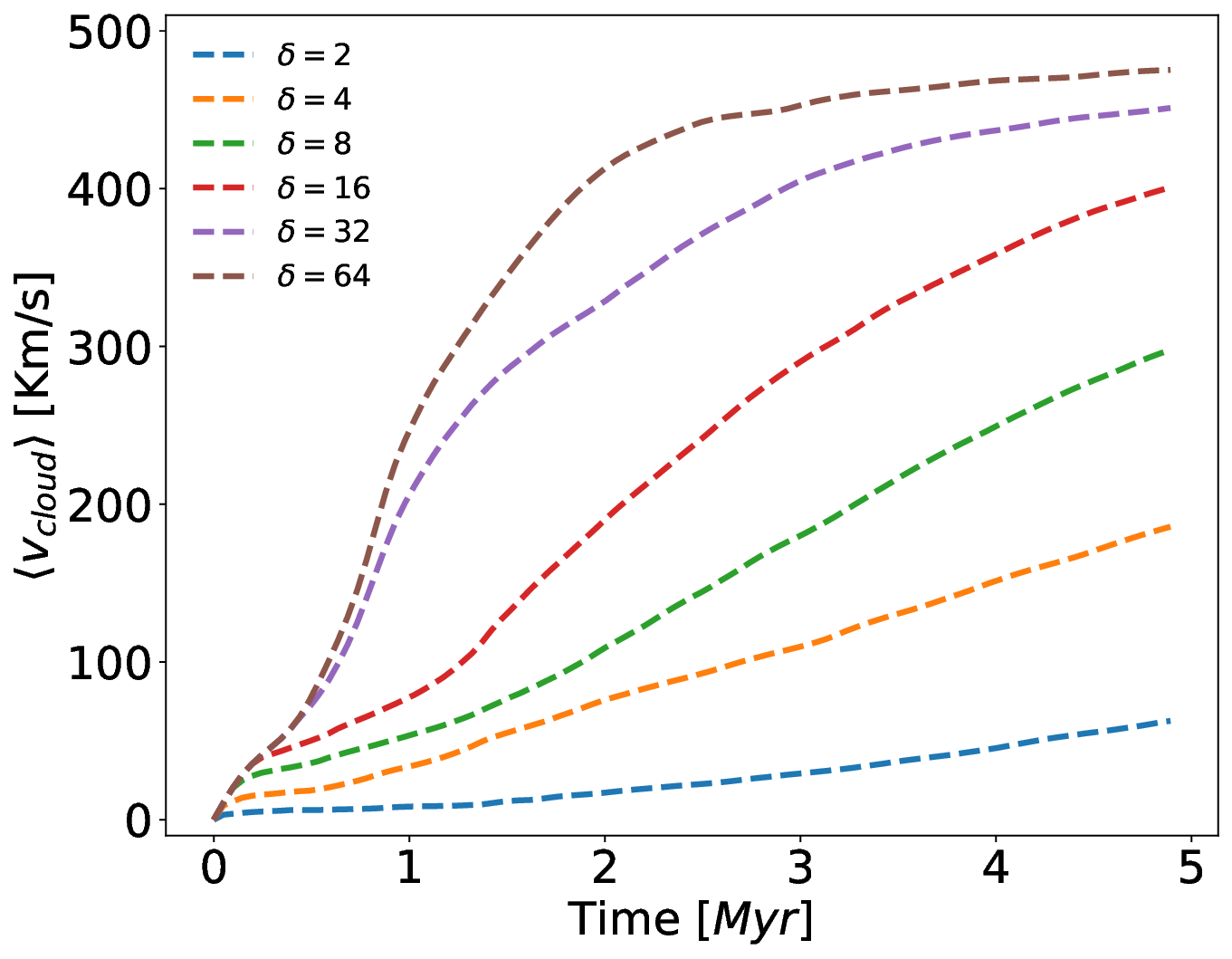}}
       \\
  \end{tabular}
  \caption{Time evolution of the mixing fraction (left column), cold gas mass fraction (middle column), and mass-weighted average velocity (right column) for different values of the parameter $\delta$ in the adiabatic (top) and radiative (bottom) models. Small $\delta$ values reduce mixing, preserve more cold gas via shielding, and decrease acceleration. Hydrodynamic shielding protects the clouds by reducing drag forces, and it is more efficient in radiative scenarios, aided by re-condensation.
  } 
  \label{evo4}
\end{center}
\end{figure*}

\subsubsection{Cloud gas dynamics}\label{hs5}

The right-hand column of Figure \ref{evo4} shows that the mass-weighted velocity of cloud gas increases as the separation distance $\delta$ increases. In other words, multi-cloud systems with large $\delta$ values can be accelerated very efficiently to end up nearly co-moving with the wind, while multi-cloud systems with small $\delta$ values take much longer to reach high velocities. This can be attributed to the fact that a smaller $\delta$ value results in the formation of gas streams with larger effective column densities. Gas with larger column densities is harder to accelerate via direct momentum transfer from the wind as it has more inertia. The mass-weighted velocity plots also reveal that adiabatic models have systematically higher velocities compared to their radiative counterparts. This has important implications for the entrainment of cold gas, which is the process by which dense gas originally from the ISM of galaxies is carried along by a much hotter outflowing galactic wind (\citealt*{2009ApJ...700L.149V}). In our models, entrainment of dense gas depends on the interplay between hydrodynamic shielding and radiative processes. Our results indicate that entrainment becomes increasingly weaker for smaller $\delta$ values and also that the re-condensation of warm gas along the stream also reduces entrainment. The latter effect is due to a reduction of the strength of the hydrodynamic drag forces acting on the clouds, which has a direct impact on the velocity of dense gas. The subsonic and transonic wind-multicloud simulations reported by \citep{Forbes_2019} show a very similar trend (cf. our results with their Figure 3).\par

In order to study the impact of hydrodynamic shielding on cloud dynamics in a quantitative manner, we present an analytical model to evaluate the late-stage velocity of the multicloud medium as a function of $\delta$. In the frame of the wind, \cite{Forbes_2019} introduces a model to represent the mixed gas region as a cylinder of radius $\xi\,r_{\rm cl}$ within which clouds can cool efficiently. The initial momentum over a given distance across the cloud distribution is $p_{0} = N_{\rm c}\,m_{\rm c}\,v_{\rm wind}$, where $N_{\rm c}$ represents the number of clouds in the volume, $m_{\rm c}$ is the mass of a single cloud, and $v_{\rm wind}$ is the wind speed. In the final stage, the momentum of the cold material is given by $N_{\rm c}\,m_{\rm c}\,(v_{\rm cl, final})$, where $v_{\rm cl, final}$ is the final velocity of cloud gas. The fraction of the momentum transferred to the cylinder is $\rho_{\rm wind}\,N_{\rm c}\,(v_{\rm cl, final})(\pi(\xi r_{\rm cl})^2 \delta r_{\rm cl} - (4/3)\pi r_{\rm cl}^3)$. As a result of employing momentum conservation, the final velocity as a function of the initial velocity, in the frame of the multicloud system, is defined as follows:
\begin{equation}
    \frac{v_{\rm cl, final}}{v_{\rm wind}}=1-\frac{\chi}{\chi+\frac{3}{4}\,\xi^2\,\delta - 1}
\end{equation}

\begin{figure}
\begin{center}
  \begin{tabular}{c}
       \hspace{-0.7cm}\resizebox{!}{80mm}{\includegraphics{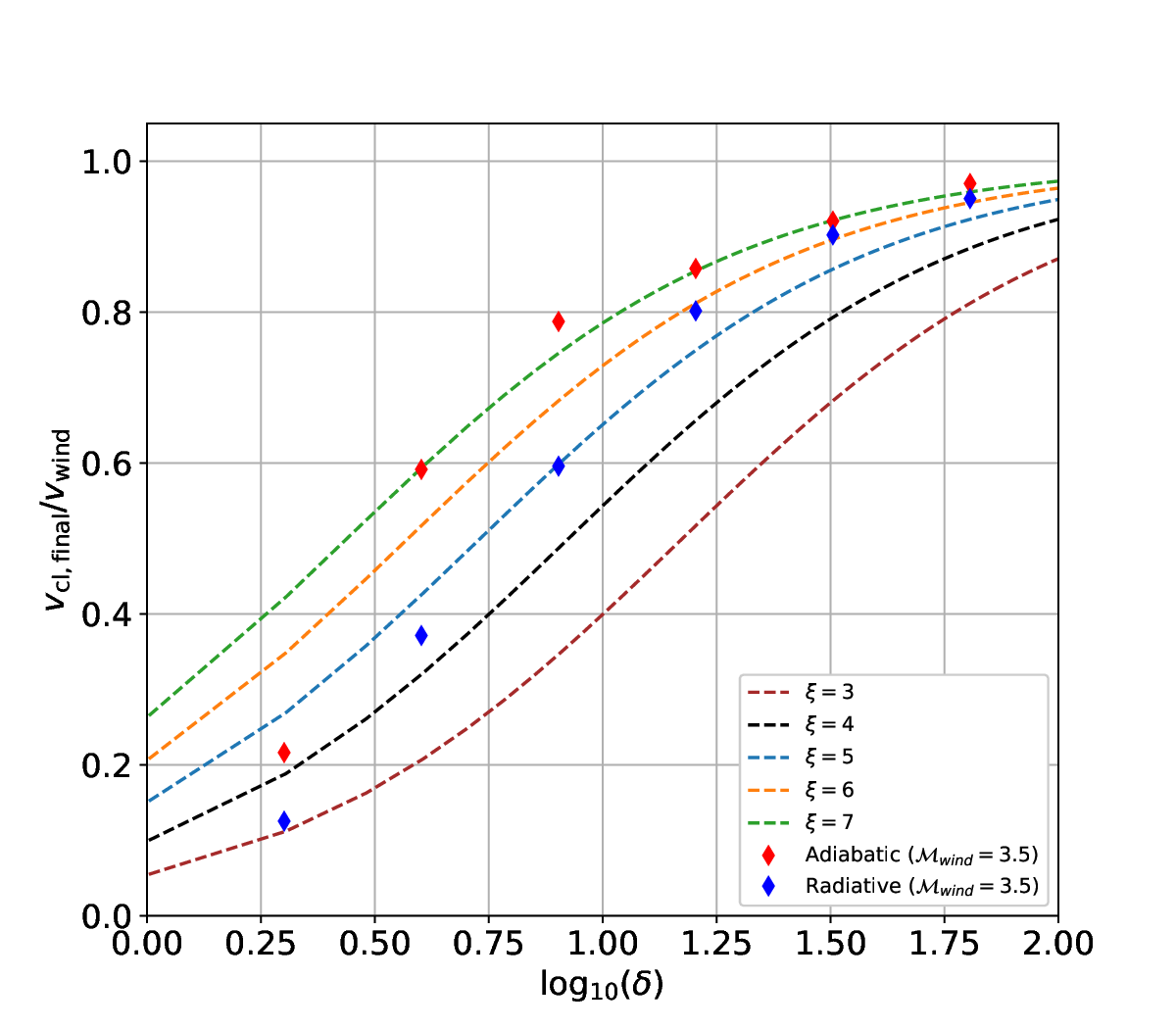}} 
  \end{tabular}
  \caption{Ratio of the final cloud velocity at $t_{\rm sim}=400 \,t_{\rm cross}$ to the wind velocity as a function of the initial separation distance, $\delta$. The cylinder model introduced by \citealt{Forbes_2019} is applied to our radiative models in the frame of the multicloud system for different values of $\xi$ ($\xi\,r_{\rm cl}$ is the cylinder radius). The markers represent the numerically-obtained ratios, $v_{\rm cl, final}/v_{\rm wind}$ in our adiabatic (red) and radiative (blue) models.} 
  \label{veloc_del}
\end{center}
\end{figure}

Figure \ref{veloc_del} shows the ratio of the velocities $v_{\rm cl, final}/v_{\rm wind}$ as a function of $\delta$, where $v_{\rm cl, final}$ is the cloud speed at the end of each simulation. This plot shows that nearly all the adiabatic markers fall in the line where $\xi = 7$. This suggests that the mixture region is a wide cylinder that is wider than the size of the computational box. On the other hand, the leftmost marker suggests that, for small separation values, the cylinder has a thin structure with a small radius. The sudden change in the radius $\xi r_{\rm cl}$ is explained by the strong hydrodynamic shielding effect that appears at $\delta^{\rm sup}_{\rm shield}\approx 4.9$ predicted by equation \ref{eq:sup_delta}. The radiative models, on the other hand, show a spread trend where simulations with different separation $\delta$ values can be explained by different $\xi$ values. The two right-most markers follow the $\xi=6$ model while the four left-most markers (for $\delta < 32$) systematically spread towards smaller $\xi$, which implies that they acquire thinner mixing regions. The reason is outlined by equation \ref{eq:sup_delta} which predicts that $\delta^{\rm sup,rad}_{\rm shield}\approx 33.8$ for the radiative case. Therefore, the increased effectiveness of hydrodynamic shielding in radiative models also compress the mixing region, which becomes systematically smaller as the separation distance between the clouds is reduced.

\section{Discussion}\label{sec:discussion}

\subsection{Implications for the physics of galactic winds}\label{implications}

In this section, we discuss the implications of our study for the physics of galactic winds. First, we compare our findings with previous numerical studies and then we discuss the relevance of our work for observations. \cite{Forbes_2019} conducted a similar set of simulations to ours, but focused on scale-free, adiabatic scenarios with subsonic and transonic winds. Our simulations, on the other hand, include radiative processes and supersonic winds consistent with the CC85 model, so our results are complementary to theirs. Both studies show that clouds, arranged in closer configurations, lead to hydrodynamic shielding, and each demonstrates that the threshold separation distance, $\delta_{\rm shield}$, varies widely with the choice of wind Mach number, cloud-to-wind density contrast, and cooling length (see Figure \ref{plot:shield}).\par

Another relevant study is that by \cite{Aluzas_12} who investigated the adiabatic interaction between shock waves and regions containing multiple, randomly-distributed clouds. They found that the lifetime of downstream clouds is reduced by the turbulence generated from interactions between the shock and upstream clouds. While the authors did not explicitly studied hydrodynamic shielding in their setups, this may point to a possible sensitivity of shielding on the initial cloud arrangement. One of the differences between our work and theirs is the initial placement of clouds: along a stream in our models versus randomly distributed in their models. Despite this, it is hard to draw conclusions because they also modelled shocks and not winds, and clouds have been shown to survive longer in winds with the same Mach numbers of shocks (see \citealt{2017MNRAS.470.2427G}). To corroborate these hypotheses, simulations of shock-multicloud systems and radiative wind-multicloud systems with more intricate cloud placements are needed.\par

Earlier shock-multicloud studies performed by our group, \cite{Banda_20,Banda_21} can also be compared to our new results. These models include fractal multicloud layers and radiative processes, and show that the disruption of cloud gas and its dynamics depend on the number of cloudlets in the layer. They found that for systems with a high number of cloudlets, turbulence and mixing are reduced. This may be indicative of hydrodynamic shielding effects, as we previously established that if the distance between clouds is small, hydrodynamic shielding reduces the effect of dynamical instabilities and lowers mixing fractions. For radiative models, \citep{Banda_21} showed that re-condensation via mixed-gas cooling may explain the presence of dense gas observed along galactic winds. We also find that dense gas can be maintained along gas streams for several $\rm Myr$ via cooling and condensation-driven shielding. In our models larger column densities are also generated in multicloud systems with smaller $\delta$ values, which also diminish cloud gas acceleration. This is in agreement with their models in that more porous cloud distributions (equivalent to high $\delta$ values) reach lower bulk speeds than more compact (low-$\delta$) cloud distributions.\par

Our work is also relevant for studies that target larger spatial scales. For instance, \cite{Schneider_18,Schneider_20} presented a set of simulations of isolated galactic disc-wind models covering scales of tens of $\rm kpc$. Their models suggest that radiative cooling and momentum mixing are the sources of fast-moving cool gas along galactic winds. On slightly smaller $\sim \rm kpc$ scales, \citealt{2017MNRAS.466.1903G,2022ApJ...936..133C} captured wind-launching in stratified galactic discs and self-consistently generated multiphase winds. While hydrodynamic shielding was not explicitly studied in these papers, in all these large-scale simulations, wind-multicloud interactions occur. Thus, hydrodynamic shielding may be acting and should play some role in preserving dense gas. Based on our results, hydrodynamic shielding would be important near the base of the wind-launching region where ISM clouds would be more naturally together, but it may also play a role in reducing the disruption of CGM clouds that reform via re-condensation and inhabit compact cloud complexes along the wind.\par

The above statement also has important implications to observational studies of galactic winds around star-forming galaxies, particularly to emission- and absorption-line studies (e.g. see \citealt{2008A&A...487..583B,2013ApJ...770L...4M,2017A&A...607A..48R} for our Galaxy, \citealt{Shopbell_98,2023ApJ...958..109L} for M82, and ), in which a cool phase can be detected via multiple tracers (\citealt{Tumlinson_17}). While our work focuses on idealised small 3D sections of such galactic winds, it highlights that strong radiative cooling facilitates hydrodynamic shielding in supersonic winds and thus provide a step forward in understanding the presence and survival of cold gas in hot CGM environments.

\subsection{Numerical resolution and limitations}\label{convergence1}

\begin{figure*}
\begin{center}
  \begin{tabular}{c c c} 
   \multicolumn{3}{l}{(a) $\delta = 8$ models}\\
       \hspace{-0.3cm}\resizebox{!}{44mm}{\includegraphics{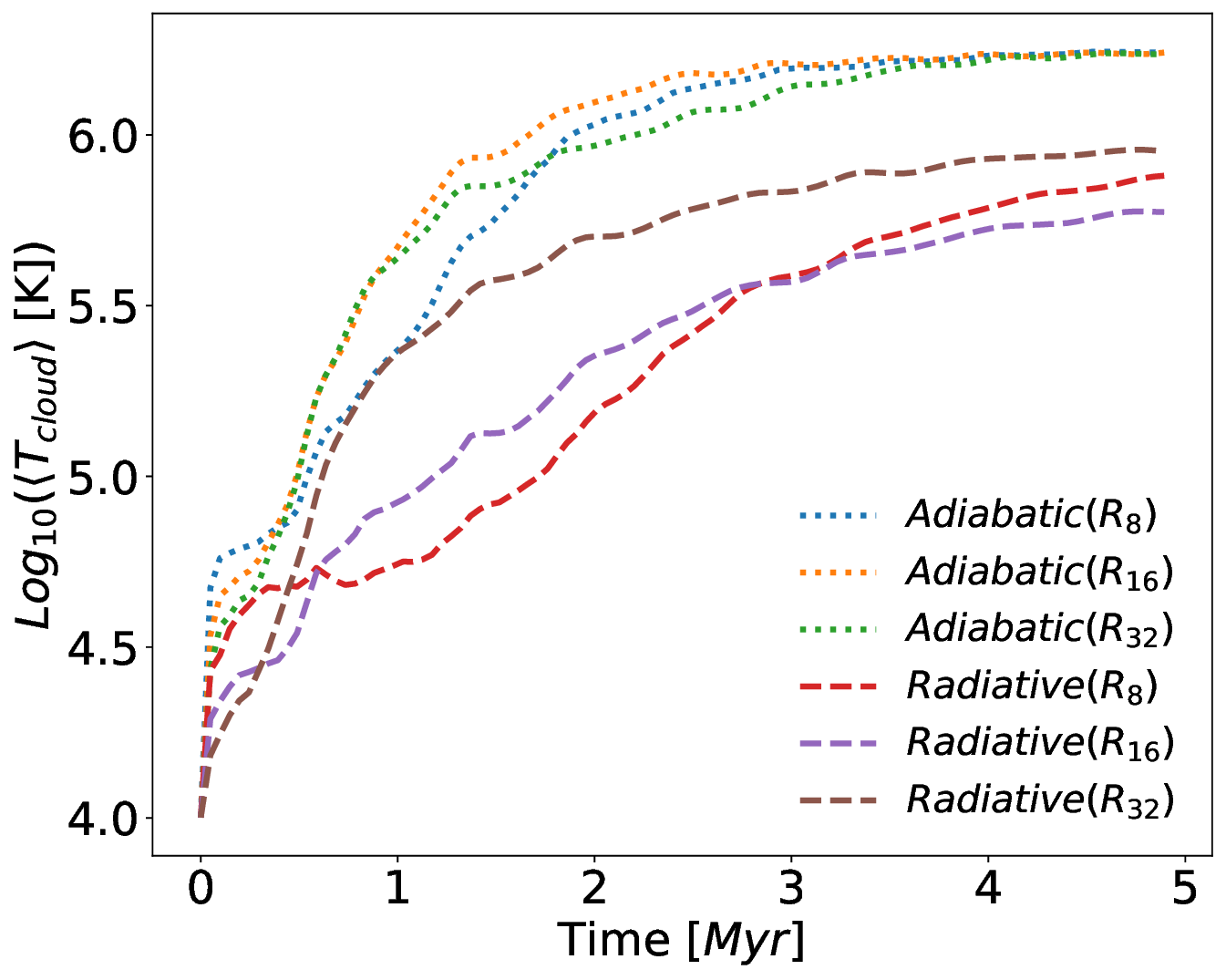}} & \hspace{-0.3cm}\resizebox{!}{44mm}{\includegraphics{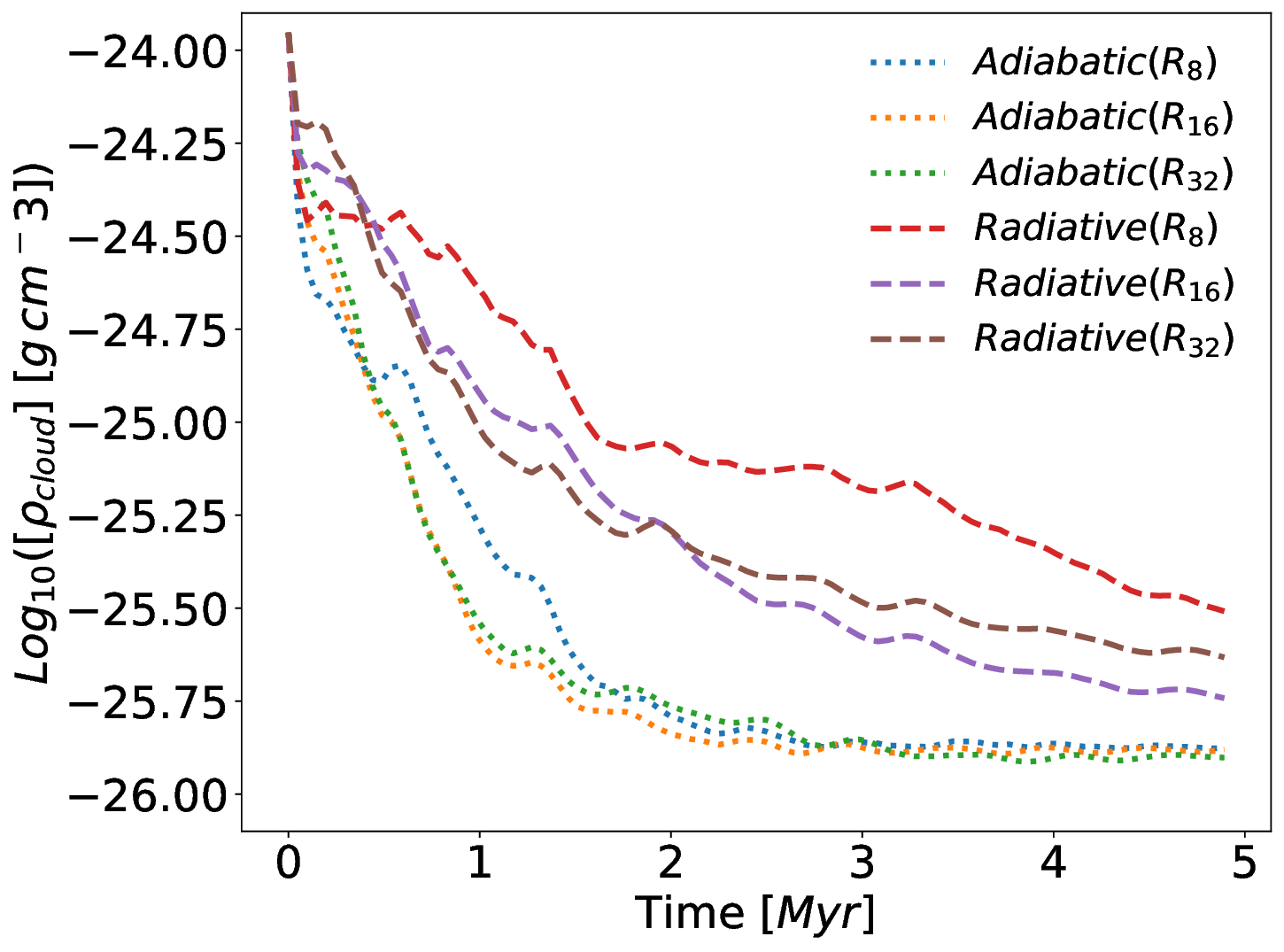}} & \hspace{-0.3cm}\resizebox{!}{44mm}{\includegraphics{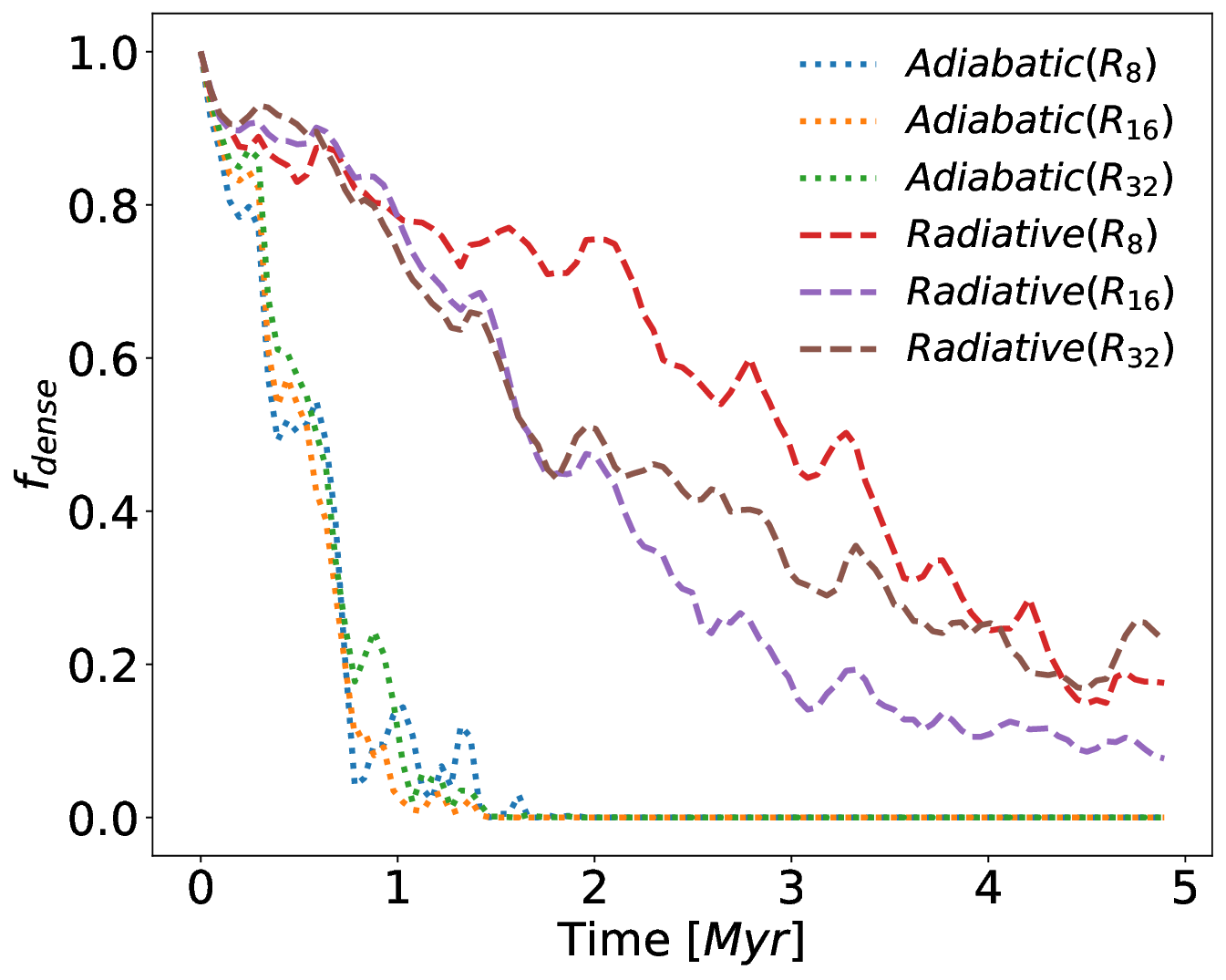}}
       \\
       \hspace{-0.3cm}\resizebox{!}{44mm}{\includegraphics{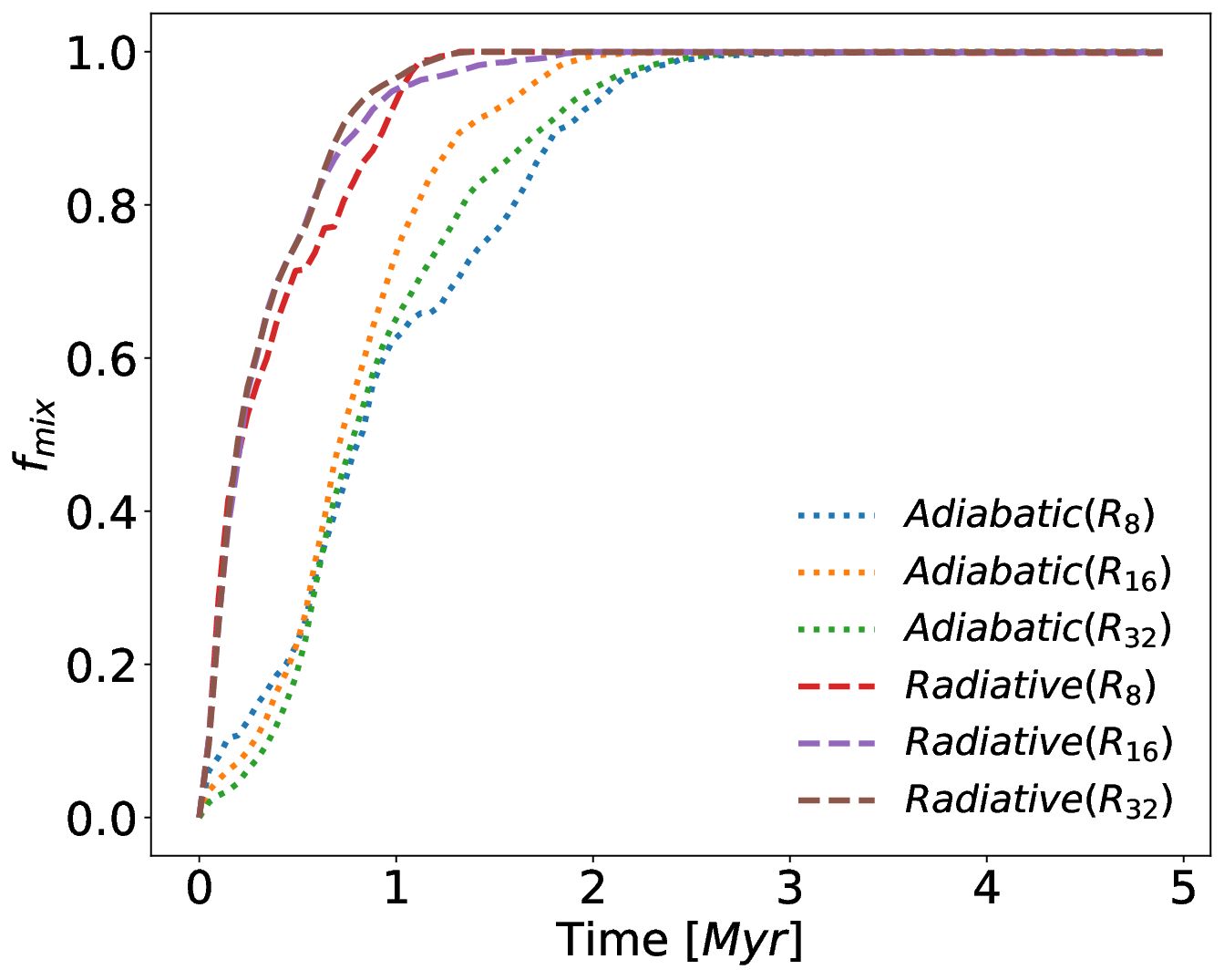}} & \hspace{0.1cm}\resizebox{!}{44mm}{\includegraphics{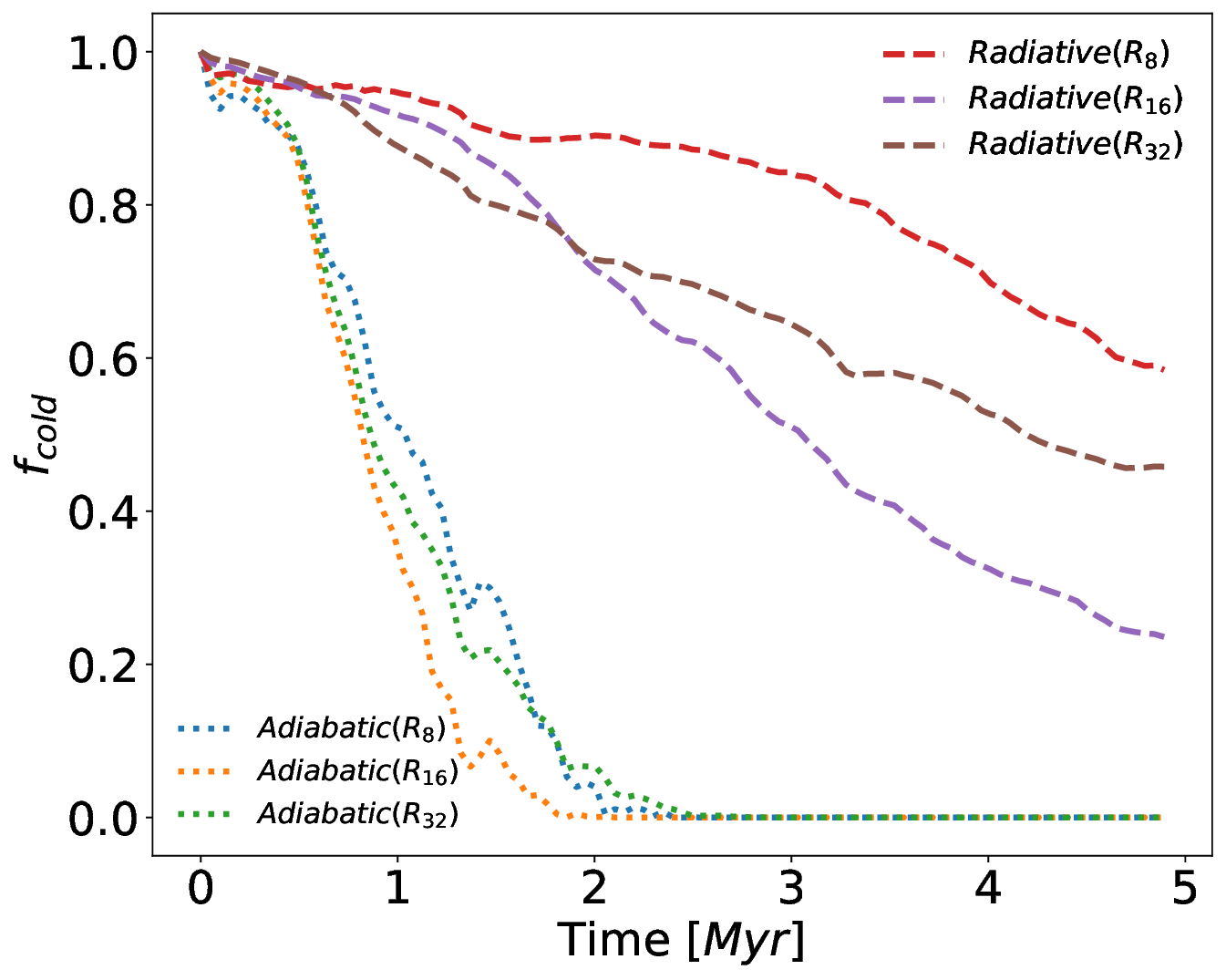}} & \hspace{-0.3cm}\resizebox{!}{44mm}{\includegraphics{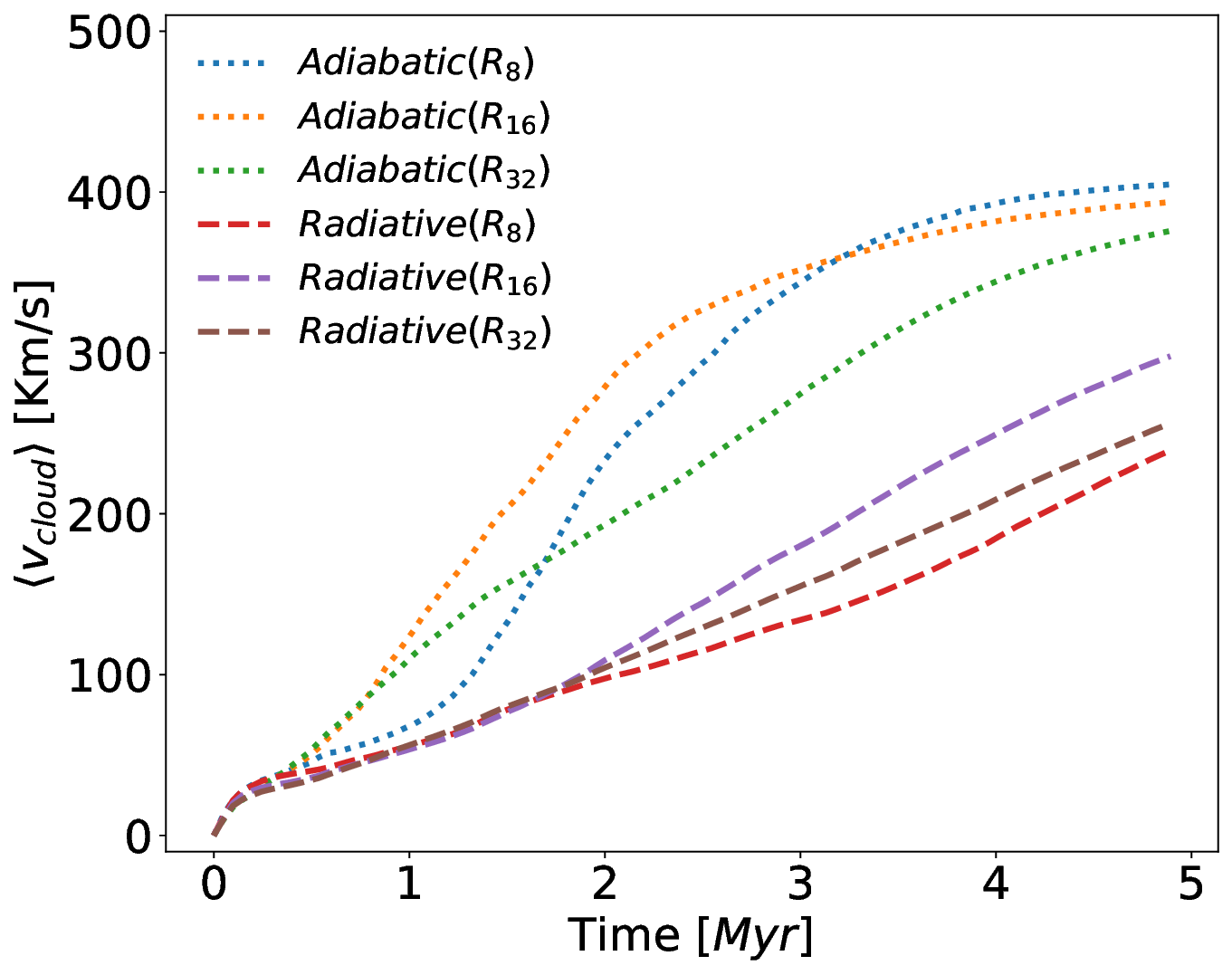}}
       \\
    \multicolumn{3}{l}{(b) $\delta = 16$ models}\\
       \hspace{-0.3cm}\resizebox{!}{44mm}{\includegraphics{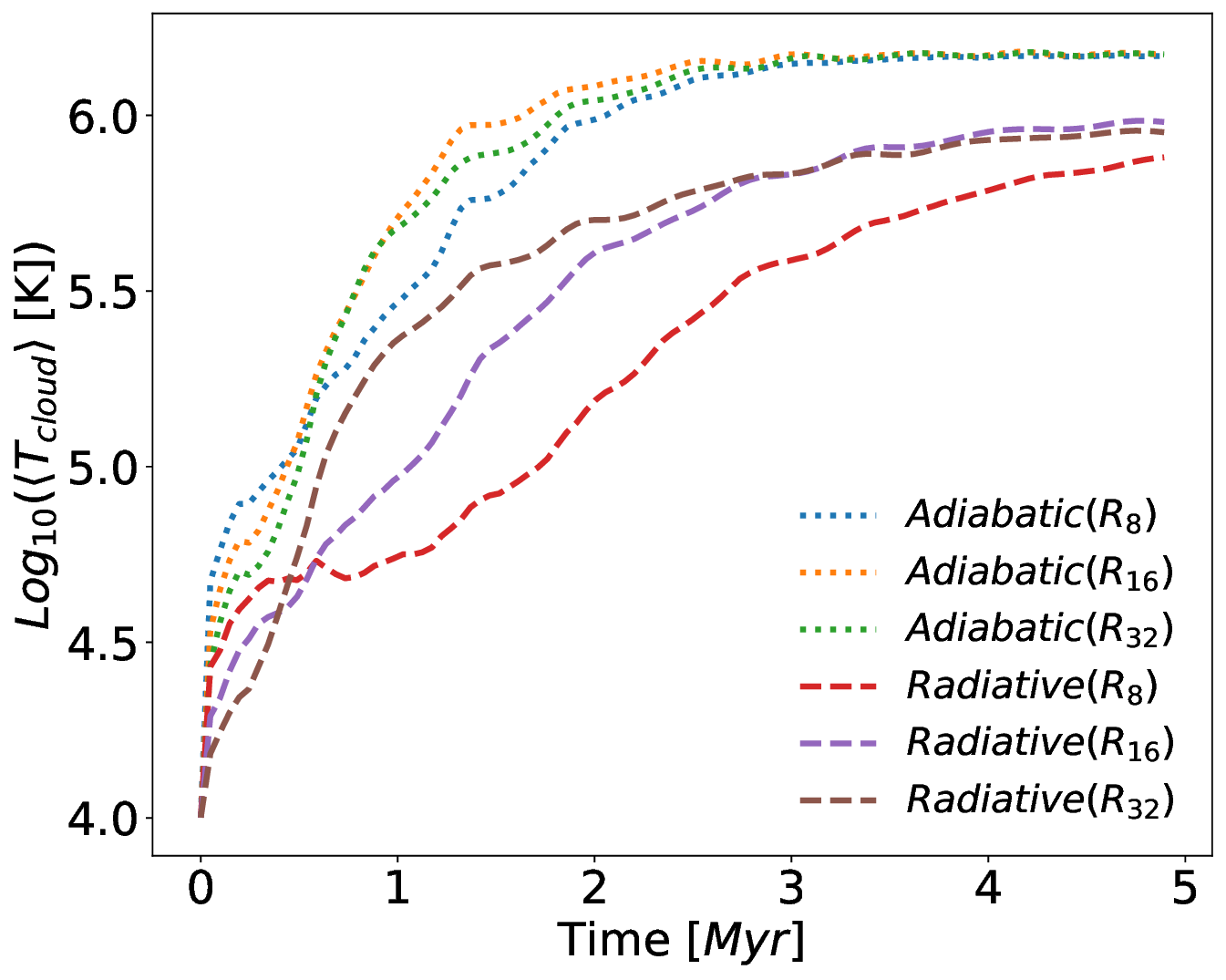}} & \hspace{-0.3cm}\resizebox{!}{44mm}{\includegraphics{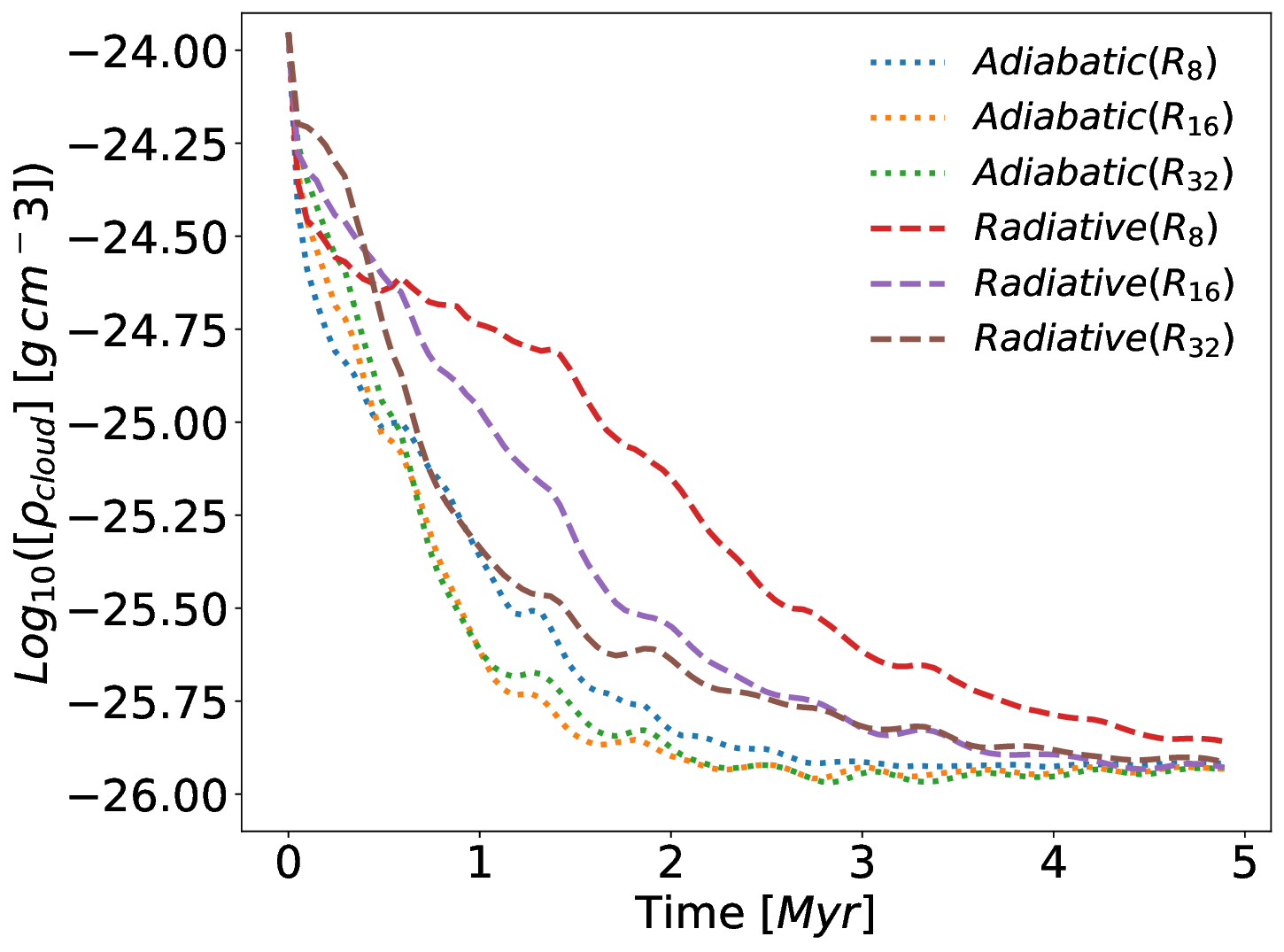}} & \hspace{-0.3cm}\resizebox{!}{44mm}{\includegraphics{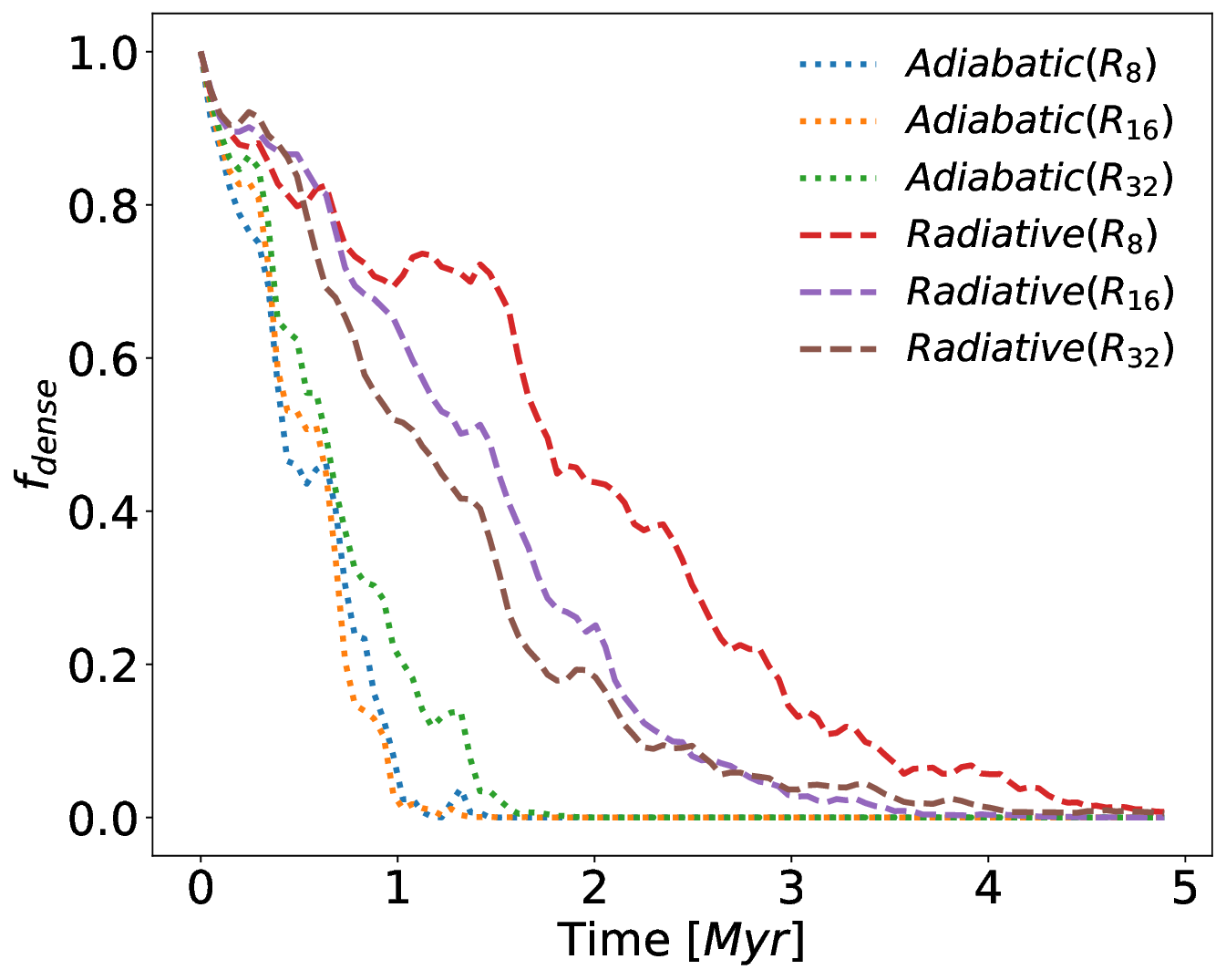}}
       \\
       \hspace{-0.3cm}\resizebox{!}{44mm}{\includegraphics{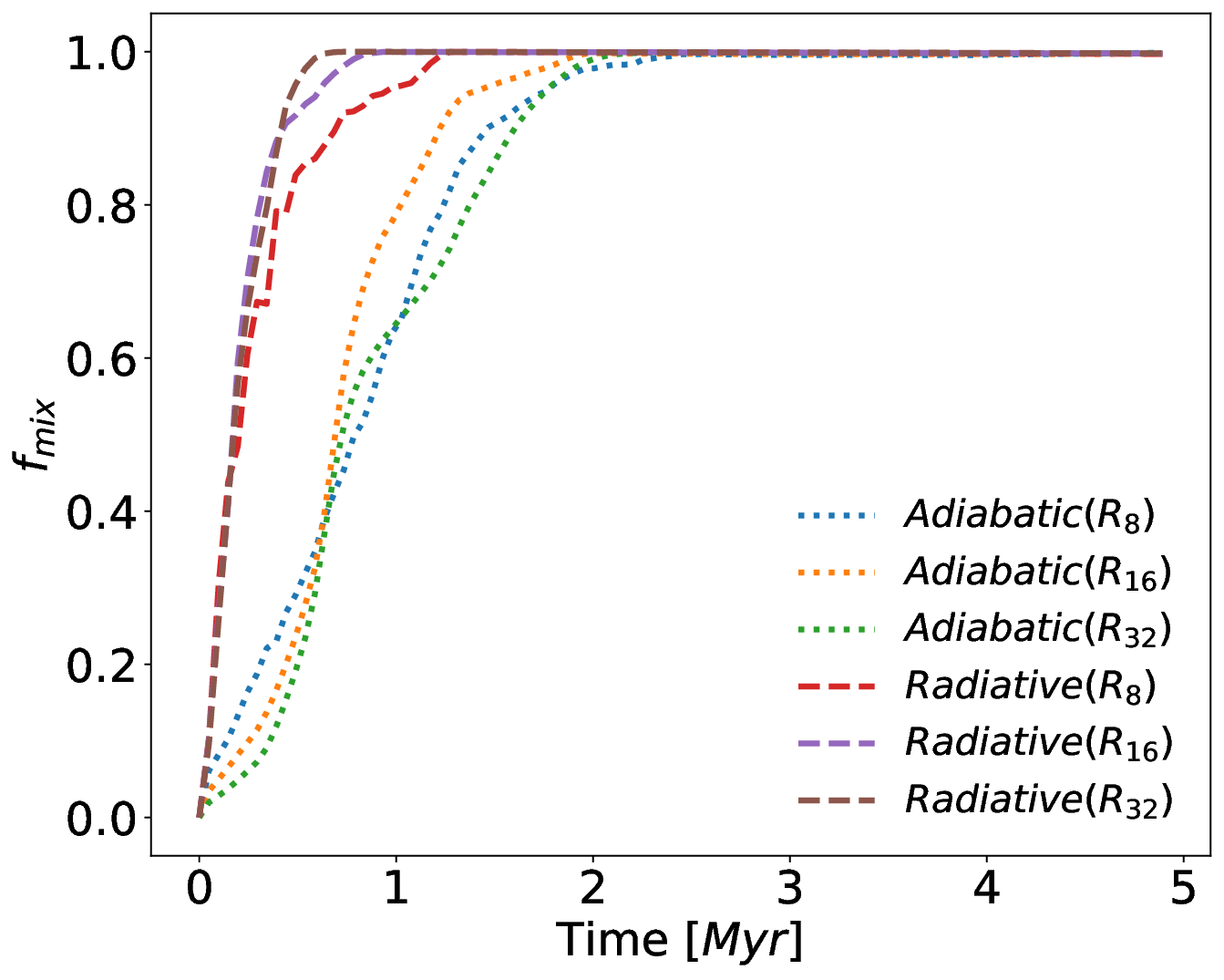}} & \hspace{0.1cm}\resizebox{!}{44mm}{\includegraphics{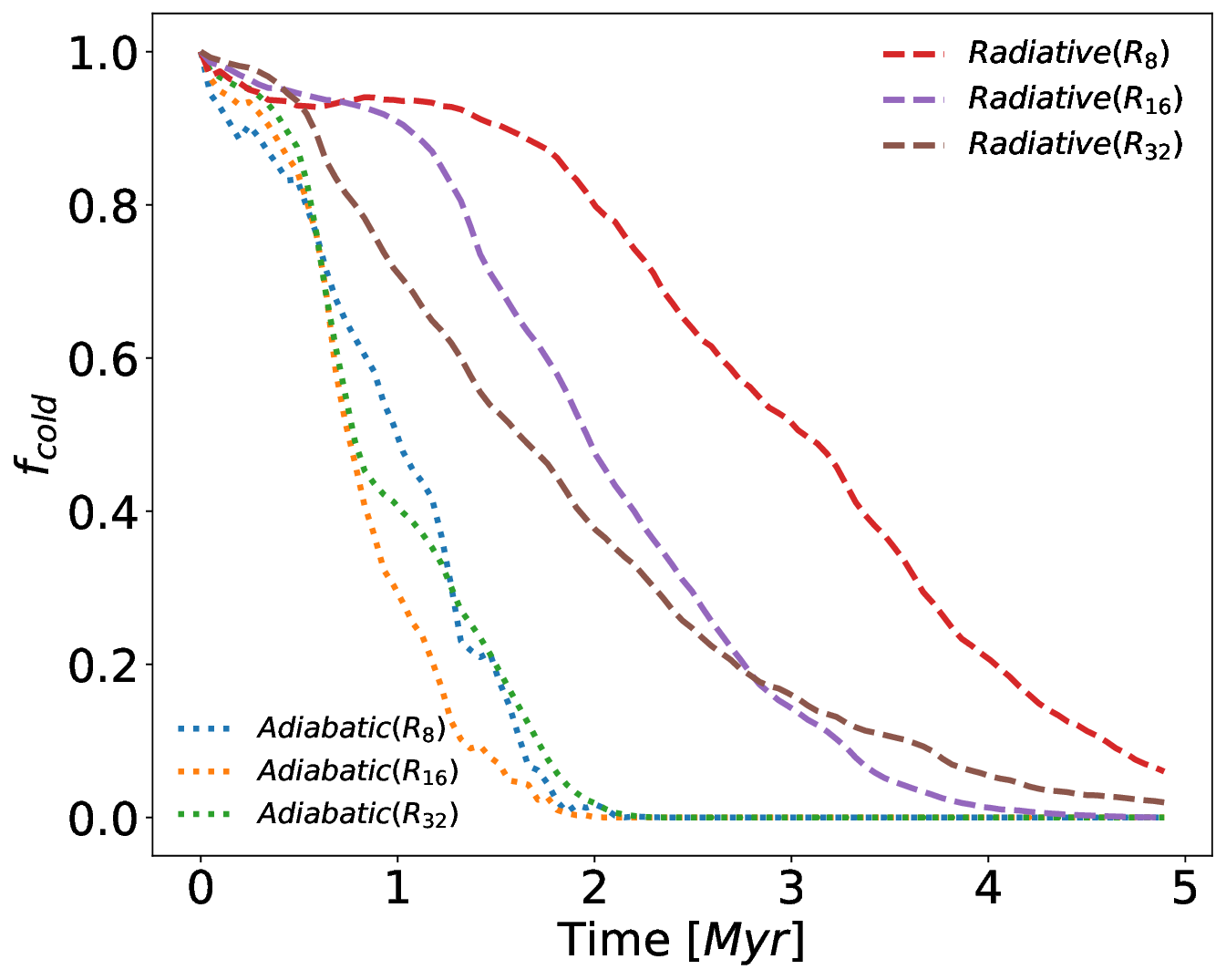}} & \hspace{-0.3cm}\resizebox{!}{44mm}{\includegraphics{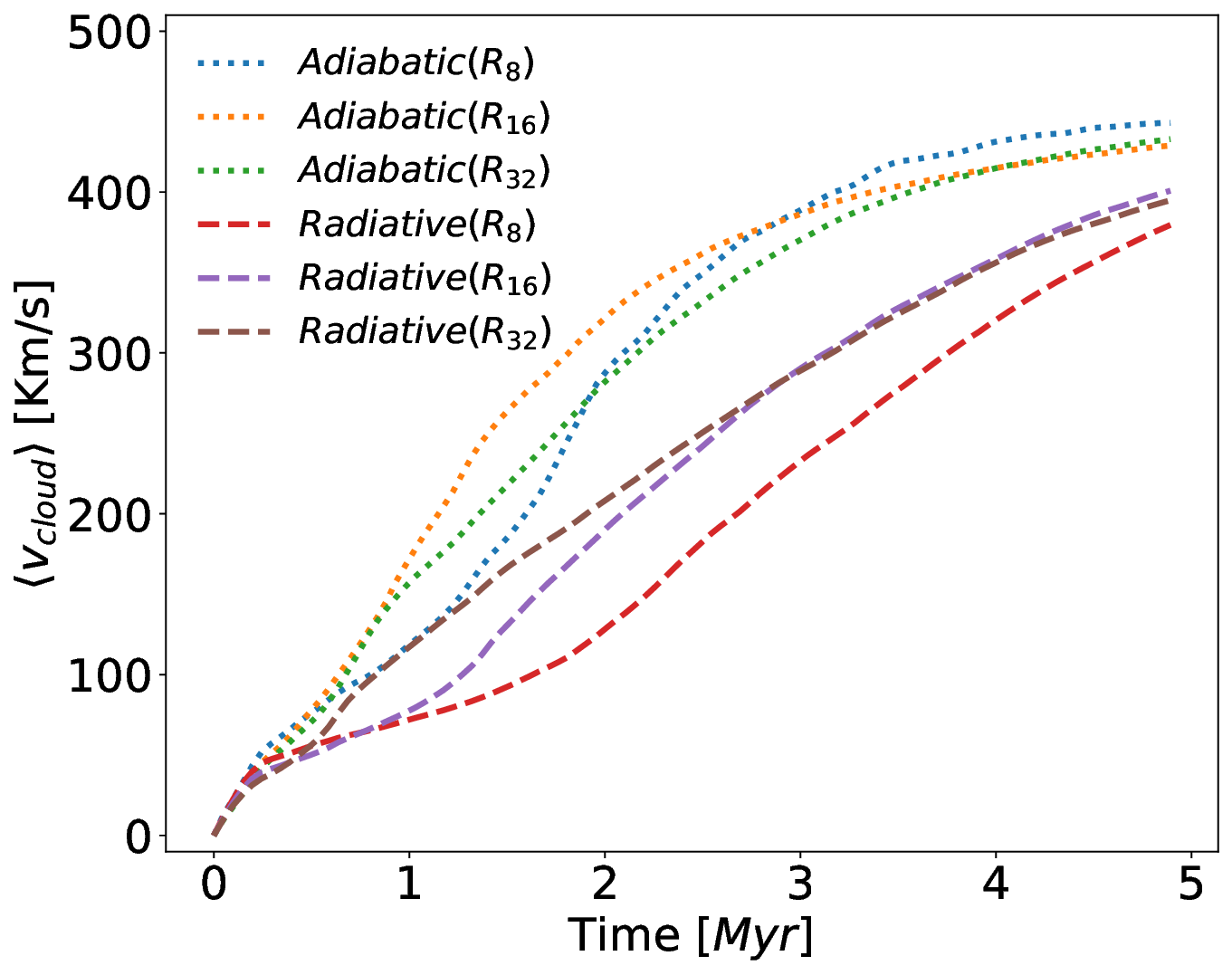}}
       \\
  \end{tabular}
  \caption{Comparison of adiabatic and radiative models at different numerical resolutions, $R_{8,16,32}$, for $\delta = 8$ (top two rows) and $\delta = 16$ (bottom two rows). All the diagnostics in adiabatic models exhibit overall convergence, regardless of the separation distance. In radiative models, $f_{\rm dense}$ and $f_{\rm cold}$, are not converged for $R_{8}$, show agreement for higher resolutions in the $\delta=16$ model, and show some spread for higher resolutions in the $\delta=8$ model. The latter may be due to turbulence induced by dense gas collisions, which the $\delta=8$ model is more prone to. All results hold for models compared at a fixed resolution.}
  \label{conv2}
\end{center}
\end{figure*}

In this section we discuss how the numerical resolution of our simulations influences the diagnostics reported in this paper and the limitations of our study. To address convergence, we present $8$ additional simulations with the standard separation values of $\delta=8$ and $\delta=16$ and low-($R_{8}$), standard-($R_{16}$), and high-($R_{32}$) numerical resolutions. Four of these models are adiabatic and the rest are radiative (see table \ref{table_simul} for details). Figure \ref{conv2} shows the time evolution of the same quantities presented in Figures \ref{evo3} and \ref{evo4}. Panel a in this figure shows the comparison for $\delta=8$ and and panel b for $\delta=16$.\par

In the adiabatic models, all the quantities appear well converged, regardless of the separation $\delta$ value being analysed. Apart from a small drop seen in the cloud velocity for the $\delta=8$ model at high resolution, convergence is reached in all diagnostics. In the radiative models, on the other hand, some parameters seem better converged than others for both $\delta$ values. For example, the cloud velocities, mean cloud densities (for $R_{\geq 16}$), and (surprisingly) the mixing fractions appear better converged, while the cloud temperature and the fractions of dense and cold gas show a higher resolution dependence. Mixing fractions depend on small-scale phenomena (e.g. vorticity deposited at shear layers), whose length scale varies with resolution, thus we would not expect it to converge, and in fact before $t<1\,\rm Myr$, $R_{32}$ does appear to produce higher $f_{\rm mix}$ (in agreement with \citealt{2019MNRAS.486.4526B}, see their figure B1). The seemingly convergent lines we see after that time just reflect the fact that mixing in these models occurs at a very fast pace (at $t\approx 1\,\rm Myr$ all cloud gas is mixed). When it comes to the fractions of dense and cold gas, we find that models with a $R_{8}$ resolution consistently overestimates these diagnostics, so it may not be adequate to study shielding quantitatively. For higher resolutions, we find that $\delta=16$ models show better convergence for these quantities than $\delta=8$ models. While we could not explore much higher resolutions in these models, it is possible that this result is due to additional turbulence induced via cloud-cloud or dense gas-dense gas collisions, which are more likely in cloud complexes with smaller $\delta$.\par

Based on the evolution of these specific variables, we conclude that resolutions of $R_{\geq 16}$ correctly capture the dynamics, mixing fractions and dense-gas fractions of the multi-cloud system for both the adiabatic and radiative systems. In the $\delta=8$ radiative case, $f_{\rm dense}$ and $f_{\rm cold}$, do not converge, but the trends and differences discussed in the main body of the paper hold for fixed resolution.\par

While the analysis of hydrodynamic interactions between hot winds and multicloud systems provides an overall view of the physics galactic winds, it is essential to acknowledge some limitations of our numerical work. The exclusion in our models of additional physics may have some impact on hydrodynamic shielding, so we briefly discuss them here. First, \citep{2005A&A...444..505O,Bruggen_16} showed that the inclusion of electron thermal conduction has a significant impact on how wind-swept clouds evolve, and can lead to e.g. thinner filaments with smaller velocities in clouds with large column densities. Given that conductive clouds evolve into denser filament suggests that hydrodynamic shielding could be enhanced in such models. Our study did not incorporate magnetic fields either. Magnetohydrodynamic (MHD) scenarios are known to significantly affect the dynamics and survival of dense gas \citep[see][]{Banda_2018}, but the effects largely depend on the strength and orientation of the field (see \citealt{2018ApJ...865...64G}). Magnetic fields transverse to the wind have been shown to produce more significant effects via magnetic draping (\citealt{Aluzas_14,Banda_16}), which could aid hydrodynamic shielding. However, predicting the actual effects based on our current models is not straighforward, so it is crucial that future studies conduct MHD simulations of wind-multicloud systems that incorporate electron thermal conduction to further investigate how hydrodynamic shielding operates in such environments. Self-gravity is another aspect not included in our models, but for our initial conditions the free fall time is small compared to the typical dynamical time-scales, so we would not expect self-gravity to largely impact our results. It may, however, be important for wind-multicloud models targeting denser ISM clouds as they would undergo compression, so some regions of them could become gravitationally unstable. This phenomenon has been observed in our specific context of cold, dense spherical clouds moving through a hot medium (e.g., see \citealt{Murray_93}, \citealt*{2014MNRAS.444.2884L}).

\section{Conclusions}\label{Sec5}

We have described a set of 3D hydrodynamical simulations of supersonic winds (with ${\cal M}_{\rm wind} = 3.5$ and $T\sim 10^{6}\,\rm K$) interacting with radiative and adiabatic multicloud systems (with $T \sim 10^{4}\,\rm K$ and $\chi=100$). In these models, clouds are placed along a vertical gas stream with different separation distances $\delta$ (in units of $r_{\rm cl}$). The goals of the paper were to study wind-multicloud systems in the supersonic regime, to evaluate how radiative cooling alters cloud gas dynamics, and to assess the efficiency of hydrodynamic shielding. Our models extend those presented by \citep{Forbes_2019}, who focused on studying shielding upon adiabatic, subsonic models. Our main conclusions are:

\begin{itemize}
    \item Wind-multicloud systems evolve into four stages: (i) Shock formation, in which refracted shocks emerge in the clouds and reflected waves are produced on the cloud fronts by the initial collision; (ii) Hydrodynamic shielding, in which the disruption of upstream clouds reduces the drag forces acting upon downstream clouds, (iii) KH shredding and mixing, in which shear instabilities promote turbulence and filament formation, and (iv) RT break-up and disruption, in which acceleration-driven instabilities break up the clouds into several cloudlets (see Section \ref{Sec4.1}). 
    \item Hydrodynamic shielding is also triggered in supersonic winds and can delay cloud disruption, but in adiabatic models it requires smaller $\delta$ values than in subsonic winds. In supersonic models, the wind speed exceeds the wind sound speed, so adiabatic clouds are required to be in very close proximity for hydrodynamic shielding to operate. \citep{Forbes_2019} reported $\delta^{\rm sub}_{\rm shield}\sim 20$ for ${\cal M}_{\rm wind}=1$ and $\delta^{\rm sub}_{\rm shield}\sim40$ for ${\cal M}_{\rm wind}=0.31$ in adiabatic scenarios, while we find a much smaller $\delta^{\rm sup}_{\rm shield}\sim5$ for ${\cal M}_{\rm wind}=3.5$ (see Figure \ref{plot:shield} in Section \ref{sec:discussion}).
    \item Hydrodynamic shielding also operates differently in adiabatic and radiative regimes. Adiabatic clouds lack an efficient energy release mechanism, so they are heated up and disrupted more effectively than radiative clouds. Radiative wind-multicloud systems have a multiphase structure characterised by the cold cloud gas, the warm mixed phase (formed via cloud KH shredding), and the hot wind phase. If the cooling length of the cold and warm phases is shorter than the cloud radius, strong cooling promotes the condensation of warm, mixed gas in between the clouds, facilitating hydrodynamic shielding by replenishing dense gas along the stream (see Section \ref{hs3}). The proximity of dense gas due to condensation increases the threshold separation for shielding to $\delta^{\rm sup, rad}_{\rm shield}\sim34$.
    \item The initial cloud separation distance, $\delta$, determines the efficiency of hydrodynamic shielding. While large separation distances facilitate mixing and dense-gas destruction via dynamical instabilities, small separation distances between clouds favour hydrodynamic shielding. Clouds in closely-spaced arrangements experience reduced drag forces (i.e. smaller accelerations) and have smaller mixing regions around them (as measured by $\xi$, see Section \ref{hs5}). The transition between shielding and no-shielding scenarios across different $\delta$ values is smooth in radiative models, as opposed to their adiabatic counterparts for which clouds need to be in close proximity.
    \item The dynamics and mixing processes of the multicloud layer are also sensitive to $\delta$. When clouds are closer together, larger column densities emerge, so the mass-weighted velocity of cloud material decreases and mixing is reduced. The farther apart the clouds initially are, the higher their acceleration and the generation of dynamic instabilities, which enhance mixing (see Section \ref{hs3} and \ref{hs4}). 
    \item Numerical resolution does not influence our results for the adiabatic models, as even the lowest resolution explored of $8$ cells per cloud radius ($R_{8}$) shows convergence. For the radiative models, we find that the fractions of dense and cold gas are sensitive to resolution, particularly for small separation $\delta$ values, which may be subjected to additional turbulence from dense gas collisions. Despite this, all our results hold when comparing models at a fixed resolution. 
\end{itemize}

Our wind-multicloud models reveal important physics relevant for galactic winds. Atop condensation, hydrodynamic shielding should also help preserve dense gas, particularly at the base of the wind-launching region, but possibly also further along the wind where gas re-condenses. In the future, we plan to incorporate magnetic fields and electron thermal conduction into our models.

\section*{Acknowledgements}

This paper is based on the thesis work of A. S. Villares (ASV) at Yachay Tech University (\url{http://repositorio.yachaytech.edu.ec/handle/123456789/657}). The authors gratefully acknowledge the Gauss Centre for Supercomputing e.V. (\url{www.gauss-centre.eu}) for funding this project by providing computing time (via grant pn34qu) on the GCS Supercomputer SuperMUC-NG at the Leibniz Supercomputing Centre (\url{www.lrz.de}). In addition, the authors thank CEDIA (\url{www.cedia.edu.ec}) for providing access to their HPC cluster as well as for their technical support. This work has made use of the VISIT visualisation software (\citealt{HPV:VisIt}). We also thank the developers of the {\sevensize PLUTO} code for making this hydrodynamic code available to the community. WEBB is supported by the National Secretariat of Higher Education, Science, Technology, and Innovation of Ecuador, SENESCYT.

\section*{Data Availability}
The data underlying this article will be shared on reasonable request to the corresponding authors.
 



\bibliographystyle{mnras}
\bibliography{mnras_template} 

\begin{thebibliography}{}
\makeatletter
\relax
\def\mn@urlcharsother{\let\do\@makeother \do\$\do\&\do\#\do\^\do\_\do\%\do\~}
\def\mn@doi{\begingroup\mn@urlcharsother \@ifnextchar [ {\mn@doi@}
  {\mn@doi@[]}}
\def\mn@doi@[#1]#2{\def\@tempa{#1}\ifx\@tempa\@empty \href
  {http://dx.doi.org/#2} {doi:#2}\else \href {http://dx.doi.org/#2} {#1}\fi
  \endgroup}
\def\mn@eprint#1#2{\mn@eprint@#1:#2::\@nil}
\def\mn@eprint@arXiv#1{\href {http://arxiv.org/abs/#1} {{\tt arXiv:#1}}}
\def\mn@eprint@dblp#1{\href {http://dblp.uni-trier.de/rec/bibtex/#1.xml}
  {dblp:#1}}
\def\mn@eprint@#1:#2:#3:#4\@nil{\def\@tempa {#1}\def\@tempb {#2}\def\@tempc
  {#3}\ifx \@tempc \@empty \let \@tempc \@tempb \let \@tempb \@tempa \fi \ifx
  \@tempb \@empty \def\@tempb {arXiv}\fi \@ifundefined
  {mn@eprint@\@tempb}{\@tempb:\@tempc}{\expandafter \expandafter \csname
  mn@eprint@\@tempb\endcsname \expandafter{\@tempc}}}

\bibitem[\protect\citeauthoryear{{Abruzzo}, {Fielding}  \& {Bryan}}{{Abruzzo}
  et~al.}{2022}]{2022arXiv221015679A}
{Abruzzo} M.~W.,  {Fielding} D.~B.,   {Bryan} G.~L.,  2022, \mn@doi [arXiv
  e-prints] {10.48550/arXiv.2210.15679}, \href
  {https://ui.adsabs.harvard.edu/abs/2022arXiv221015679A} {p. arXiv:2210.15679}

\bibitem[\protect\citeauthoryear{{Al{\={u}}zas}, {Pittard}, {Hartquist},
  {Falle}  \& {Langton}}{{Al{\={u}}zas} et~al.}{2012}]{Aluzas_12}
{Al{\={u}}zas} R.,  {Pittard} J.~M.,  {Hartquist} T.~W.,  {Falle} S.~A.~E.~G.,
   {Langton} R.,  2012, \mn@doi [\mnras] {10.1111/j.1365-2966.2012.21598.x},
  \href {https://ui.adsabs.harvard.edu/abs/2012MNRAS.425.2212A} {425, 2212}

\bibitem[\protect\citeauthoryear{{Al{\={u}}zas}, {Pittard}, {Falle}  \&
  {Hartquist}}{{Al{\={u}}zas} et~al.}{2014}]{Aluzas_14}
{Al{\={u}}zas} R.,  {Pittard} J.~M.,  {Falle} S.~A.~E.~G.,   {Hartquist} T.~W.,
   2014, \mn@doi [\mnras] {10.1093/mnras/stu1501}, \href
  {https://ui.adsabs.harvard.edu/abs/2014MNRAS.444..971A} {444, 971}

\bibitem[\protect\citeauthoryear{{Banda-Barrag{\'a}n}, {Parkin}, {Federrath},
  {Crocker}  \& {Bicknell}}{{Banda-Barrag{\'a}n} et~al.}{2016}]{Banda_16}
{Banda-Barrag{\'a}n} W.~E.,  {Parkin} E.~R.,  {Federrath} C.,  {Crocker} R.~M.,
    {Bicknell} G.~V.,  2016, \mn@doi [\mnras] {10.1093/mnras/stv2405}, \href
  {https://ui.adsabs.harvard.edu/abs/2016MNRAS.455.1309B} {455, 1309}

\bibitem[\protect\citeauthoryear{{Banda-Barrag{\'a}n}, {Federrath}, {Crocker}
  \& {Bicknell}}{{Banda-Barrag{\'a}n} et~al.}{2018}]{Banda_2018}
{Banda-Barrag{\'a}n} W.~E.,  {Federrath} C.,  {Crocker} R.~M.,   {Bicknell}
  G.~V.,  2018, \mn@doi [\mnras] {10.1093/mnras/stx2541}, \href
  {https://ui.adsabs.harvard.edu/abs/2018MNRAS.473.3454B} {473, 3454}

\bibitem[\protect\citeauthoryear{{Banda-Barrag{\'a}n}, {Zertuche}, {Federrath},
  {Garc{\'\i}a Del Valle}, {Br{\"u}ggen}  \& {Wagner}}{{Banda-Barrag{\'a}n}
  et~al.}{2019}]{2019MNRAS.486.4526B}
{Banda-Barrag{\'a}n} W.~E.,  {Zertuche} F.~J.,  {Federrath} C.,  {Garc{\'\i}a
  Del Valle} J.,  {Br{\"u}ggen} M.,   {Wagner} A.~Y.,  2019, \mn@doi [\mnras]
  {10.1093/mnras/stz1040}, \href
  {https://ui.adsabs.harvard.edu/abs/2019MNRAS.486.4526B} {486, 4526}

\bibitem[\protect\citeauthoryear{Banda-Barragán, Brüggen, Federrath, Wagner,
  Scannapieco  \& Cottle}{Banda-Barragán et~al.}{2020}]{Banda_20}
Banda-Barragán W.~E.,  Brüggen M.,  Federrath C.,  Wagner A.~Y.,  Scannapieco
  E.,   Cottle J.,  2020, \mn@doi [Monthly Notices of the Royal Astronomical
  Society] {10.1093/mnras/staa2904}, 499, 2173

\bibitem[\protect\citeauthoryear{Banda-Barragán, Brüggen, Heesen,
  Scannapieco, Cottle, Federrath  \& Wagner}{Banda-Barragán
  et~al.}{2021}]{Banda_21}
Banda-Barragán W.~E.,  Brüggen M.,  Heesen V.,  Scannapieco E.,  Cottle J.,
  Federrath C.,   Wagner A.~Y.,  2021, \mn@doi [Monthly Notices of the Royal
  Astronomical Society] {10.1093/mnras/stab1884}, 506, 5658

\bibitem[\protect\citeauthoryear{{Ben Bekhti}, {Richter}, {Westmeier}  \&
  {Murphy}}{{Ben Bekhti} et~al.}{2008}]{2008A&A...487..583B}
{Ben Bekhti} N.,  {Richter} P.,  {Westmeier} T.,   {Murphy} M.~T.,  2008,
  \mn@doi [\aap] {10.1051/0004-6361:20079067}, \href
  {https://ui.adsabs.harvard.edu/abs/2008A&A...487..583B} {487, 583}

\bibitem[\protect\citeauthoryear{{Borthakur} et~al.,}{{Borthakur}
  et~al.}{2015}]{2015ApJ...813...46B}
{Borthakur} S.,  et~al., 2015, \mn@doi [\apj] {10.1088/0004-637X/813/1/46},
  \href {https://ui.adsabs.harvard.edu/abs/2015ApJ...813...46B} {813, 46}

\bibitem[\protect\citeauthoryear{{Br{\"u}ggen} \& {Scannapieco}}{{Br{\"u}ggen}
  \& {Scannapieco}}{2016}]{Bruggen_16}
{Br{\"u}ggen} M.,  {Scannapieco} E.,  2016, \mn@doi [\apj]
  {10.3847/0004-637X/822/1/31}, \href
  {https://ui.adsabs.harvard.edu/abs/2016ApJ...822...31B} {822, 31}

\bibitem[\protect\citeauthoryear{{Casavecchia}, {Banda-Barrag{\'a}n},
  {Br{\"u}ggen}  \& {Brighenti}}{{Casavecchia} et~al.}{2023}]{Casavecchia_23}
{Casavecchia} B.,  {Banda-Barrag{\'a}n} W.~E.,  {Br{\"u}ggen} M.,   {Brighenti}
  F.,  2023, \mn@doi [IAU Symposium] {10.1017/S1743921322001211}, \href
  {https://ui.adsabs.harvard.edu/abs/2023IAUS..362...56C} {362, 56}

\bibitem[\protect\citeauthoryear{Chandrasekhar}{Chandrasekhar}{1961}]{chandrasekhar1961}
Chandrasekhar S.,  1961, Hydrodynamic and Hydromagnetic Stability, By S.
  Chandrasekhar.
International series of monographs on physics, Oxford University Press'
  International Series of Monographs on Physics, \url
  {https://books.google.com.ec/books?id=nyfucQAACAAJ}

\bibitem[\protect\citeauthoryear{{Chevalier} \& {Clegg}}{{Chevalier} \&
  {Clegg}}{1985}]{Chevalier}
{Chevalier} R.~A.,  {Clegg} A.~W.,  1985, \mn@doi [\nat] {10.1038/317044a0},
  \href {https://ui.adsabs.harvard.edu/abs/1985Natur.317...44C} {317, 44}

\bibitem[\protect\citeauthoryear{Childs et~al.,}{Childs
  et~al.}{2012}]{HPV:VisIt}
Childs H.,  et~al., 2012, in , {High Performance Visualization--Enabling
  Extreme-Scale Scientific Insight}.
"Lawrence Berkeley National Laboratory Scientific Data", pp 357--372

\bibitem[\protect\citeauthoryear{{Choi}, {Kim}  \& {Chung}}{{Choi}
  et~al.}{2022}]{2022ApJ...936..133C}
{Choi} W.,  {Kim} C.-G.,   {Chung} A.,  2022, \mn@doi [\apj]
  {10.3847/1538-4357/ac82ba}, \href
  {https://ui.adsabs.harvard.edu/abs/2022ApJ...936..133C} {936, 133}

\bibitem[\protect\citeauthoryear{{Cooper}, {Bicknell}, {Sutherland}  \&
  {Bland-Hawthorn}}{{Cooper} et~al.}{2008}]{2008ApJ...674..157C}
{Cooper} J.~L.,  {Bicknell} G.~V.,  {Sutherland} R.~S.,   {Bland-Hawthorn} J.,
  2008, \mn@doi [\apj] {10.1086/524918}, \href
  {https://ui.adsabs.harvard.edu/abs/2008ApJ...674..157C} {674, 157}

\bibitem[\protect\citeauthoryear{{Cooper}, {Bicknell}, {Sutherland}  \&
  {Bland-Hawthorn}}{{Cooper} et~al.}{2009}]{Cooper_2009}
{Cooper} J.~L.,  {Bicknell} G.~V.,  {Sutherland} R.~S.,   {Bland-Hawthorn} J.,
  2009, \mn@doi [\apj] {10.1088/0004-637X/703/1/330}, \href
  {https://ui.adsabs.harvard.edu/abs/2009ApJ...703..330C} {703, 330}

\bibitem[\protect\citeauthoryear{{Cottle}, {Scannapieco}, {Br{\"u}ggen},
  {Banda-Barrag{\'a}n}  \& {Federrath}}{{Cottle} et~al.}{2020}]{Cottle_20}
{Cottle} J.,  {Scannapieco} E.,  {Br{\"u}ggen} M.,  {Banda-Barrag{\'a}n} W.,
  {Federrath} C.,  2020, \mn@doi [\apj] {10.3847/1538-4357/ab76d1}, \href
  {https://ui.adsabs.harvard.edu/abs/2020ApJ...892...59C} {892, 59}

\bibitem[\protect\citeauthoryear{{Di Teodoro}, {McClure-Griffiths}, {Lockman}
  \& {Armillotta}}{{Di Teodoro} et~al.}{2020}]{2020Natur.584..364D}
{Di Teodoro} E.~M.,  {McClure-Griffiths} N.~M.,  {Lockman} F.~J.,
  {Armillotta} L.,  2020, \mn@doi [\nat] {10.1038/s41586-020-2595-z}, \href
  {https://ui.adsabs.harvard.edu/abs/2020Natur.584..364D} {584, 364}

\bibitem[\protect\citeauthoryear{{Faucher-Gigu{\`e}re} \&
  {Oh}}{{Faucher-Gigu{\`e}re} \& {Oh}}{2023}]{2023ARA&A..61..131F}
{Faucher-Gigu{\`e}re} C.-A.,  {Oh} S.~P.,  2023, \mn@doi [\araa]
  {10.1146/annurev-astro-052920-125203}, \href
  {https://ui.adsabs.harvard.edu/abs/2023ARA&A..61..131F} {61, 131}

\bibitem[\protect\citeauthoryear{{Ferland}, {Korista}, {Verner}, {Ferguson},
  {Kingdon}  \& {Verner}}{{Ferland} et~al.}{1998}]{1998PASP..110..761F}
{Ferland} G.~J.,  {Korista} K.~T.,  {Verner} D.~A.,  {Ferguson} J.~W.,
  {Kingdon} J.~B.,   {Verner} E.~M.,  1998, \mn@doi [\pasp] {10.1086/316190},
  \href {https://ui.adsabs.harvard.edu/abs/1998PASP..110..761F} {110, 761}

\bibitem[\protect\citeauthoryear{{Fielding}, {Ostriker}, {Bryan}  \&
  {Jermyn}}{{Fielding} et~al.}{2020}]{2020ApJ...894L..24F}
{Fielding} D.~B.,  {Ostriker} E.~C.,  {Bryan} G.~L.,   {Jermyn} A.~S.,  2020,
  \mn@doi [\apjl] {10.3847/2041-8213/ab8d2c}, \href
  {https://ui.adsabs.harvard.edu/abs/2020ApJ...894L..24F} {894, L24}

\bibitem[\protect\citeauthoryear{Forbes \& Lin}{Forbes \&
  Lin}{2019}]{Forbes_2019}
Forbes J.~C.,  Lin D. N.~C.,  2019, \mn@doi [The Astronomical Journal]
  {10.3847/1538-3881/ab3230}, 158, 124

\bibitem[\protect\citeauthoryear{{Fragile}, {Murray}, {Anninos}  \& {van
  Breugel}}{{Fragile} et~al.}{2004}]{2004ApJ...604...74F}
{Fragile} P.~C.,  {Murray} S.~D.,  {Anninos} P.,   {van Breugel} W.,  2004,
  \mn@doi [\apj] {10.1086/381726}, \href
  {https://ui.adsabs.harvard.edu/abs/2004ApJ...604...74F} {604, 74}

\bibitem[\protect\citeauthoryear{{Gatto} et~al.,}{{Gatto}
  et~al.}{2017}]{2017MNRAS.466.1903G}
{Gatto} A.,  et~al., 2017, \mn@doi [\mnras] {10.1093/mnras/stw3209}, \href
  {https://ui.adsabs.harvard.edu/abs/2017MNRAS.466.1903G} {466, 1903}

\bibitem[\protect\citeauthoryear{{Goldsmith} \& {Pittard}}{{Goldsmith} \&
  {Pittard}}{2017}]{2017MNRAS.470.2427G}
{Goldsmith} K.~J.~A.,  {Pittard} J.~M.,  2017, \mn@doi [\mnras]
  {10.1093/mnras/stx1431}, \href
  {https://ui.adsabs.harvard.edu/abs/2017MNRAS.470.2427G} {470, 2427}

\bibitem[\protect\citeauthoryear{Gregori, Miniati, Ryu  \& Jones}{Gregori
  et~al.}{2000}]{Gregori_2000}
Gregori G.,  Miniati F.,  Ryu D.,   Jones T.~W.,  2000, \mn@doi [The
  Astrophysical Journal] {10.1086/317130}, 543, 775

\bibitem[\protect\citeauthoryear{{Gronke} \& {Oh}}{{Gronke} \&
  {Oh}}{2018}]{2018MNRAS.480L.111G}
{Gronke} M.,  {Oh} S.~P.,  2018, \mn@doi [\mnras] {10.1093/mnrasl/sly131},
  \href {https://ui.adsabs.harvard.edu/abs/2018MNRAS.480L.111G} {480, L111}

\bibitem[\protect\citeauthoryear{{Gronke} \& {Oh}}{{Gronke} \&
  {Oh}}{2020a}]{2020MNRAS.492.1970G}
{Gronke} M.,  {Oh} S.~P.,  2020a, \mn@doi [\mnras] {10.1093/mnras/stz3332},
  \href {https://ui.adsabs.harvard.edu/abs/2020MNRAS.492.1970G} {492, 1970}

\bibitem[\protect\citeauthoryear{{Gronke} \& {Oh}}{{Gronke} \&
  {Oh}}{2020b}]{2020MNRAS.494L..27G}
{Gronke} M.,  {Oh} S.~P.,  2020b, \mn@doi [\mnras] {10.1093/mnrasl/slaa033},
  \href {https://ui.adsabs.harvard.edu/abs/2020MNRAS.494L..27G} {494, L27}

\bibitem[\protect\citeauthoryear{{Gronke}, {Oh}, {Ji}  \& {Norman}}{{Gronke}
  et~al.}{2022}]{2022MNRAS.511..859G}
{Gronke} M.,  {Oh} S.~P.,  {Ji} S.,   {Norman} C.,  2022, \mn@doi [\mnras]
  {10.1093/mnras/stab3351}, \href
  {https://ui.adsabs.harvard.edu/abs/2022MNRAS.511..859G} {511, 859}

\bibitem[\protect\citeauthoryear{{Gr{\o}nnow}, {Tepper-Garc{\'\i}a},
  {Bland-Hawthorn}  \& {McClure-Griffiths}}{{Gr{\o}nnow}
  et~al.}{2017}]{2017ApJ...845...69G}
{Gr{\o}nnow} A.,  {Tepper-Garc{\'\i}a} T.,  {Bland-Hawthorn} J.,
  {McClure-Griffiths} N.~M.,  2017, \mn@doi [\apj] {10.3847/1538-4357/aa7ed2},
  \href {https://ui.adsabs.harvard.edu/abs/2017ApJ...845...69G} {845, 69}

\bibitem[\protect\citeauthoryear{{Gr{\o}nnow}, {Tepper-Garc{\'\i}a}  \&
  {Bland-Hawthorn}}{{Gr{\o}nnow} et~al.}{2018}]{2018ApJ...865...64G}
{Gr{\o}nnow} A.,  {Tepper-Garc{\'\i}a} T.,   {Bland-Hawthorn} J.,  2018,
  \mn@doi [\apj] {10.3847/1538-4357/aada0e}, \href
  {https://ui.adsabs.harvard.edu/abs/2018ApJ...865...64G} {865, 64}

\bibitem[\protect\citeauthoryear{Heyer et~al.,}{Heyer
  et~al.}{2022}]{Heyer_2022}
Heyer M.,  et~al., 2022, \mn@doi [The Astrophysical Journal]
  {10.3847/1538-4357/ac67ea}, 930, 170

\bibitem[\protect\citeauthoryear{{Ji}, {Oh}  \& {Masterson}}{{Ji}
  et~al.}{2019}]{2019MNRAS.487..737J}
{Ji} S.,  {Oh} S.~P.,   {Masterson} P.,  2019, \mn@doi [\mnras]
  {10.1093/mnras/stz1248}, \href
  {https://ui.adsabs.harvard.edu/abs/2019MNRAS.487..737J} {487, 737}

\bibitem[\protect\citeauthoryear{{Johansson} \& {Ziegler}}{{Johansson} \&
  {Ziegler}}{2013}]{2013ApJ...766...45J}
{Johansson} E. P.~G.,  {Ziegler} U.,  2013, \mn@doi [\apj]
  {10.1088/0004-637X/766/1/45}, \href
  {https://ui.adsabs.harvard.edu/abs/2013ApJ...766...45J} {766, 45}

\bibitem[\protect\citeauthoryear{{Jung}, {Gr{\o}nnow}  \&
  {McClure-Griffiths}}{{Jung} et~al.}{2023}]{2023MNRAS.522.4161J}
{Jung} S.~L.,  {Gr{\o}nnow} A.,   {McClure-Griffiths} N.~M.,  2023, \mn@doi
  [\mnras] {10.1093/mnras/stad1236}, \href
  {https://ui.adsabs.harvard.edu/abs/2023MNRAS.522.4161J} {522, 4161}

\bibitem[\protect\citeauthoryear{{Kanjilal}, {Dutta}  \& {Sharma}}{{Kanjilal}
  et~al.}{2021}]{2021MNRAS.501.1143K}
{Kanjilal} V.,  {Dutta} A.,   {Sharma} P.,  2021, \mn@doi [\mnras]
  {10.1093/mnras/staa3610}, \href
  {https://ui.adsabs.harvard.edu/abs/2021MNRAS.501.1143K} {501, 1143}

\bibitem[\protect\citeauthoryear{{Kim} et~al.,}{{Kim}
  et~al.}{2020}]{2020ApJ...903L..34K}
{Kim} C.-G.,  et~al., 2020, \mn@doi [\apjl] {10.3847/2041-8213/abc252}, \href
  {https://ui.adsabs.harvard.edu/abs/2020ApJ...903L..34K} {903, L34}

\bibitem[\protect\citeauthoryear{{Klein}, {McKee}  \& {Colella}}{{Klein}
  et~al.}{1994}]{Klein_94}
{Klein} R.~I.,  {McKee} C.~F.,   {Colella} P.,  1994, \mn@doi [\apj]
  {10.1086/173554}, \href
  {https://ui.adsabs.harvard.edu/abs/1994ApJ...420..213K} {420, 213}

\bibitem[\protect\citeauthoryear{{Lecoanet} et~al.,}{{Lecoanet}
  et~al.}{2016}]{2016MNRAS.455.4274L}
{Lecoanet} D.,  et~al., 2016, \mn@doi [\mnras] {10.1093/mnras/stv2564}, \href
  {https://ui.adsabs.harvard.edu/abs/2016MNRAS.455.4274L} {455, 4274}

\bibitem[\protect\citeauthoryear{{Levy} et~al.,}{{Levy}
  et~al.}{2023}]{2023ApJ...958..109L}
{Levy} R.~C.,  et~al., 2023, \mn@doi [\apj] {10.3847/1538-4357/acff6e}, \href
  {https://ui.adsabs.harvard.edu/abs/2023ApJ...958..109L} {958, 109}

\bibitem[\protect\citeauthoryear{{Li}, {Frank}  \& {Blackman}}{{Li}
  et~al.}{2014}]{2014MNRAS.444.2884L}
{Li} S.,  {Frank} A.,   {Blackman} E.~G.,  2014, \mn@doi [\mnras]
  {10.1093/mnras/stu1571}, \href
  {https://ui.adsabs.harvard.edu/abs/2014MNRAS.444.2884L} {444, 2884}

\bibitem[\protect\citeauthoryear{{Li}, {Hopkins}, {Squire}  \& {Hummels}}{{Li}
  et~al.}{2020}]{2020MNRAS.492.1841L}
{Li} Z.,  {Hopkins} P.~F.,  {Squire} J.,   {Hummels} C.,  2020, \mn@doi
  [\mnras] {10.1093/mnras/stz3567}, \href
  {https://ui.adsabs.harvard.edu/abs/2020MNRAS.492.1841L} {492, 1841}

\bibitem[\protect\citeauthoryear{{Li} et~al.,}{{Li}
  et~al.}{2022}]{2022ApJ...933..139L}
{Li} F.,  et~al., 2022, \mn@doi [\apj] {10.3847/1538-4357/ac7526}, \href
  {https://ui.adsabs.harvard.edu/abs/2022ApJ...933..139L} {933, 139}

\bibitem[\protect\citeauthoryear{{Marinacci}, {Binney}, {Fraternali}, {Nipoti},
  {Ciotti}  \& {Londrillo}}{{Marinacci} et~al.}{2010}]{2010MNRAS.404.1464M}
{Marinacci} F.,  {Binney} J.,  {Fraternali} F.,  {Nipoti} C.,  {Ciotti} L.,
  {Londrillo} P.,  2010, \mn@doi [\mnras] {10.1111/j.1365-2966.2010.16352.x},
  \href {https://ui.adsabs.harvard.edu/abs/2010MNRAS.404.1464M} {404, 1464}

\bibitem[\protect\citeauthoryear{{McClure-Griffiths}, {Green}, {Hill},
  {Lockman}, {Dickey}, {Gaensler}  \& {Green}}{{McClure-Griffiths}
  et~al.}{2013}]{2013ApJ...770L...4M}
{McClure-Griffiths} N.~M.,  {Green} J.~A.,  {Hill} A.~S.,  {Lockman} F.~J.,
  {Dickey} J.~M.,  {Gaensler} B.~M.,   {Green} A.~J.,  2013, \mn@doi [\apjl]
  {10.1088/2041-8205/770/1/L4}, \href
  {https://ui.adsabs.harvard.edu/abs/2013ApJ...770L...4M} {770, L4}

\bibitem[\protect\citeauthoryear{{McClure-Griffiths}
  et~al.,}{{McClure-Griffiths} et~al.}{2018}]{2018NatAs...2..901M}
{McClure-Griffiths} N.~M.,  et~al., 2018, \mn@doi [Nature Astronomy]
  {10.1038/s41550-018-0608-8}, \href
  {https://ui.adsabs.harvard.edu/abs/2018NatAs...2..901M} {2, 901}

\bibitem[\protect\citeauthoryear{{McCourt}, {Oh}, {O'Leary}  \&
  {Madigan}}{{McCourt} et~al.}{2018}]{2018MNRAS.473.5407M}
{McCourt} M.,  {Oh} S.~P.,  {O'Leary} R.,   {Madigan} A.-M.,  2018, \mn@doi
  [\mnras] {10.1093/mnras/stx2687}, \href
  {https://ui.adsabs.harvard.edu/abs/2018MNRAS.473.5407M} {473, 5407}

\bibitem[\protect\citeauthoryear{{McPherson} et~al.,}{{McPherson}
  et~al.}{2023}]{2023MNRAS.525.6170M}
{McPherson} D.~K.,  et~al., 2023, \mn@doi [\mnras] {10.1093/mnras/stad2685},
  \href {https://ui.adsabs.harvard.edu/abs/2023MNRAS.525.6170M} {525, 6170}

\bibitem[\protect\citeauthoryear{{Mignone}}{{Mignone}}{2014}]{Mignone_14}
{Mignone} A.,  2014, \mn@doi [Journal of Computational Physics]
  {10.1016/j.jcp.2014.04.001}, \href
  {https://ui.adsabs.harvard.edu/abs/2014JCoPh.270..784M} {270, 784}

\bibitem[\protect\citeauthoryear{Mignone, Bodo, Massaglia, Matsakos, Tesileanu,
  Zanni  \& Ferrari}{Mignone et~al.}{2007}]{Mignone_2007}
Mignone A.,  Bodo G.,  Massaglia S.,  Matsakos T.,  Tesileanu O.,  Zanni C.,
  Ferrari A.,  2007, \mn@doi [The Astrophysical Journal Supplement Series]
  {10.1086/513316}, 170, 228

\bibitem[\protect\citeauthoryear{{Murray}, {White}, {Blondin}  \&
  {Lin}}{{Murray} et~al.}{1993}]{Murray_93}
{Murray} S.~D.,  {White} S. D.~M.,  {Blondin} J.~M.,   {Lin} D. N.~C.,  1993,
  \mn@doi [\apj] {10.1086/172540}, \href
  {https://ui.adsabs.harvard.edu/abs/1993ApJ...407..588M} {407, 588}

\bibitem[\protect\citeauthoryear{Nakamura, McKee, Klein  \& Fisher}{Nakamura
  et~al.}{2006}]{Nakamura_2006}
Nakamura F.,  McKee C.~F.,  Klein R.~I.,   Fisher R.~T.,  2006, \mn@doi [The
  Astrophysical Journal Supplement Series] {10.1086/501530}, 164, 477

\bibitem[\protect\citeauthoryear{{Orlando}, {Peres}, {Reale}, {Bocchino},
  {Rosner}, {Plewa}  \& {Siegel}}{{Orlando} et~al.}{2005}]{2005A&A...444..505O}
{Orlando} S.,  {Peres} G.,  {Reale} F.,  {Bocchino} F.,  {Rosner} R.,  {Plewa}
  T.,   {Siegel} A.,  2005, \mn@doi [\aap] {10.1051/0004-6361:20052896}, \href
  {http://adsabs.harvard.edu/abs/2005A%26A...444..505O} {444, 505}

\bibitem[\protect\citeauthoryear{{Pfuhl} et~al.,}{{Pfuhl}
  et~al.}{2015}]{Pfuhl_2015}
{Pfuhl} O.,  et~al., 2015, \mn@doi [\apj] {10.1088/0004-637X/798/2/111}, \href
  {https://ui.adsabs.harvard.edu/abs/2015ApJ...798..111P} {798, 111}

\bibitem[\protect\citeauthoryear{{Pittard}}{{Pittard}}{2022}]{2022MNRAS.515.1815P}
{Pittard} J.~M.,  2022, \mn@doi [\mnras] {10.1093/mnras/stac1954}, \href
  {https://ui.adsabs.harvard.edu/abs/2022MNRAS.515.1815P} {515, 1815}

\bibitem[\protect\citeauthoryear{{Pittard} \& {Parkin}}{{Pittard} \&
  {Parkin}}{2016}]{2016MNRAS.457.4470P}
{Pittard} J.~M.,  {Parkin} E.~R.,  2016, \mn@doi [\mnras]
  {10.1093/mnras/stw025}, \href
  {https://ui.adsabs.harvard.edu/abs/2016MNRAS.457.4470P} {457, 4470}

\bibitem[\protect\citeauthoryear{{Ponti}, {Morris}, {Churazov}, {Heywood}  \&
  {Fender}}{{Ponti} et~al.}{2021}]{2021A&A...646A..66P}
{Ponti} G.,  {Morris} M.~R.,  {Churazov} E.,  {Heywood} I.,   {Fender} R.~P.,
  2021, \mn@doi [\aap] {10.1051/0004-6361/202039636}, \href
  {https://ui.adsabs.harvard.edu/abs/2021A&A...646A..66P} {646, A66}

\bibitem[\protect\citeauthoryear{{Rathjen}, {Naab}, {Walch}, {Seifried},
  {Girichidis}  \& {W{\"u}nsch}}{{Rathjen} et~al.}{2023}]{2023MNRAS.522.1843R}
{Rathjen} T.-E.,  {Naab} T.,  {Walch} S.,  {Seifried} D.,  {Girichidis} P.,
  {W{\"u}nsch} R.,  2023, \mn@doi [\mnras] {10.1093/mnras/stad1104}, \href
  {https://ui.adsabs.harvard.edu/abs/2023MNRAS.522.1843R} {522, 1843}

\bibitem[\protect\citeauthoryear{{Reichardt Chu} et~al.,}{{Reichardt Chu}
  et~al.}{2022}]{2022ApJ...941..163R}
{Reichardt Chu} B.,  et~al., 2022, \mn@doi [\apj] {10.3847/1538-4357/aca1bd},
  \href {https://ui.adsabs.harvard.edu/abs/2022ApJ...941..163R} {941, 163}

\bibitem[\protect\citeauthoryear{{Richter} et~al.,}{{Richter}
  et~al.}{2017}]{2017A&A...607A..48R}
{Richter} P.,  et~al., 2017, \mn@doi [\aap] {10.1051/0004-6361/201630081},
  \href {https://ui.adsabs.harvard.edu/abs/2017A&A...607A..48R} {607, A48}

\bibitem[\protect\citeauthoryear{{Roberts-Borsani}, {Saintonge}, {Masters}  \&
  {Stark}}{{Roberts-Borsani} et~al.}{2020}]{2020MNRAS.493.3081R}
{Roberts-Borsani} G.~W.,  {Saintonge} A.,  {Masters} K.~L.,   {Stark} D.~V.,
  2020, \mn@doi [\mnras] {10.1093/mnras/staa464}, \href
  {https://ui.adsabs.harvard.edu/abs/2020MNRAS.493.3081R} {493, 3081}

\bibitem[\protect\citeauthoryear{{Rubin} et~al.,}{{Rubin}
  et~al.}{2022}]{2022ApJ...936..171R}
{Rubin} K. H.~R.,  et~al., 2022, \mn@doi [\apj] {10.3847/1538-4357/ac7b88},
  \href {https://ui.adsabs.harvard.edu/abs/2022ApJ...936..171R} {936, 171}

\bibitem[\protect\citeauthoryear{{Rupke}}{{Rupke}}{2018}]{2018Galax...6..138R}
{Rupke} D.,  2018, \mn@doi [Galaxies] {10.3390/galaxies6040138}, \href
  {https://ui.adsabs.harvard.edu/abs/2018Galax...6..138R} {6, 138}

\bibitem[\protect\citeauthoryear{Salak, Tomiyasu, Nakai, Kuno, Miyamoto  \&
  Kaneko}{Salak et~al.}{2018}]{Salak_18}
Salak D.,  Tomiyasu Y.,  Nakai N.,  Kuno N.,  Miyamoto Y.,   Kaneko H.,  2018,
  \mn@doi [The Astrophysical Journal] {10.3847/1538-4357/aab2ac}, 856, 97

\bibitem[\protect\citeauthoryear{{Scannapieco} \& {Br{\"u}ggen}}{{Scannapieco}
  \& {Br{\"u}ggen}}{2015}]{2015ApJ...805..158S}
{Scannapieco} E.,  {Br{\"u}ggen} M.,  2015, \mn@doi [\apj]
  {10.1088/0004-637X/805/2/158}, \href
  {https://ui.adsabs.harvard.edu/abs/2015ApJ...805..158S} {805, 158}

\bibitem[\protect\citeauthoryear{{Schneider} \& {Robertson}}{{Schneider} \&
  {Robertson}}{2017}]{Schneider_17}
{Schneider} E.~E.,  {Robertson} B.~E.,  2017, \mn@doi [\apj]
  {10.3847/1538-4357/834/2/144}, \href
  {https://ui.adsabs.harvard.edu/abs/2017ApJ...834..144S} {834, 144}

\bibitem[\protect\citeauthoryear{{Schneider}, {Robertson}  \&
  {Thompson}}{{Schneider} et~al.}{2018}]{Schneider_18}
{Schneider} E.~E.,  {Robertson} B.~E.,   {Thompson} T.~A.,  2018, \mn@doi
  [\apj] {10.3847/1538-4357/aacce1}, \href
  {https://ui.adsabs.harvard.edu/abs/2018ApJ...862...56S} {862, 56}

\bibitem[\protect\citeauthoryear{{Schneider}, {Ostriker}, {Robertson}  \&
  {Thompson}}{{Schneider} et~al.}{2020}]{Schneider_20}
{Schneider} E.~E.,  {Ostriker} E.~C.,  {Robertson} B.~E.,   {Thompson} T.~A.,
  2020, \mn@doi [\apj] {10.3847/1538-4357/ab8ae8}, \href
  {https://ui.adsabs.harvard.edu/abs/2020ApJ...895...43S} {895, 43}

\bibitem[\protect\citeauthoryear{Shopbell \& Bland-Hawthorn}{Shopbell \&
  Bland-Hawthorn}{1998}]{Shopbell_98}
Shopbell P.~L.,  Bland-Hawthorn J.,  1998, \mn@doi [The Astrophysical Journal]
  {10.1086/305108}, 493, 129

\bibitem[\protect\citeauthoryear{{Sparre}, {Pfrommer}  \& {Ehlert}}{{Sparre}
  et~al.}{2020}]{2020MNRAS.499.4261S}
{Sparre} M.,  {Pfrommer} C.,   {Ehlert} K.,  2020, \mn@doi [\mnras]
  {10.1093/mnras/staa3177}, \href
  {https://ui.adsabs.harvard.edu/abs/2020MNRAS.499.4261S} {499, 4261}

\bibitem[\protect\citeauthoryear{{Strickland} \& {Stevens}}{{Strickland} \&
  {Stevens}}{2000}]{2000MNRAS.314..511S}
{Strickland} D.~K.,  {Stevens} I.~R.,  2000, \mn@doi [\mnras]
  {10.1046/j.1365-8711.2000.03391.x}, \href
  {https://ui.adsabs.harvard.edu/abs/2000MNRAS.314..511S} {314, 511}

\bibitem[\protect\citeauthoryear{{Tan} \& {Fielding}}{{Tan} \&
  {Fielding}}{2024}]{2024MNRAS.527.9683T}
{Tan} B.,  {Fielding} D.~B.,  2024, \mn@doi [\mnras] {10.1093/mnras/stad3793},
  \href {https://ui.adsabs.harvard.edu/abs/2024MNRAS.527.9683T} {527, 9683}

\bibitem[\protect\citeauthoryear{{Tchernyshyov} et~al.,}{{Tchernyshyov}
  et~al.}{2022}]{2022ApJ...927..147T}
{Tchernyshyov} K.,  et~al., 2022, \mn@doi [\apj] {10.3847/1538-4357/ac450c},
  \href {https://ui.adsabs.harvard.edu/abs/2022ApJ...927..147T} {927, 147}

\bibitem[\protect\citeauthoryear{{Te{\c{s}}ileanu}, {Mignone}  \&
  {Massaglia}}{{Te{\c{s}}ileanu} et~al.}{2008}]{2008A&A...488..429T}
{Te{\c{s}}ileanu} O.,  {Mignone} A.,   {Massaglia} S.,  2008, \mn@doi [\aap]
  {10.1051/0004-6361:200809461}, \href
  {https://ui.adsabs.harvard.edu/abs/2008A&A...488..429T} {488, 429}

\bibitem[\protect\citeauthoryear{Teodoro, McClure-Griffiths, Lockman, Denbo,
  Endsley, Ford  \& Harrington}{Teodoro et~al.}{2018}]{Di_18}
Teodoro E. M.~D.,  McClure-Griffiths N.~M.,  Lockman F.~J.,  Denbo S.~R.,
  Endsley R.,  Ford H.~A.,   Harrington K.,  2018, \mn@doi [The Astrophysical
  Journal] {10.3847/1538-4357/aaad6a}, 855, 33

\bibitem[\protect\citeauthoryear{Toro, Spruce  \& Speares}{Toro
  et~al.}{1994}]{toro1994}
Toro E.~F.,  Spruce M.,   Speares W.,  1994, Shock waves, 4, 25

\bibitem[\protect\citeauthoryear{Tripp et~al.,}{Tripp et~al.}{2011}]{Tripp_11}
Tripp T.~M.,  et~al., 2011, \mn@doi [Science] {10.1126/science.1209850}, 334,
  952

\bibitem[\protect\citeauthoryear{Tumlinson et~al.,}{Tumlinson
  et~al.}{2011}]{Tumlinson_11}
Tumlinson J.,  et~al., 2011, \mn@doi [Science] {10.1126/science.1209840}, 334,
  948

\bibitem[\protect\citeauthoryear{{Tumlinson}, {Peeples}  \& {Werk}}{{Tumlinson}
  et~al.}{2017}]{Tumlinson_17}
{Tumlinson} J.,  {Peeples} M.~S.,   {Werk} J.~K.,  2017, \mn@doi [\araa]
  {10.1146/annurev-astro-091916-055240}, \href
  {https://ui.adsabs.harvard.edu/abs/2017ARA&A..55..389T} {55, 389}

\bibitem[\protect\citeauthoryear{{Veena} et~al.,}{{Veena}
  et~al.}{2023}]{2023A&A...674L..15V}
{Veena} V.~S.,  et~al., 2023, \mn@doi [\aap] {10.1051/0004-6361/202346702},
  \href {https://ui.adsabs.harvard.edu/abs/2023A&A...674L..15V} {674, L15}

\bibitem[\protect\citeauthoryear{{Veilleux}, {Rupke}  \& {Swaters}}{{Veilleux}
  et~al.}{2009}]{2009ApJ...700L.149V}
{Veilleux} S.,  {Rupke} D. S.~N.,   {Swaters} R.,  2009, \mn@doi [\apjl]
  {10.1088/0004-637X/700/2/L149}, \href
  {https://ui.adsabs.harvard.edu/abs/2009ApJ...700L.149V} {700, L149}

\bibitem[\protect\citeauthoryear{{Veilleux}, {Maiolino}, {Bolatto}  \&
  {Aalto}}{{Veilleux} et~al.}{2020}]{2020A&ARv..28....2V}
{Veilleux} S.,  {Maiolino} R.,  {Bolatto} A.~D.,   {Aalto} S.,  2020, \mn@doi
  [\aapr] {10.1007/s00159-019-0121-9}, \href
  {https://ui.adsabs.harvard.edu/abs/2020A&ARv..28....2V} {28, 2}

\bibitem[\protect\citeauthoryear{{Veilleux} et~al.,}{{Veilleux}
  et~al.}{2021}]{2021MNRAS.508.4902V}
{Veilleux} S.,  et~al., 2021, \mn@doi [\mnras] {10.1093/mnras/stab2881}, \href
  {https://ui.adsabs.harvard.edu/abs/2021MNRAS.508.4902V} {508, 4902}

\bibitem[\protect\citeauthoryear{Villares}{Villares}{2023}]{Villares_2023}
Villares A.~S.,  2023, Unravelling the multi-scale and turbulent structure of
  galactic winds.
"Universidad Yachay Tech", \url {http://repositorio.yachaytech.
  edu.ec/handle/123456789/657}

\bibitem[\protect\citeauthoryear{{Westmoquette}, {Smith}  \&
  {Gallagher}}{{Westmoquette} et~al.}{2008}]{2008MNRAS.383..864W}
{Westmoquette} M.~S.,  {Smith} L.~J.,   {Gallagher} J.~S.,  2008, \mn@doi
  [\mnras] {10.1111/j.1365-2966.2007.12628.x}, \href
  {https://ui.adsabs.harvard.edu/abs/2008MNRAS.383..864W} {383, 864}

\bibitem[\protect\citeauthoryear{{Xu} \& {Stone}}{{Xu} \&
  {Stone}}{1995}]{1995ApJ...454..172X}
{Xu} J.,  {Stone} J.~M.,  1995, \mn@doi [\apj] {10.1086/176475}, \href
  {https://ui.adsabs.harvard.edu/abs/1995ApJ...454..172X} {454, 172}

\bibitem[\protect\citeauthoryear{{Yirak}, {Frank}  \& {Cunningham}}{{Yirak}
  et~al.}{2010}]{2010ApJ...722..412Y}
{Yirak} K.,  {Frank} A.,   {Cunningham} A.~J.,  2010, \mn@doi [\apj]
  {10.1088/0004-637X/722/1/412}, \href
  {https://ui.adsabs.harvard.edu/abs/2010ApJ...722..412Y} {722, 412}

\bibitem[\protect\citeauthoryear{{Zhang}, {Thompson}, {Quataert}  \&
  {Murray}}{{Zhang} et~al.}{2017}]{2017MNRAS.468.4801Z}
{Zhang} D.,  {Thompson} T.~A.,  {Quataert} E.,   {Murray} N.,  2017, \mn@doi
  [\mnras] {10.1093/mnras/stx822}, \href
  {https://ui.adsabs.harvard.edu/abs/2017MNRAS.468.4801Z} {468, 4801}

\bibitem[\protect\citeauthoryear{{Zhang}, {Li}  \& {Morris}}{{Zhang}
  et~al.}{2021}]{2021ApJ...913...68Z}
{Zhang} M.,  {Li} Z.,   {Morris} M.~R.,  2021, \mn@doi [\apj]
  {10.3847/1538-4357/abf927}, \href
  {https://ui.adsabs.harvard.edu/abs/2021ApJ...913...68Z} {913, 68}

\makeatother
\end{thebibliography}




\appendix

\section{CC85 Model}\label{a1}

The equations that describe the steady-state hydrodynamics of a hot wind are given by:

\begin{equation}\label{s1}
    \frac{1}{r^{2}} \frac{d}{dr} (\rho v r^{2}) = q,
\end{equation}
\begin{equation}\label{s2}
    v \frac{dv}{dr} = - \frac{1}{\rho} \frac{dP}{dr} - \frac{qv}{\rho},
\end{equation}
\begin{equation}\label{s3}
    \frac{1}{r^{2}} \frac{d}{dr}\left[\rho v r^{2} \left(\frac{1}{2}v^{2} + \frac{\gamma}{\gamma - 1} \frac{P}{\rho} \right) \right] = Q,
\end{equation}
where $\rho$ is the density, r is the radial radius, $v$ is the wind velocity, $\gamma$ is the adiabatic index, and $P$ is the pressure.

The solutions obtained from equations (\ref{s1}) to (\ref{s3}) can be expressed using the Mach number of the wind $M$, as follows:

\begin{equation}
  \begin{aligned}
    \left( \frac{3\gamma + 1/M^{2}}{1+3\gamma} \right)^{-(3\gamma+1)/(5\gamma +1)}
    \left(\frac{\gamma -1+2/M^{2}}{1+\gamma} \right)^{(\gamma+1)/\left[2(5\gamma+1)\right]}  =\\   \frac{r}{R}  \hspace{.25cm} \left(r<R\right),
  \end{aligned}
\end{equation}

\begin{equation}\label{2_5}
    M^{2/(\gamma-1)} \left(\frac{\gamma-1+2/M^{2}}{1+\gamma}\right)^{(\gamma+1)/\left[2(\gamma-1)\right]}   =   \left(\frac{r}{R}\right)^{2} \hspace{.25cm} (r\geq R),
\end{equation}

The solution for the wind velocity as a function of the radius of the wind $r$ can be represented in a dimensionless form by using the dimensionless radius, $r^{*} = r/R$. Additionally, the dimensionless velocity $u_{*}$, density $\rho_{*}$, and pressure $P_{*}$, are also employed in this representation as follows:

\begin{equation}\label{2_6}
    u   =   u_{*} \dot{M}_{hot}^{-1/2} \dot{E}_{hot}^{1/2},
\end{equation}
\begin{equation}
    \rho    =   \rho_{*} \dot{M}_{hot}^{3/2} \dot{E}_{hot}^{-1/2} R^{-2},
\end{equation}
\begin{equation}
    P   =   P_{*} \dot{M}_{hot}^{1/2} \dot{E}_{hot}^{1/2} R^{-2},
\end{equation}

where the injection rates of total mass and energy into the wind are represented by $\dot{M}_{\text{hot}}$ and $\dot{E}_{\text{hot}}$, respectively. The reader is referred to \cite{Villares_2023} for additional details on the analytical modelling.

\begin{figure}
\begin{center}
  \begin{tabular}{c}
    \multicolumn{1}{l}{\hspace{2mm}A1a) Mach Number}
    \\
       \hspace{-0.3cm}\resizebox{!}{50mm}{\includegraphics{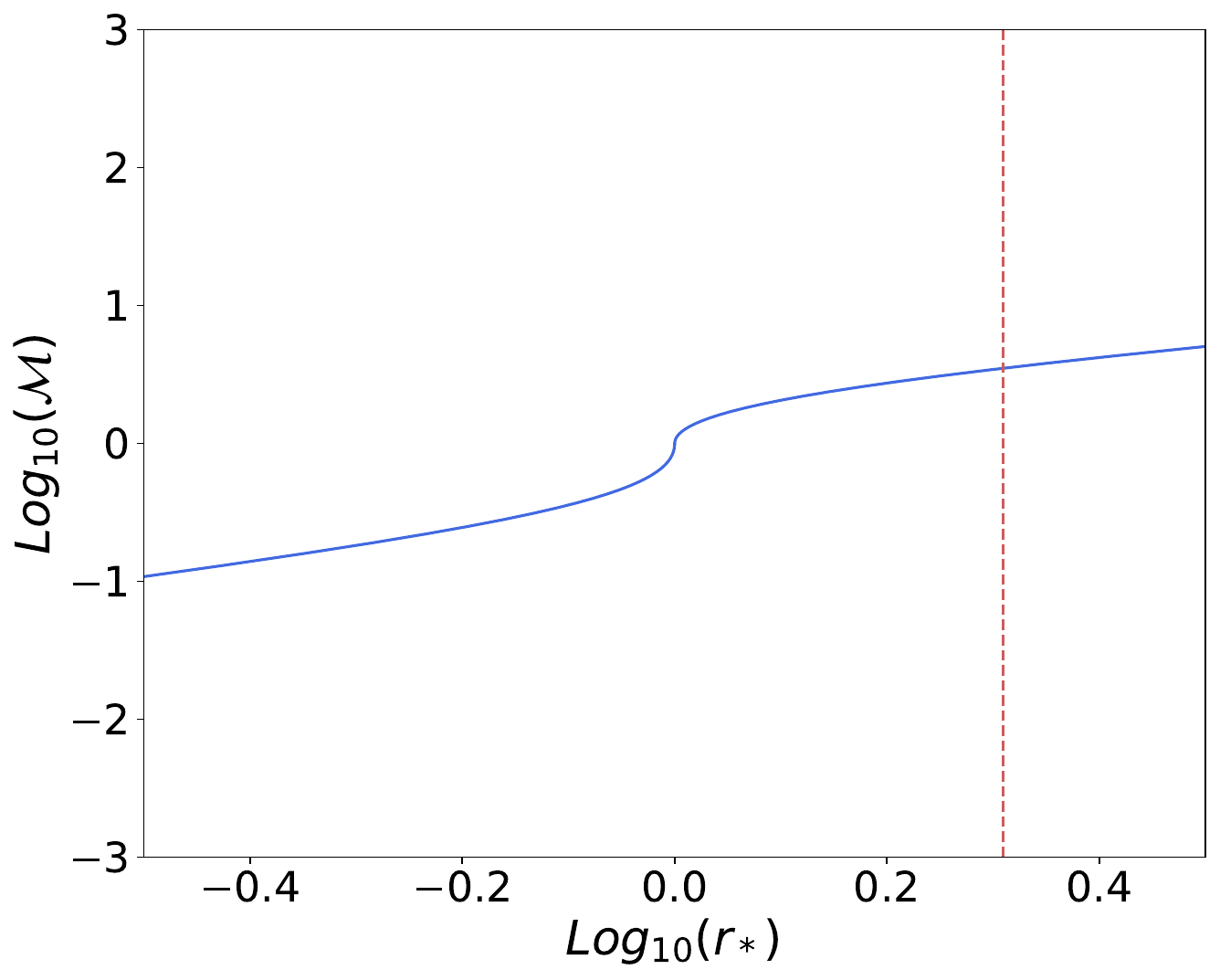}} 
       \\
    \multicolumn{1}{l}{\hspace{2mm}A1b) Dimensionless velocity}
      \\
       \hspace{-0.3cm}\resizebox{!}{50mm}{\includegraphics{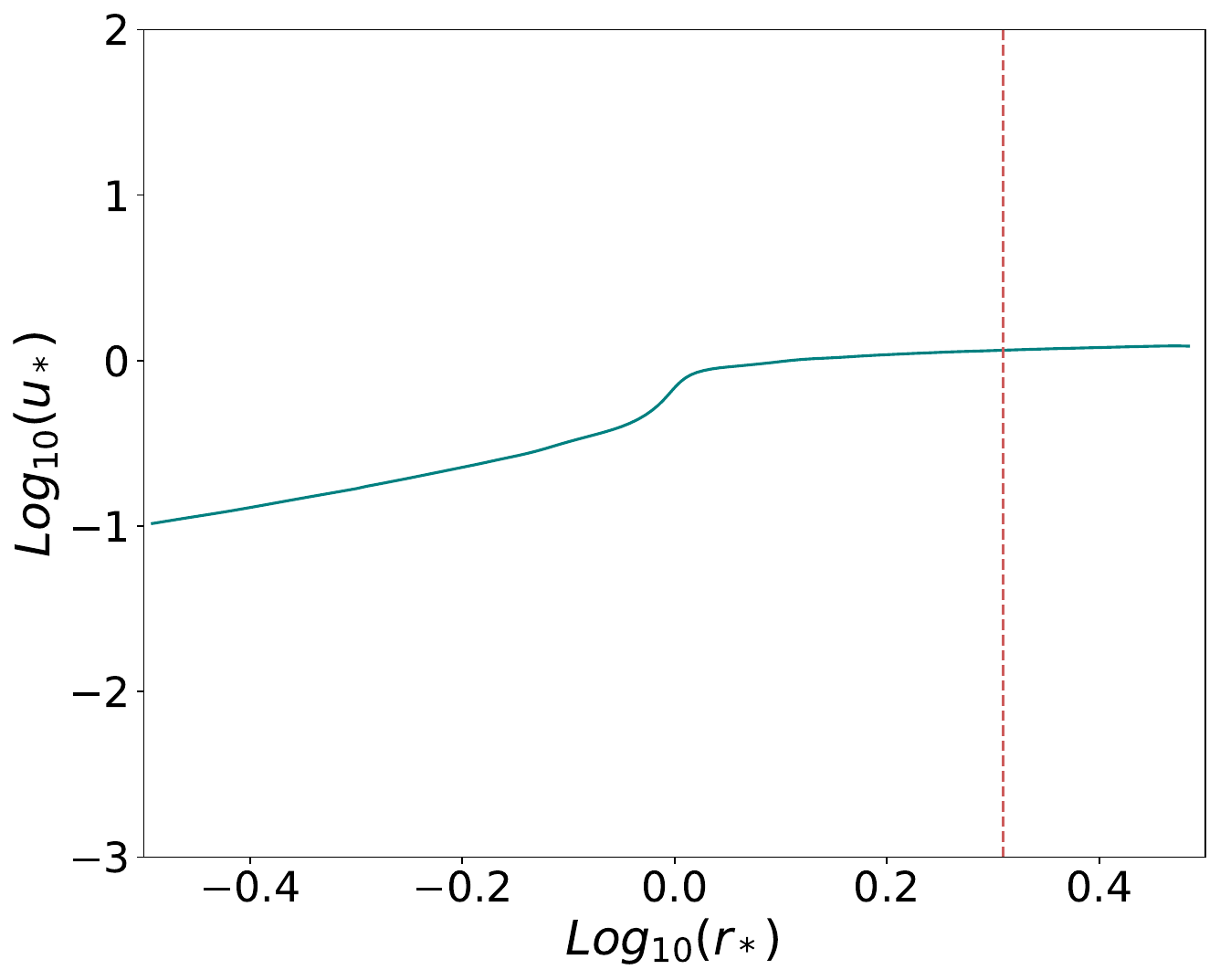}}
       \\
    \multicolumn{1}{l}{\hspace{2mm}A1c) Dimensionless density}
       \\
       \hspace{-0.3cm}\resizebox{!}{50mm}{\includegraphics{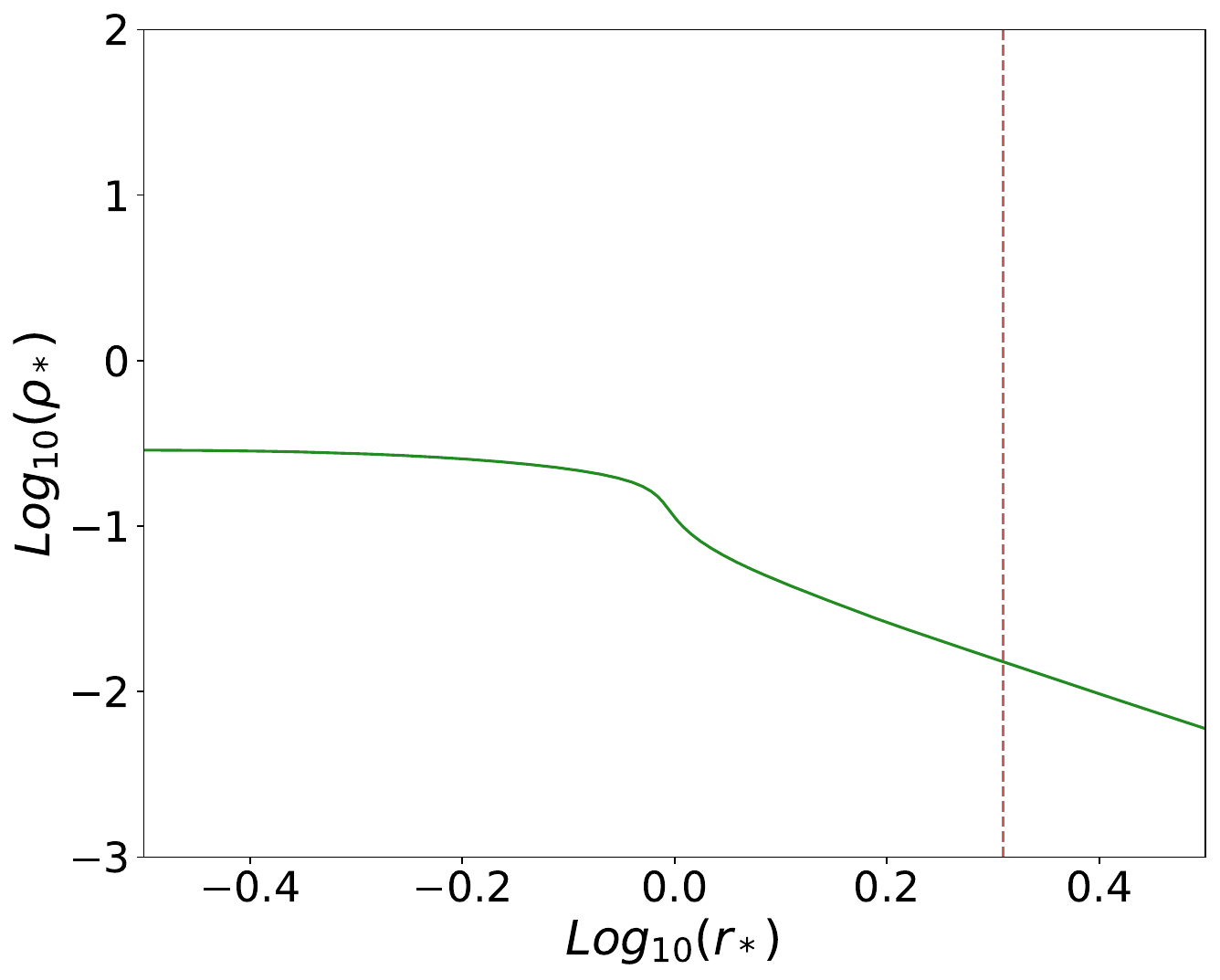}} 
  \end{tabular}
  \caption{Analytical solutions of the CC85 model for the wind Mach number (panel A1a), the dimensionless wind velocity (panel A1b), and the dimensionless wind density (panel A1d).} 
  \label{fa1}
\end{center}
\end{figure}

The panels of Figure \ref{fa1} show the steady-state wind solution for the CC85 model as a function of distance from the central starburst. The intersection between the dimensionless parameters and the red vertical dashed lines $(\rm \log_{10}(\rm r^{*})\,=\,0.3)$ represent the initial conditions chosen for the simulations (see Section \ref{models}).

\bsp	
\label{lastpage}
\end{document}